\documentclass[a4paper,fleqn,usenatbib]{mnras}

\usepackage{newtxtext,newtxmath}

\usepackage[T1]{fontenc}
\usepackage{ae,aecompl}

\usepackage{graphicx}	
\usepackage{amsmath}	
\usepackage{amssymb}	
\graphicspath{{./plots/}}

\newcommand{\msun}{M$_{\odot}$}

\newcommand{\kms}{km~s$^{-1}$}
\newcommand{\ergs}{erg s$^{-1}$}
\newcommand{\Ha}{H$\alpha$}

\newcommand{\HeI}{\ion{He}{I}}

\newcommand{\OI}{O~{\sc i}}
\newcommand{\Oneb}{\ion{[O}{I]}}
\newcommand{\CII}{C~{\sc ii}}
\newcommand{\AlII}{Al~{\sc ii}}
\newcommand{\NaI}{Na~{\sc i}}
\newcommand{\MgII}{Mg~{\sc ii}}
\newcommand{\MgI}{Mg~{\sc i}}
\newcommand{\SiI}{Si~{\sc i}}
\newcommand{\SiII}{Si~{\sc ii}}

\newcommand{\SI}{S~{\sc i}}
\newcommand{\SII}{S~{\sc ii}}
\newcommand{\CaII}{Ca~{\sc ii}}

\newcommand{\FeII}{Fe~{\sc ii}}

\newcommand{\CoII}{Co~{\sc ii}}

\newcommand{\Fefs}{$^{56}$Fe}
\newcommand{\Cofs}{$^{56}$Co}
\newcommand{\Nifs}{$^{56}$Ni}
\newcommand{\mej}{$M_\mathrm{ej}$}
\newcommand{\ke}{$E_\mathrm{k}$}
\newcommand{\ek}{$E_\mathrm{k}$}
\newcommand{\vph}{$v_\mathrm{ph}$}
\newcommand{\lp}{$L_\mathrm{p}$}
\newcommand{\trise}{$t_{-1/2}$}
\newcommand{\tdecay}{$t_{+1/2}$}
\newcommand{\eom}{$E_\mathrm{k}/M_\mathrm{ej}$}
\newcommand{\lam}{$\lambda$}
\newcommand{\loglp}{$\mathrm{log}_{10}\left( L_\mathrm{p}/\mathrm{erg~s}^{-1}\right)$}
\newcommand{\tmax}{$t_\mathrm{max}$}
\newcommand{\Eh}{$E\left(B-V\right)_\mathrm{host}$}
\newcommand{\Emw}{$E\left(B-V\right)_\mathrm{MW}$}
\newcommand{\Etot}{$E\left(B-V\right)_\mathrm{tot}$}

\newcommand{\mni}{$M_\mathrm{Ni}$}
\newcommand{\Mp}{$M_\mathrm{peak}$}

\newcommand{\tp}{$t_\mathrm{p}$}
\newcommand{\mzams}{$M_\mathrm{ZAMS}$}

\newcommand{\mN}{$\left<N\right>$}

\title[2016coi/ASASSN-16fp]{SN 2016coi/ASASSN-16fp: An example of residual helium in a type Ic supernova?}

\author[S. J. Prentice]{S. J. Prentice$^{1}$\thanks{E-mail: sipren.astro@gmail.com}, C. Ashall$^{1,17}$, P. A. Mazzali$^{1,2}$,  J.-J. Zhang$^{3,4,5}$, P. A. James$^{1}$,    
\newauthor X.-F. Wang$^{6}$, J. Vink\'o$^{10,12,13}$, S. Percival$^{1}$, L. Short$^{1}$, A. Piascik$^{1}$, F. Huang$^{6}$,  
\newauthor J. Mo$^{6}$, L.-M. Rui$^{6}$, J.-G. Wang$^{3,4,5}$, D.-F. Xiang$^{6}$, Y.-X. Xin$^{3,4,5}$, W.-M. Yi$^{3,4,5}$, 
\newauthor X.-G. Yu$^{3,4,5}$, Q. Zhai$^{3,4,5}$, T.-M. Zhang$^{7}$,   
G. Hosseinzadeh$^{8,9}$, D. A. Howell$^{8,9}$, C. \newauthor McCully$^{8,9}$, 
 S. Valenti$^{14}$,
B. Cseh$^{10}$, 
O. Hanyecz$^{10}$, 
L. Kriskovics$^{10}$, 
A. P\'al$^{10}$, 
\newauthor K. S\'arneczky$^{10}$, 
 \'A. S\'odor$^{10}$, 
R. Szak\'ats$^{10}$,
P. Sz\'ekely$^{11}$,
E. Varga-Vereb\'elyi$^{10}$,
\newauthor K. Vida$^{10}$, M. Bradac$^{14}$, D. E. Reichart$^{15}$, D. Sand$^{16}$, L. Tartaglia$^{14,16}$
\\
$^{1}$Astrophysics Research Institute, Liverpool John Moores University, IC2, Liverpool Science Park, 146 Brownlow Hill, \\  Liverpool L3 5RF, UK\\
$^{2}$Max-Planck-Institut f{\"u}r Astrophysik, Karl-Schwarzschild-Str. 1, D-85748 Garching, Germany\\
$^{3}$Yunnan Observatories (YNAO), Chinese Academy of Sciences, Kunming, 650216, China\\
$^{4}$Key Laboratory for the Structure and Evolution of Celestial Objects, Chinese Academy of Sciences, Kunming, 650216, China\\
$^{5}$Center for Astronomical Mega-Science, Chinese Academy of Sciences, 20A Datun Road, Chaoyang District, Beijing, 100012, China\\
$^{6}$Physics Department and Tsinghua Center for Astrophysics (THCA), Tsinghua University, Beijing 100084, China\\
$^{7}$National Astronomical Observatories of China (NAOC), Chinese Academy of Sciences, Beijing 100012, China\\
$^{8}$Las Cumbres Observatory, 6740 Cortona Dr. Suite 102, Goleta, CA, USA 93117\\
$^{9}$University of California, Santa Barbara, Department of Physics, Broida Hall, Santa Barbara, CA, USA 93111\\
$^{10}${Konkoly Observatory, Research Centre for Astronomy and Earth
Sciences, Budapest, Konkoly-Thege \'ut 15-17, 1121, Hungary}\\
$^{11}${Department of Experimental Physics, University of Szeged,
D\'om t\'er 9, Szeged, 6720 Hungary}\\
$^{12}${Department of Optics and Quantum Electronics, University of
Szeged, D\'om t\'er 9, Szeged, 6720 Hungary}\\
$^{13}${Department of Astronomy, University of Texas at Austin, 2515 Speedway, Austin, TX, 78712-1205 USA}\\
$^{14}$Department of Physics, University of California, Davis, CA 95
616, USA\\
$^{15}$University of North Carolina  269 Phillips Hall, CB $\#$3255  Chapel Hill, NC 27599 \\
$^{16}$Department of Astronomy/Steward Observatory, 933 North Cherry Avenue, Room N204, Tucson, AZ 85721-0065, USA\\
$^{17}$Department of Physics, Florida State University, Tallahassee, FL 32306, USA\\
}

\date{Accepted XXX. Received YYY; in original form ZZZ}

\pubyear{2018}

\begin{document}
\label{firstpage}
\pagerange{\pageref{firstpage}--\pageref{lastpage}}
\maketitle

\begin{abstract}
The optical observations of Ic-4 supernova (SN) 2016coi/ASASSN-16fp, from $\sim
2$ to $\sim450$ days after explosion, are presented along with analysis of its
physical properties.  The SN shows the broad lines associated with SNe Ic-3/4
but with a key difference. The early spectra display a strong absorption feature
at $\sim 5400$ \AA\ which is not seen in other SNe~Ic-3/4 at this epoch. This
feature has been attributed to \HeI\ in the literature.  Spectral modelling of
the SN in the early photospheric phase suggests the presence of residual He in a
C/O dominated shell. However, the behaviour of the \HeI\ lines is unusual when
compared with He-rich SNe, showing relatively low velocities and weakening rather than strengthening over
time.  The SN is found to rise to peak $\sim 16$ d after core-collapse reaching
a bolometric luminosity of \lp$\sim 3\times10^{42}$ \ergs.  Spectral models,
including the nebular epoch, show that the SN ejected $2.5-4$ \msun\ 
of material, with $\sim 1.5$ \msun\ below 5000 \kms, and with a kinetic energy
of $(4.5-7)\times10^{51}$ erg. The explosion synthesised $\sim 0.14$ \msun\ of
\Nifs. There are significant uncertainties in \Eh\ and the distance however, which will affect \lp\ and \mni.
SN 2016coi exploded in a host similar to the Large Magellanic Cloud
(LMC) and away from star-forming regions.  The properties of the SN and the
host-galaxy suggest that the progenitor had $M_\mathrm{ZAMS}$ of $23-28$ \msun\
and was stripped almost entirely down to its C/O core at explosion.\\

\end{abstract}

\begin{keywords}
supernovae: individual
\end{keywords}


\newpage
\section{Introduction}

In order for the death of a massive star to result in a stripped-envelope supernova (SE-SN) event \citep{Clocchiatti1997} the progenitor star must undergo a period of severe envelope stripping but how this mass loss occurs is not fully understood. There are three favoured mechanisms for envelope stripping. The first is through strong stellar winds, which are both metallicity and rotation dependent, and can results in mass-loss rates of $10^{-4} - 10^{-5} M_\odot$ yr$^{-1}$ \citep[e.g.,][]{Maeda2015,Langer2012}. However, stellar evolution models of single stars struggle to remove He to the upper limit given by \citet{Hachinger2012} for a He-poor SN. 

The second mechanism is episodic mass loss during a luminous blue variable (LBV) stage, where periodic pulsations eject a few \msun\ of stellar material \citep{Smith2003}. This requires a progenitor star of many 10s \msun \citep{Foley2011} and such stars may not lose enough of their H-/He envelopes before core-collapse \citep{EliasRosa2016}. 

The third is through binary interaction when a star transfers much of its mass to a donor through Roche-lobe overflow or the outer envelope is expelled during a common-envelope phase\citep{Nomoto1994,Pod1992}. This third mechanism is the most likely route for all but the most massive  progenitors of SE-SNe and allows progenitors of lower mass, which as single stars may explode as SNe IIP, to explode as SE-SN events. This reconciles the discrepancy between the relative rates of core-collapse SNe and mass-driven rates \cite[e.g,][]{Shivvers2017}.  

There have been detections of progenitor stars for a few SE-SNe, these are all He-rich \citep[See, for example,][]{Arcavi2011,VanDyk2014, Eldridge2015,Eldridge2016,Kilpatrick2017, Tartaglia2017}. In the case of SN 1993J \citep{Maund2004,Fox2014}, SN 2011dh \citep{Folatelli2014,Maund2015}, and SN 2001ig \citep{Ryder2018}, late time imaging of the explosion site has revealed evidence for companion stars. It is unknown whether these companions would have been close enough to affect the evolution of the SN progenitor. The progenitors of SNe Ic are not well understood. From single star evolution models they are expected to be Wolf-Rayet stars \citep[e.g.,][]{Georgy2012}. However, no confirmed progenitor has yet been seen in archival images placing strict limits on massive WR progenitors \citep{Smartt2009,Yoon2012}. It has been found that ejecta masses for these He-poor SNe range from $\sim 1$ \msun\
\citep{Sauer2006,Mazzali2010} to $\sim 13$ \msun\ \citep{Mazzali2006b}. This translates into a range of progenitor masses from $\sim15-50$ \msun. 
The lower end of this distribution is in the range of progenitors of SNe IIP, where observations suggest that \mzams $\sim 10-16$ \msun \citep{Smartt2009,Valenti2016} and theory predicts progenitors of up to 25 \msun. The discrepancy may be caused by an underestimate in amount of circumstellar extinction \citep{Beasor2016}.

In this work we present optical photometric and spectroscopic observations of
the nearby SN 2016coi/ASASSN-16fp. The SN is densely sampled between $\sim 20 -
200$ d after explosion with 55 spectroscopic observations making SN 2016coi one
of the best sampled SE-SNe to date, a consequence of its early discovery and
proximity. This SN was originally classified as a ``broad-lined'' type Ic SN but
\cite{Yamanaka2017} presented a case for the presence of He in the ejecta. Based
upon analytical analysis of the observational data they proposed a new
classification of SN 2016coi as a broad-lined Ib.  There have been previous
discussions of He in SNe Ic \citep[See, for example,][]{Filippenko1995,Taubenberger2006,Modjaz2014} and some claimed detections (e.g. SN 2012ap
\citep{Mili2015}, SN 2009bb \citep{Pignata2011}).  However, none of these claims
have provided conclusive proof, indeed, it is possible to infer a detection of
He in some SNe Ic from the coincidental alignment of Doppler shifted He lines
and absorption features but these do not behave as He-lines do in He-rich SNe.
Additionally, there are many more examples where similar features in other SNe
Ic are not compatible with He lines. 
Recent analysis has suggested that SNe Ic with highly blended lines are He-free \citep{Modjaz2016}.

The classification of SE-SNe was revisited in \cite{Prentice2017} in order to
link the taxonomic scheme with physical parameters. For the SN sample used in
that work, it was found that when He was obviously present in the ejecta of a SN
it formed strong lines (e.g., there were no examples of weak He lines).  This
allowed a natural division between He-rich and He-poor SNe. For the He-rich SNe,
classification was based upon characterising the presence and strength of H and led to
the sub division of type Ib and IIb into Ib, Ib(II), IIb(I), and IIb for weakest
to strongest H lines in the spectra. For He-poor SNe, classification was based
upon line blending and lead to subdivision of the SN Ic category into
Ic-$\left<N\right>$ where $\left<N\right>$ is the mean number of absorption
features in the pre-peak spectra from a set list of line transitions and takes
an integer value between 3 and 7. The lower the value of $\left<N\right>$ the
more severe the line blending and a higher specific kinetic energy. Such SNe
show high kinetic energies, broad lines, and significant line blending, e.g., SN
1998bw \citep{Iwamoto1998}, SN 1997ef \citep{Mazzali2000}, SN 2002ap
\citep{Mazzali2002}, SN 2003dh \citep{Mazzali2003}, SN 2010ah
\citep{Corsi2011,Mazzali2013}, SN 2016jca \citep{Ashall2017}. The most energetic
of these SNe are also associated with gamma-ray bursts (GRB) (e.g., SN
1998bw/GRB 980425, SN 2003dh/GRB 030329, SN2016jca/GRB 161219B). An injection of
$\sim 10^{52}$ erg of energy into the ejecta likely requires some contribution
from a rapidly rotating compact object, either a magnetar \citep{Mazzali2014} or
a black hole \citep{Woosley1994}. The maximum rotational energy of these compact
objects is a few $10^{53}$ erg \citep{Metzger2015}, and would have to be
injected on a short time-scale in order to influence the SN dynamically but not
to influence the light curve. 

Some SE-SNe are classified as Ib/c owing to the ambiguity of the presence of He in the spectra \citep[For example, SN 2013ge][]{Drout2016}, or lack of spectral coverage. However, a supernova that is genuinely a transitional event between SNe Ic and SNe Ib would be an important discovery and may help to explain why SNe Ic should show no clear indication of He in their spectra and why there is such a sharp distinction between SNe with He and SNe without He. In this work, we use analytical methods and spectral modelling to investigate the physical properties and elemental structure of the ejecta of SN 2016coi. 

In Section~\ref{sec:obs} we detail the observations and data reduction. In Section~\ref{sec:host} the host-galaxy of the SN, UGC 11868, is analysed. Sections~\ref{sec:lc} and \ref{sec:bol} present the light curves and associated properties for the multi-band photometry and the pseudo-bolometric light curve respectively. We examine the spectra analytically in Section~\ref{sec:spec} and model the early spectra and nebular spectra in Section~\ref{sec:mod}. We briefly discuss the SN in Section~\ref{sec:discussion} before presenting our conclusions in Section~\ref{sec:conclusions}.

\section{Observations and data reduction} \label{sec:obs}
\begin{table*}
	\centering
	\caption{Properties of the environment towards the SN}
	\begin{tabular}{lccccccc}
    \hline
	SN   &   $\alpha$ (J2000)    & $\delta$    &   Host   & $z$   & $\mu$  & \Emw & \Eh \\
	       &           &           &             &          & [mag] &[mag] & [mag] \\
    \hline
	2016coi &    21:59:04.14 & +18:11:10.46 & UGC 11868 & 0.0036 & 31.00  & 0.08 & $0.125\pm{0.025}$ \\
    \hline
	\end{tabular}       
	\label{tab:props}
\end{table*}

SN 2016coi/ASASSN-16fp was discovered on 2016-05-27.55 UT by the All Sky Automated Survey for Supernovae (ASAS-SN) \citep[See][]{Shappee2014} and was located in the galaxy UGC 11868, $z=0.0036$, at $\alpha$ = $21^{\mathrm{h}}59^{\mathrm{m}}04.14^{\mathrm{s}}$ $\delta$ $= +18\degr 11'10.46''$ (J2000), the last non-detection had been 6 days prior \citep{Holoien2016Atel}. It was subsequently classified as a pre-maximum ``broad lined'' Type Ic SN on 2016-05-28.52 UT. 
 
Our first observations were taken prior to this on 2016-05-28.20 UT using the Spectrograph for the Rapid Acquisition of Transients (SPRAT) \citep{Piascik2014} on the 2.0~m Liverpool Telescope (LT) \citep{Steele2004}, based at the Roque de los Muchachos Observatory. 
Subsequent photometric and spectroscopic follow up observations were conducted with by several different facilities around the world:

\begin{itemize}
	\item{Photometry and spectroscopy using the optical wide-field camera IO:O and SPRAT on the LT.}

    \item{Photometry and spectroscopy via the Spectral cameras and Floyds spectrograph on the Las Cumbres Observatory (LCO) network 2.0~m telescopes at the Haleakala Observatory and the Siding Spring Observatory (SSO), the Sinistro cameras on the LCO 1~m telescopes at the South African Astronomical Observatory (SAAO), the McDonald Observatory, and the Cerro Tololo Inter-American Observatory (CTIO) \citep{Brown2013}.}    
    
	\item{Photometric and spectroscopic observations from the Li-Jiang 2.4 m telescope (LJT, \citealp{Fan15}) at Li-Jiang Observatory of Yunnan  Observatories (YNAO) using the Yunnan Faint Object Spectrograph and Camera (YFOSC; \citealp{Zhang14}), the Xing-Long 2.16 m telescope (XLT) at Xing-Long Observation of National Astronomical Observatories (NAOC) with Bei-Jing Faint Object Spectrograph and Camera (BFOSC). The spectra of LJT and XLT  were reduced using standard IRAF long-slit spectra routines. The flux calibration was done with the standard spectrophotometric flux stars observed at a similar airmass on the same night. 
Optical photometry were obtained in the Johnson $UBV$ and Kron-Cousins $RI$ bands by Tsinghua-NAOC 0.8 m telescope (TNT;\citealp{Wang08,Huang12}); Johnson $BV$,  Kron-Cousins $R$ and SDSS $ugriz$ bands by LJT with YFOSC.}  

	\item{Photometry from the 0.6/0.9m Schmidt telescope, equipped with a liquid-cooled Apogee Alta U16 $4096 \times 4096$ CCD camera
(field-of-view $70 \times 70$ arcmin$^2$)
and Bessell $BVRI$ filters, at Piszk\'estet{\H o} Station of Konkoly Observatory, Hungary. The CCD frames were bias-, dark- and flatfield-corrected by applying standard IRAF routines.}

	\item{Photometric data from the $0.4\,\rm{m}$ PROMPT~5 telescope that monitors luminous, nearby ($\rm{D}<40\,\rm{Mpc}$) galaxies (DLT40, Tartaglia et al in prep). These data were reduced using aperture photometry on difference images \citep[with HOTPANTS;][]{Becker2015}. }
    
	\item{Two spectra using the Kast Double Spectrograph on the Shane 3~m telescope at the Lick Observatory. These were reduced through standard {\sc iraf} routines.}

	\item{A single spectrum was obtained using the Intermediate Dispersion Spectrograph (IDS), on the 2.5~m Issac Newton Telescope (INT), at the Roque de los Muchachos Observatory in La Palma.  
The EEV10 detector was used, along with the R400V grating.
Data reduction was performed using standard routines within the Starlink software packages Figaro and Kappa, and flux calibrated using custom software.}

	\item{A single spectrum from the Deep Imaging Multi-Object Spectrograph (DEIMOS) spectrograph \citep{Faber2003} on the W. M. Keck Observatory, Haleakala. }

\end{itemize}

Much of the LCO data, in addition to the Lick spectra, were obtained as part of the LCO Key Supernova Project.
For the photometry obtained at Konkoly Observatory, the magnitudes for the SN and some local comparison stars were obtained via PSF-photometry using {\sc iraf/daophot}. The instrumental magnitudes were transformed to the standard system using linear colour terms and zero-points. The zero-points were tied to the PS1-photometry\footnote{http://archive.stsci.edu/panstarrs/search.php} of the local comparison stars after converting their $g_p$, $r_p$, $i_p$ magnitudes to BVRI \citep{Tonry12}. 
Aperture photometry was performed on the remaining photometric data using a custom script utilising {\sc pyraf} as part of the {\sc ureka}\footnote{http://ssb.stsci.edu/ureka/} package. 
The instrumental magnitudes were calibrated relative to Sloan Digital Sky Survey (SDSS) stars in the field for the $ugriz$ bands.  The equations of \cite{Jordi2006} were used to convert American Association of Variable Star Observers Photometric All-Sky Survey (APASS) standard star $BVgri$ photometry into $BVRI$ and SDSS $ugriz$ photometry into $U$. 
A series of apertures were used to derive the instrumental magnitudes and the median value taken as the calibrated magnitude. The uncertainty was taken to be either the standard deviation of the photometric equation fit or the standard deviation of the calibrated magnitudes, whichever was larger. 
SPRAT spectra were reduced and flux calibrated using the LT pipeline \citep{Barnsley2012} and a custom {\sc python} script. LCO/Floyds spectra were reduced using the publicly available LCO pipeline\footnote{https://lco.global/}.

\section{Host-galaxy - UGC 11868} \label{sec:host}

\subsection{Line-of-sight attenuation}
The Galactic extinction in the direction of the SN is \Emw\ $=0.08$ mag \citep{Schlafly2011}.  
\Eh\ can be calculated through a number of methods including 
 measurement of the equivalent width of the rest-frame \NaI\ D lines \citep{Poznanski2012}, and by assessing the colour curve of 
 the SN relative to the bulk of the population where an offset from the mean implies some amount of extinction \citep[e.g.,][]{Drout2011,Stritzinger2018}. Throughout the following methods it is assumed that $R_V=3.1$.
 
We find no indication of strong host \NaI\ D lines in our low-resolution
spectra, the Galactic \NaI\ D lines are the dominant component in this regard.
The upper limit set by measuring the equivalent width of this weak feature is
approximately \Eh\ $=0.03$ mag using the method of \cite{Poznanski2012}. It is acknowledged that the low resolution of
the spectra may affect the measurement here \citep{Poznanski2011} and that \Eh\
may be greater than this, although it could not be significantly
larger as experience shows that the host \NaI\ D lines would become prominent in the
spectra. In Section~\ref{sec:LCcomp} the $g-r$ colour curve of SN~2016coi is
examined in relation to other SE-SNe. We find that some small to moderate
extinction $\sim 0.1-0.2$ mag in $E(B-V)$ is required to move the colour curve
of SN 2016coi into the host-corrected distribution. A correction for \Eh\ $\sim 0.4$ mag is required to place the colour curve at the bottom of the distribution.
Given the potential for uncertainty we apply a third test. In
Section~\ref{sec:mod} we use spectral models to examine a range of values for
\Eh, and determine that it could be anywhere from \Eh=0.1--0.15\,mag. 
Considering the results from the different methods here, we adopt a \Eh\ of
$0.125\pm{0.025}$ mag, and an \Etot\ $=0.205\pm{0.025}$ mag. The upper limit is
constrained by the weakness of the host \NaI\ D absorption lines in the spectra.

\subsection{The properties of UGC 11868}

The host galaxy of SN 2016coi is UGC~11868, also known as II~Zw~158 and MCG~+03-56-001.  
The distance to UGC 11868 is somewhat uncertain (see NASA/IPAC Extragalactic Database \footnote{http://ned.ipac.caltech.edu/} for more details) and in this work we adopt a distance modulus of 31.00\,mag. This value is taken as absolute with no uncertainty included in order to enable easy conversion of the intrinsic light curve properties, including uncertainties derived from the photometry and reddening, for different distances. 

UGC~11868 is a low-surface-brightness, low-luminosity galaxy of quite irregular morphology, classified as SBm in the RC3 \citep{devac1991}.  UGC~11868 was included in the H$\alpha$ Galaxy Survey \citep{James2004} which included $R$-band and narrow-band H$\alpha$ imaging.  
Correcting the data from that study to an adopted distance of 15.8~Mpc, distance modulus 31.00, UGC~11868 has an apparent $R$-band magnitude of 13.10, an absolute $R$-band magnitude of --17.90, and a star formation rate of 0.078~M$_{\odot}$~yr$^{-1}$.  
The latter value has been corrected for internal extinction using the absolute-magnitude dependent extinction formula of \citealt{Helmboldt2004}, since the global correction of 1.1~mag applied by \citealt{James2004} is almost certainly an over-estimate for such a low-luminosity system.  The Magellanic Clouds provide useful reference points for UGC~11868, with star formation rates of 0.054 and 0.23~M$_{\odot}$~yr$^{-1}$, and absolute $R$-band magnitudes of --17.10 and --18.50 for the SMC and LMC respectively.  The $\mu_B=25$ isophotal diameter from RC3 corresponds to 9.0~kpc for our adopted distance, similar to the 9.5~kpc value for the LMC, which also shares its SBm classification with UGC~11868.  Thus, host of SN2016coi is very similar to the LMC overall, but slightly more diffuse and lower in surface brightness.  The somewhat higher star formation rate for the LMC can be entirely attributed to the unusually powerful 30~Doradus complex; the star formation properties of UGC~11868 appear entirely normal for its type.

SN 2016coi lies well away from the centre of UGC~11868.  There is no well-defined nucleus, just a somewhat elongated general region of higher surface brightness that gives rise to the barred classification.  Defining the highest surface brightness region from the $R$-band image as the galaxy centre, we determine that SN 2016coi occurred 34$^{\prime \prime}$ or 2.6~kpc from this location.  There is no detectable H$\alpha$ emission at the location of the SN; a moderately bright region is located 5$^{\prime \prime}$ or 375~pc away.

There are no direct metallicity measurements for UGC~11868, but again the comparison with the Magellanic Clouds and other dwarf galaxies can be used to give some indications of likely values.  \citet{Berg2012} calibrate a correlation between absolute magnitude and oxygen abundance for star forming dwarf galaxies, from which we derive a value of 12$+$log(O/H) $=$ 8.21, very close to the measured value of 8.26 for the LMC from the same study.  In terms of [Fe/H], \citet{Cioni2009} show a central value of $\sim-1.0$ for the LMC, but this falls to --1.3 in the outer regions, matching the global value derived for SMC which has no detectable radial gradient.  Thus, inferred values of 12$+$log(O/H) $=$ 8.21, [Fe/H] $=$ --1.3 are plausible estimates for the location of SN 2016coi.

To conclude, SN 2016coi occurred in the outer regions of a low-luminosity, LMC-like host galaxy, in a location with almost certainly low metallicity and nearby but not coincident ongoing star formation. The sub-solar metallicity is consistent with studies of the local environments of non-GRB ``broad-lined'' SNe Ic \citep[See][]{Modjaz2008,Modjaz2011,Sanders2012}

\begin{figure}
	\centering
	\includegraphics[scale=0.31]{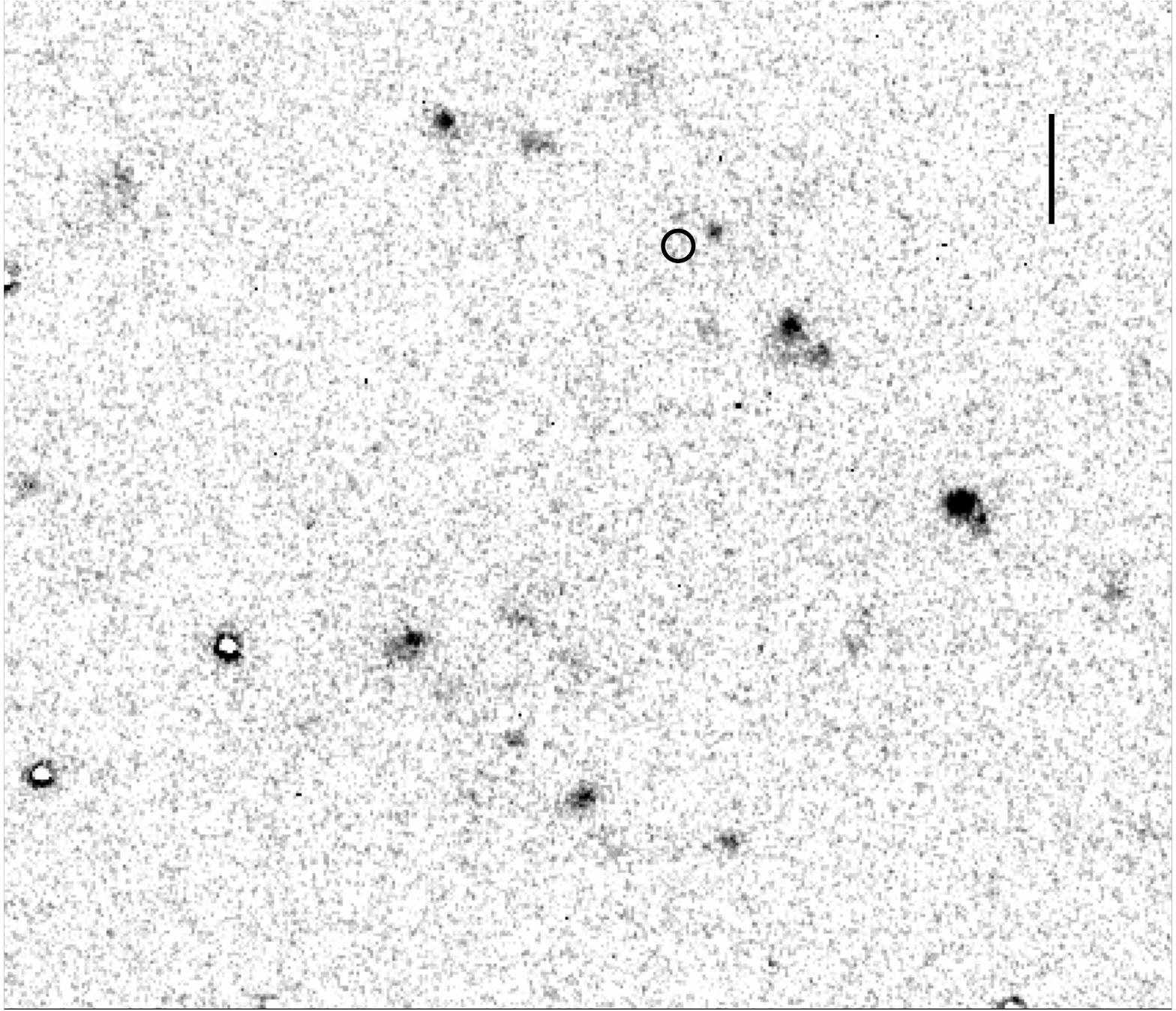}
	\caption{A pre-explosion \Ha\ image of UGC 11868. The position of SN 2016coi is designated by the circle. There is no strong emission at the location of the SN suggesting a region of little star-formation. North is up and East is left. The scale bars, located in the upper right corner of the image, corresponds to 13.0 arcseconds, or 1 kpc at the adopted distance of 15.85 Mpc, $\mu = 31.0$ mag.}
	\label{fig:Ha}
\end{figure}

\begin{figure}
	\centering
	\includegraphics[scale=0.31]{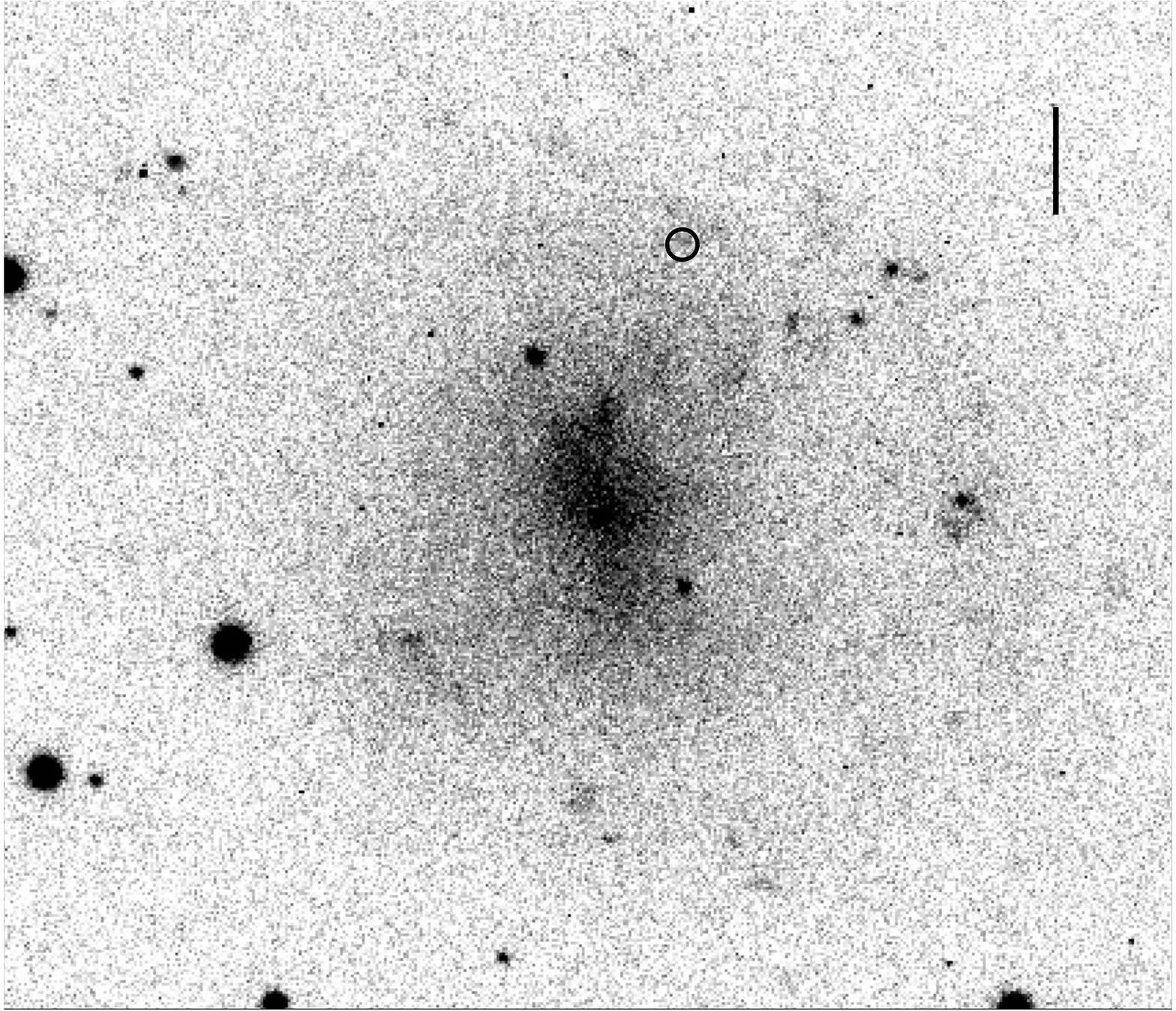}
	\caption{A pre-explosion $r$-band image of UCG 11868 taken on 6 July 2000. The position of the SN is shown by the circle. See Figure~\ref{fig:Ha} for orientation and scale.}
	\label{fig:pre}
\end{figure}

\section{Light curves} \label{sec:lc}
\begin{figure*}
	\centering
	\includegraphics[scale=0.5]{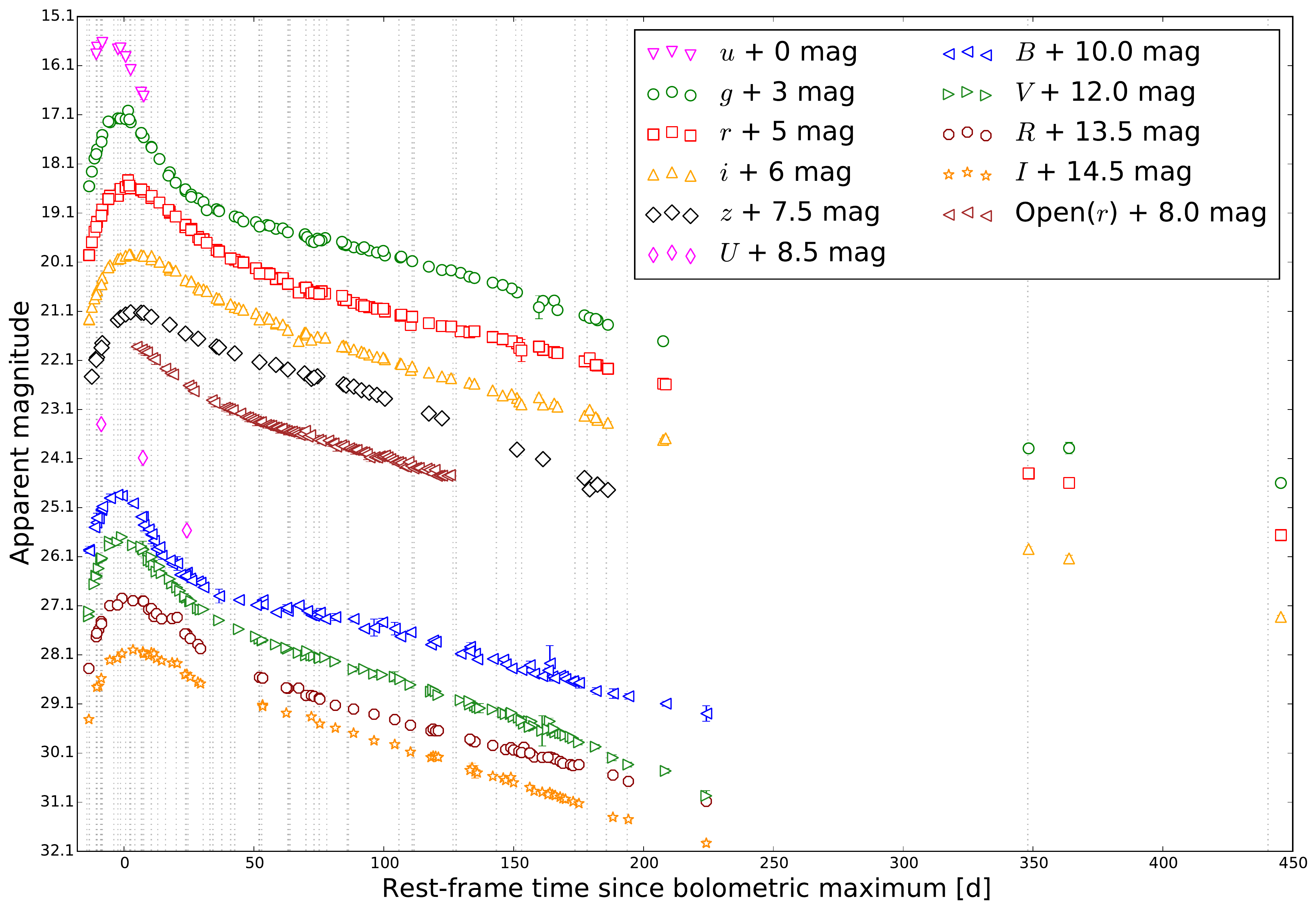}
	\caption{The LT, LCO, TNT, LJT, Konkoly, and DLT40 $ugriz$, $UBVRI$, and Open($r$) light curves of SN 2016coi shifted to rest-frame time. This period covers from shortly after explosion to the nebular phase. The grey dotted lines represent dates of spectroscopic observations.}
	\label{fig:LCs}
\end{figure*}

Figure~\ref{fig:LCs} shows the $ugriz$ (Table~\ref{tab:ugriztable}), $UBVRI$ (Table~\ref{tab:UBVRItable}), and DLT40 Open$r$ (Table~\ref{tab:DLT40table}) light curves of SN 2016coi, these are not corrected for extinction but are presented in rest-frame time.

We fit the extinction corrected light curves to find the peak magnitude \Mp, and three characteristic time-scales. These are \trise, \tdecay, and the late linear decay rate $\Delta m_\mathrm{late}$. \trise\ and \tdecay\ measure the time taken for the light curve to evolve between the maximum luminosity \lp\ and \lp/2 on the rise and decline respectively. 

Light curve parameters were measured by fitting a low order spline from {\sc UnivariateSpline} as part of the {\sc scipy} package. Errors were estimated to be within the confines of the uncertainty in the photometry and the order of the spline was allowed to vary to estimate the effect on the fit. We find that variations in the spline fit produced either a negligible effect or were clearly wrong. For $M$ the errors are small and dominated by the uncertainty in \Etot.

The time for the light curve to rise from explosion to peak \tp\ is, with the
exception of GRB-SNe where the detection of the high-energy transient signals
the moment of core collapse, an unknown quantity.  Comparatively, \trise\ is a
measurable quantity for many SNe so allows for comparison.  \tp, and by proxy
\trise, is affected by the ejecta mass \mej\, the ejecta velocity, and the
distribution of \Nifs\ within the ejecta \citep[e.g.,][]{Arnett1982}. Mass
increases the photon diffusion time while the ejecta velocity decreases it.
Placing small amounts of \Nifs\ in the outer ejecta causes the light curve to
rise quicker as the photons are able to diffuse through the low density, high
velocity, outer ejecta more rapidly. The post peak decay time is more affected
by the core mass as the outer layers are now optically thin. The late decay
follows a linear decline which is the result of energy injection from the decay
of \Cofs\ to \Fefs\ and is affected by the efficiency of the ejecta to trap
$\gamma$-rays and positrons. In the vast majority of SE-SNe the late decay rate
exceeds that of \Cofs, implying that the ejecta are not 100$\%$ efficient at
trapping gamma rays.

Table~\ref{tab:multiprops} gives the properties of the multi-band light curves, corrected for \Etot\ using the reddening law of \cite{CCM}. As is typical of SE-SNe the redder bands evolve more slowly \citep[see][]{Drout2011,Bianco2014,Taddia2015,Taddia2018} with values ranging between $\sim 8$ and $14$ d. 
It is noticeable that the light curves are more asymmetric for the redder bands. The late-time decay rates are in the range $0.015$ to $0.023$ mag d$^{-1}$.

\begin{table}
	\caption{Multi-band light curve properties}
	\begin{tabular}{lccccc}
    \hline
	Band &  \Mp & $t_{-1/2}$ & $t_{+1/2}$ & $\Delta m_\mathrm{late}$   \\
		 &[mag] & [days] & [days] & [mag d$^{-1}$]  \\
    \hline
	{\it u}  & $-16.4\pm{0.1}$ & - & 11$\pm{0.5}$ & - \\
	{\it g}  &$-17.61\pm{0.09}$ & 8.7$\pm{0.1}$ & 13.3$\pm{0.5}$ & 0.015\\
	{\it r}  & $-17.97\pm{0.07}$ & 11.8$\pm{0.1}$ & 21.0$\pm{0.5}$ & 0.014\\
	{\it i}  & $-17.46\pm{0.06}$ & 13.1$\pm{0.1}$ & 28.4$\pm{0.5}$ & 0.016\\
	{\it z}  & $-17.66\pm{0.04}$  & 13.2$\pm{0.5}$ & 34$\pm{1}$ & 0.019\\
	{\it B}  &$-18.3\pm{0.1}$   & 9.6$\pm{0.5}$ & 12.5$\pm{0.5}$ & 0.014  \\
	{\it V} &$-17.93\pm{0.07}$   & 10.3$\pm{0.5}$ & 14.4$\pm{0.4}$ & 0.016 \\
	{\it R}  &$-18.10\pm{0.06}$& 12.3$\pm{0.5}$ & 20.1$\pm{0.5}$ & 0.013 \\
	{\it I} & $-17.9\pm{0.05}$&13.7$\pm{0.5}$  &26.5$\pm{0.7}$  & 0.016 \\
    \hline
	\end{tabular}
	\label{tab:multiprops}
\end{table}

\subsection{Colours} \label{sec:col}
\begin{figure}
	\centering
	\includegraphics[scale=0.43]{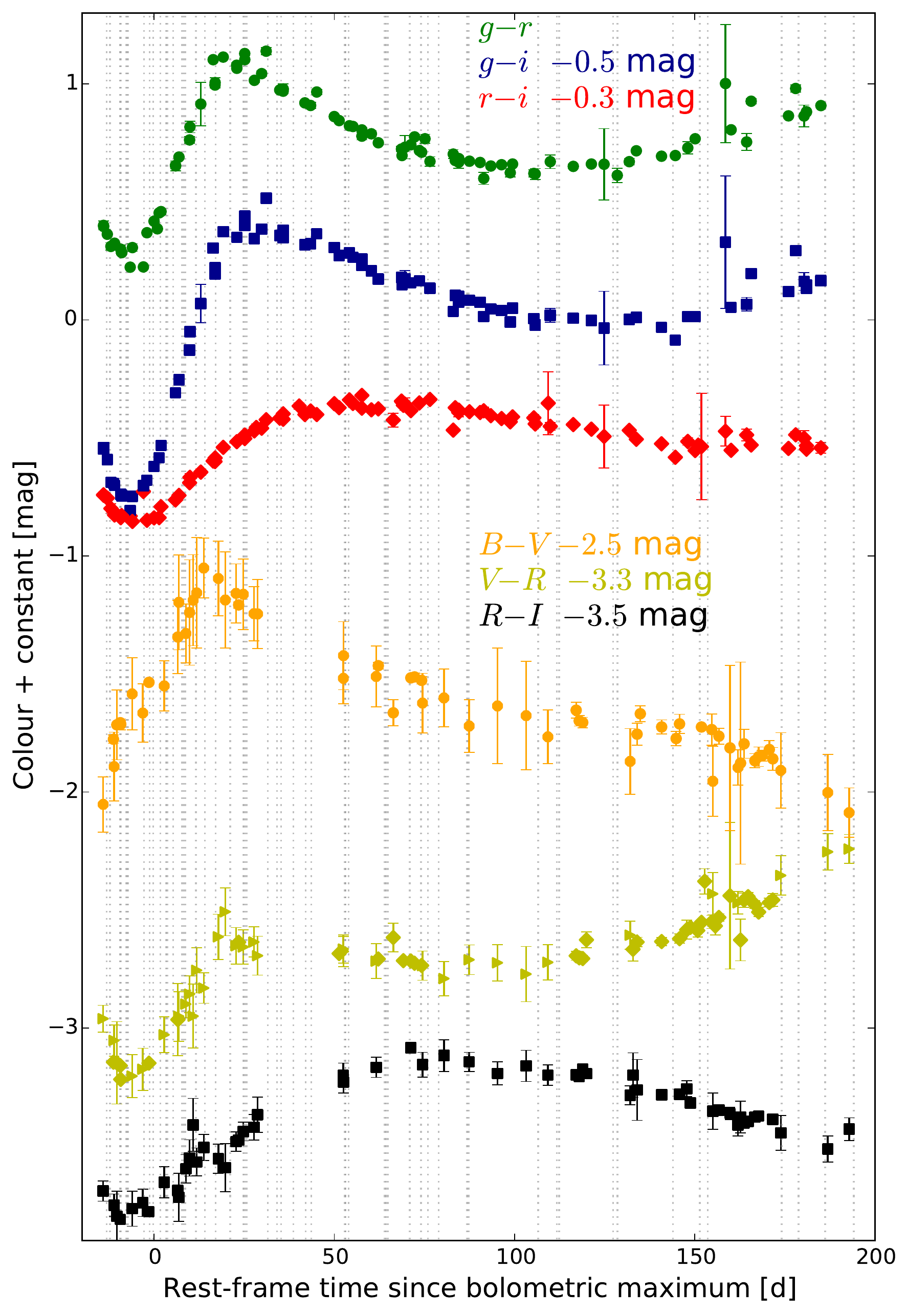}
	\caption{The colour evolution of $g-r$, $g-i$, $r-i$, $B-V$, $V-R$, and $R-I$ to $+200$ d. The photometry is corrected for \Etot\ and is presented in the rest frame. The dotted lines represent spectroscopic observations.}
	\label{fig:colours}
\end{figure}

In Figure~\ref{fig:colours} we present the colour curves of $g-r$, $g-i$, $r-i$, $B-V$, $V-R$, and $R-I$. 
The behaviour of the curves shows many features that are typical to SE-SNe. 
In all colours there is an initial blue-ward evolution as the energy deposited from the decay of \Nifs\ into the ejecta exceeds radiative losses and the photosphere recedes towards the heat source, so the SN gets bluer and more luminous \citep{Hoeflich2017}. 
Around the time when $E_\mathrm{out}>E_\mathrm{in}$ the SN ejecta expands adiabatically, cooling the photosphere, which leads to a decrease in luminosity. The cooling photosphere results in a red-ward turn in the colour curves.
After $\sim +20$ days $g-r$ and $g-i$ turn blue again as the flux in the red decreases due to the loss of the photosphere. At $\sim+100$ d the SN is in the early nebular phase and in $g-r$ and $g-i$ there is then a turn back towards the red. 
This is driven by the appearance of the \Oneb \lam \lam\ 6300, 6364 \AA\ emission line which dominates the flux in the $r$-band and the \ion{Ca}{II]} \lam \lam\ 7291, 7324, \OI\ \lam\ 7773, and \ion{[S}{I} \lam\ 7722 emission lines which dominate the flux in the $i$-band. 
Comparatively, there are few strong emission lines around the effective wavelength of $g$ ($\sim$ 4770 \AA), the strongest emission feature being a blend of \ion{Mg}{I]} \lam\ 4570 and \ion{[S}{I]} \lam\ 4589. 

$r-i$ slowly evolves to the blue from about $\sim+50$ d as the emission line flux in $r$ is greater than that in $i$. Without the dramatic change in flux seen in $g$ the curve does not show a late blue turn during the time of our observations.

\subsubsection{Comparison of $g-r$ with He-poor SNe}\label{sec:LCcomp}
\begin{figure}
	\centering
	\includegraphics[scale=0.43]{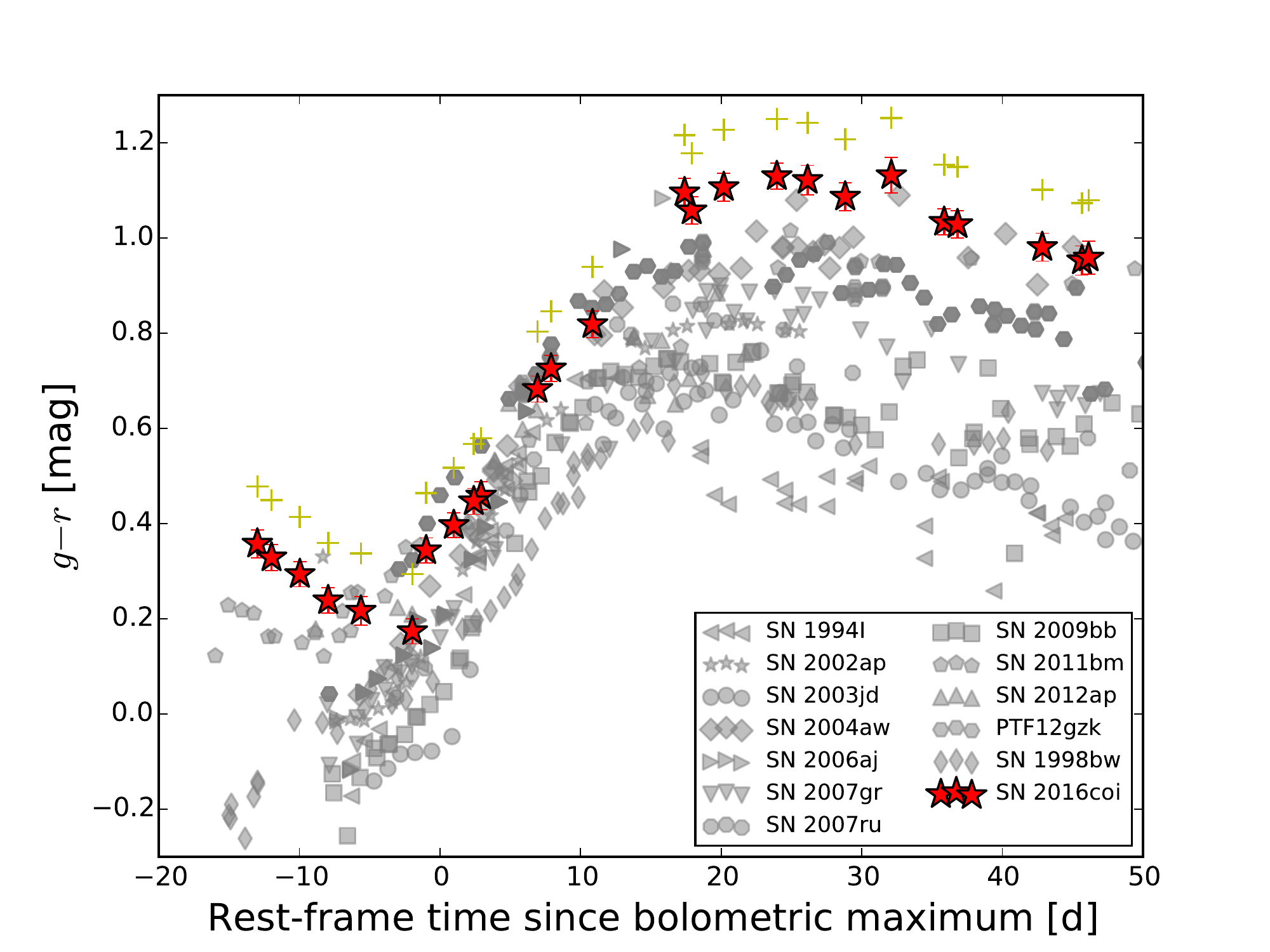}
	\caption{The $g-r$ colour evolution of SN 2016coi in relation to other SNe Ic from the sample of \citet{Prentice2016}. The photometry has been corrected for \Etot\ and our errors have been included on the plot so as to show the possible movement of the colour curve in relation to the distribution. In the absence of \Eh\ (yellow) the colour curve sits on the very top of the distribution. Our analysis suggests that 2016coi is intrinsically red. }
	\label{fig:g-r}
\end{figure}

The $g-r$ colour curve of SN 2016coi is shown in relation to He-poor SNe in Figure~\ref{fig:g-r}. The application of an extinction correction \Emw\ $=0.08$ mag places the colour curve at the upper edge of the distribution. To place the curve in the middle of the distribution requires a host-extinction of \Eh\ $\sim 0.25$ mag, which is incompatible with the weak the host \NaI\ D lines. A moderate correction of \Eh$=0.05-0.15$ places the SN at the upper edge of the distribution. The Figure is shown with our total \Etot$=0.205\pm{0.025}$ mag applied. This suggests that the SN is intrinsically red.

\section{Bolometric light curve} \label{sec:bol}
\begin{table*}
	\centering
	\caption{Statistics derived from the pseudo-bolometric light curves}
	\begin{tabular}{lccccccc}
    \hline
	Bol. type &\loglp	& \trise & \tdecay & width & \tmax & $\Delta m_\mathrm{late}$ &\mni \\
	 & & [d] & [d] & [d] & [d]& [mag d$^{-1}$] & \msun \\
     \hline
	$griz$ & 42.29$\pm{0.02}$ & $13\pm{1}$ & $23\pm{1}$ &  $36\pm{2}$& $17\pm{1}$ &0.015 & $0.09\pm{0.01}$\\
	$ugriz$ & 42.38$\pm{0.01}$ & $11\pm{1}$ & $19\pm{2}$ &  $32\pm{2}$& $16\pm^{1.3}_{0.7}$ &- & $0.104\pm^{0.02}_{0.008}$\\
    UVOIR* & $\sim$42.51 & - & - & - & - &- &$\sim0.14$\\
    \hline
    *Estimated - see text & & & & & & & \\

	\end{tabular}		
	\label{tab:bolstats}
\end{table*}

\begin{figure}
	\centering
	\includegraphics[scale=0.43]{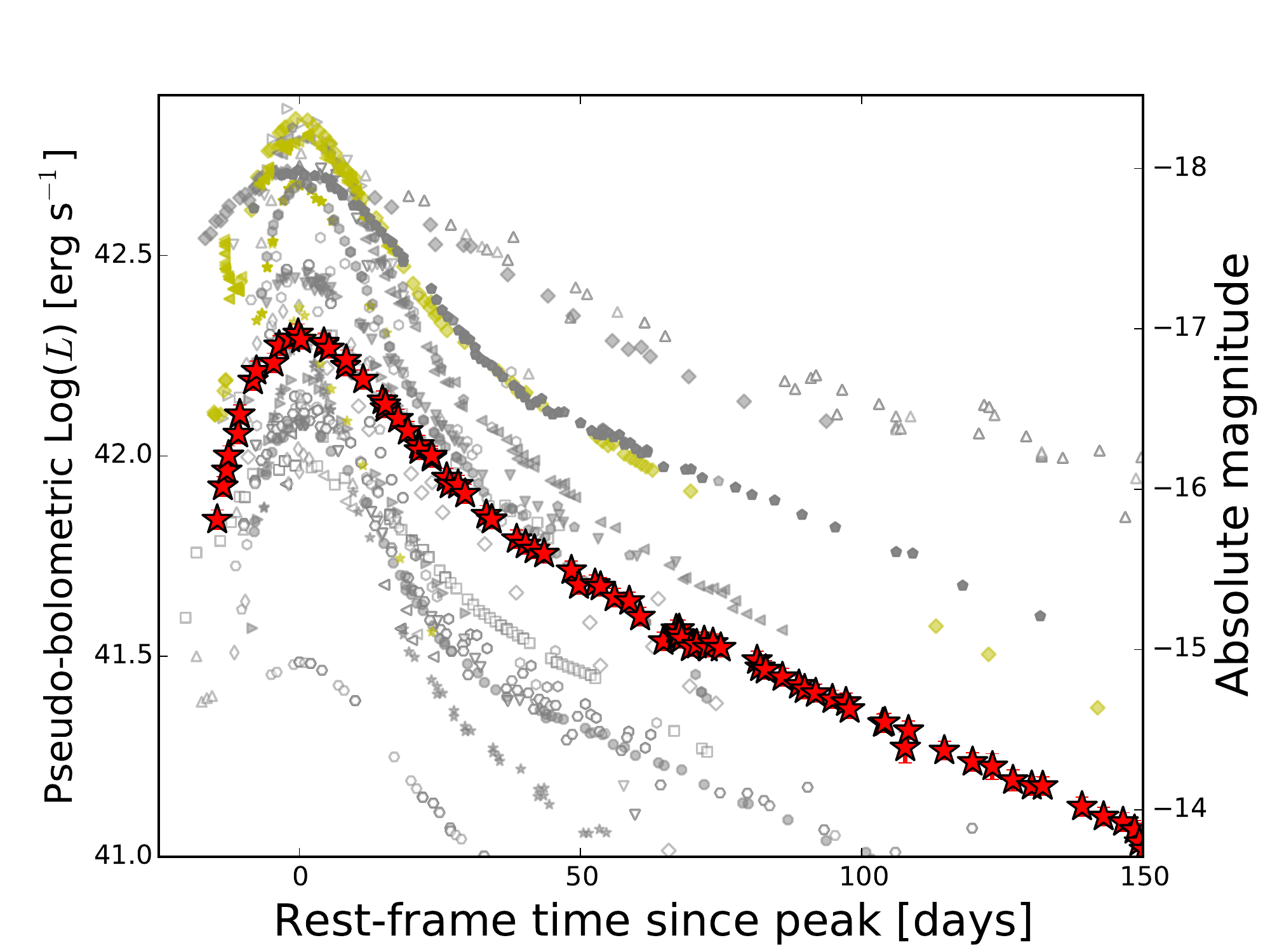}
	\caption{The $griz$ optical pseudo-bolometric light curve of SN 2016coi (red) set in context against SNe Ic \citep{Prentice2016}. The open symbols represent SNe with no correction for host extinction applied, GRB-SNe are shown in yellow. SN 2016coi is not extreme in either luminosity or temporal evolution.}
	\label{fig:bol}
\end{figure}

The pseudo-bolometric light curve (henceforth ``bolometric'') was constructed using the de-reddened $griz$ photometry converted to monochromatic flux \citep{Fukugita1996}. The resulting spectral energy distribution (SED) was integrated over the range 4000 to 10000 \AA\ and then converted to luminosity using $\mu =31.00$ mag. The bolometric light curve is presented in comparison with He-poor SE-SNe in Figure~\ref{fig:bol}.
The $u$-band data were not included at this point as the 4000 -- 10000 \AA\ range allows direct comparison of the pseudo-bolometric light curve properties of a large number of SE-SNe. However, in Table~\ref{tab:bolstats} we present the statistics obtained from the pseudo-bolometric light curve using the $u$-band data (integrating over $3000 - 10000$ \AA), and estimating the fully bolometric \lp\ as per the method given \citet{Prentice2016}. 
This latter method utilises a relation between the $ugriz$ integrated light curves of various SE-SNe and their $ugriz$ plus near-infrared (NIR) $10000 - 24000$ \AA\ light curves in order to estimate the missing NIR flux. This value is then increased by 10 percent to account for flux outside the wavelength regime to give an estimate of the UVOIR \lp.

SN 2016coi reached a peak luminosity ($griz$) of \loglp $=42.29\pm{0.02}$ or, alternatively, \lp\ $=(1.9\pm{0.1})\times10^{42}$ \ergs. The temporal values, calculated by fitting a low order spline to the light curve, were found to be \trise\ $=12.4\pm{0.5}$ d and \tdecay\ $= 23\pm{1}$ d, which equates to a width of $35\pm{1}$ d. Extrapolation from a simple quadratic fit to the pre-peak light curve reveals that the progenitor exploded approximately 2$\pm1$ days before discovery of the SN, which is consistent with the explosion date found from spectral modelling (see Section~\ref{sec:mod}). For this light curve, the time from rise to peak \tp\ $=17\pm{1}$ days.
The late time decay rate is calculated by fitting a linear function to the decline; this returns a decay rate of $\sim 0.015$ mag d$^{-1}$.

When including the $u$-band, the peak luminosity increases by 23 percent to \lp\ $=(2.4\pm{0.1})\times10^{42}$ \ergs. Inclusion of the $u$-band makes the light curve rise and decay slightly quicker (\trise\ $=11\pm{1}$ d, \tdecay\ $=19\pm{2}$ d), as would be expected from the increased energy at early times. It is calculated that the time from explosion to \lp, \tp $ = 16\pm^{1.3}_{0.7}$ d, which includes an estimated $\sim 1$ day of ``dark time'' \citep{Corsi2012}.

Finally, the estimated UVOIR bolometric luminosity is $\sim 3\times10^{42}$ \ergs. This is commensurate with that found through spectral modelling, see Section~\ref{sec:mod}.

\subsection{Estimating \mni}
To estimate the amount of \Nifs\ synthesised during explosive silicon burning in the first few seconds following core collapse we utilise the following equation from \citet{Stritzinger2005}

\begin{equation}
\begin{split}
\frac{M_{\mathrm{Ni}}}{{\rm M}_\odot}= & L_{\mathrm{p}}\times\left(10^{43} \textrm{erg s}^{-1} \right)^{-1} \\ & \times \left(6.45\times e^{-t{\mathrm{p}}/8.8}+1.45\times e^{-t{\mathrm{p}}/111.3}\right)^{-1} 
\end{split}
\label{eq:Mni}
\end{equation}

which is based upon the formulation for \mni\ given in \citet{Arnett1982} and assumes that the luminosity of the SN at peak is approximately equal to the energy emitted by the decay-chain of \Nifs\ at that time. Using a rise time from explosion to peak of \tp\ $17$ d and 16 d, for the $griz$ bolometric and $ugriz$ bolometric \lp\ respectively, we find that \mni$_{,griz}$ $=0.09\pm{0.01}$ \msun\ and \mni$_{,ugriz}$ $=0.104\pm^{0.02}_{0.008}$ \msun. The nickel mass derived from the estimated fully bolomteric luminosity corresponds to \mni$_\mathrm{,UVOIR}$ $\sim 0.14$ \msun. 
These values are based on the assumption that all the \Nifs\ is located centrally. In reality there will be some distribution of \Nifs\ throughout the ejecta, which causes the light curve to rise faster than in the centrally located case, and some degree of asphericity \citep{Maeda2008,Stevance2017}. The result is that a lower \mni\ can achieve the same results.

\subsection{Comparison of bolometric properties with SNe Ic}

Table~\ref{tab:bolstats} gives the properties of SN 2016coi derived from the
$griz$ and $ugriz$ light curves.  SN 2016coi is quite typical in luminosity
($4000-10000$ \AA) compared to He-poor SNe where the mean for non GRB-SNe is
\loglp$= 42.4\pm{0.2}$. Figure~\ref{fig:bol} plots the $griz$ light curves
of SN 2016coi and SNe Ic.  

\begin{figure}
	\centering
	\includegraphics[scale=0.43]{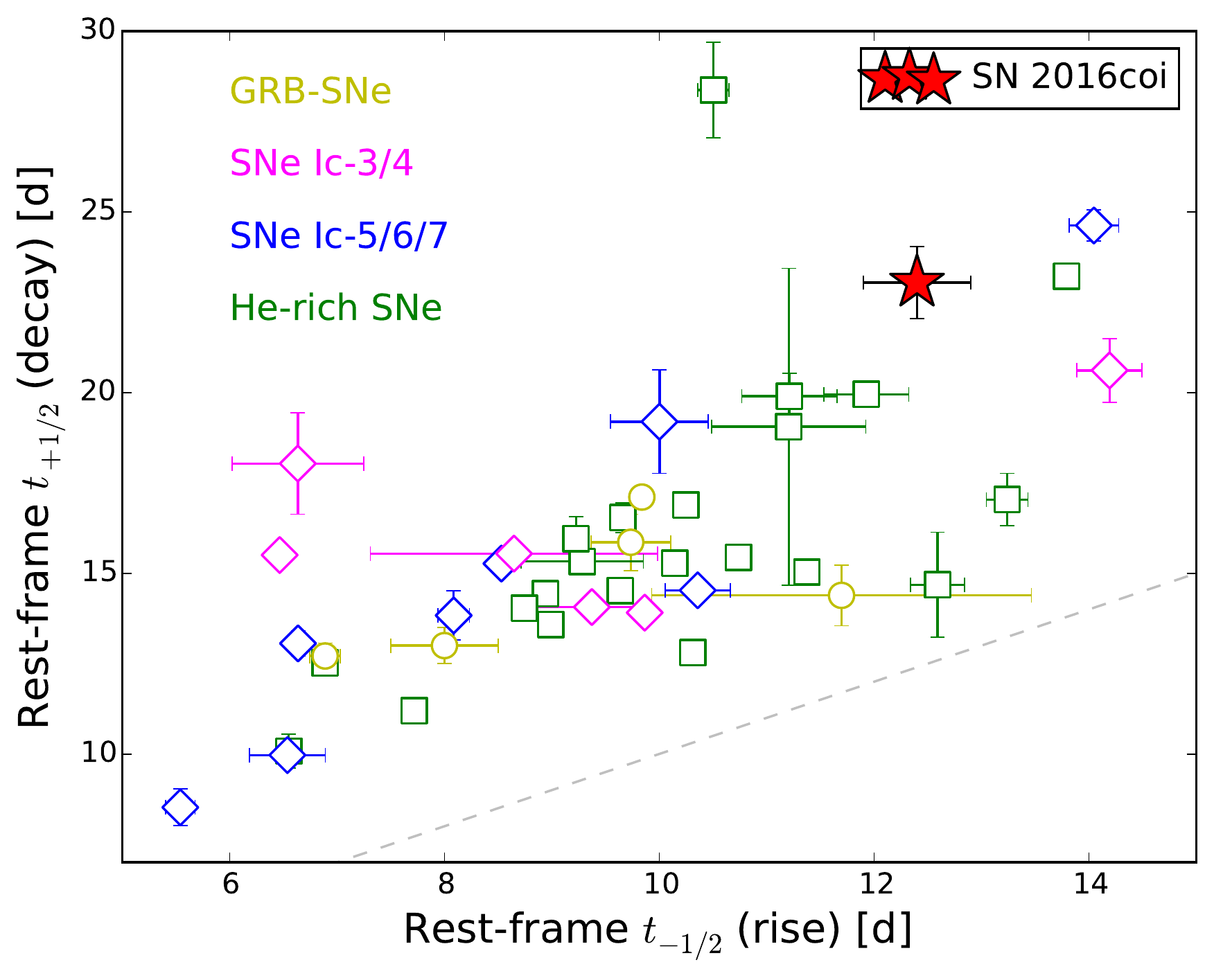}
	\caption{\trise\ against \tdecay, as calculated from the 4000 -- 10000 \AA\ light curve, for SE-SNe where the SN has a measurement of both values \citep{Prentice2016}. SN 2016coi is outside the bulk of the population in both properties but it is still within one sigma of the mean. The most extreme SNe Ic, those with long rise and decay times, are outside the field of view of this plot.}
	\label{fig:risedecay}
\end{figure}

The mean \trise\ and \tdecay\ of SNe Ic are
$10\pm{3}$ d and $20\pm{9}$ d respectively, while the mean width for those SNe
where it can be calculated is $30\pm{11}$ d.  The values derived for SN 2016coi
suggest it to be above average but within one sigma of the mean.  Figure~\ref{fig:risedecay} plots \trise\
against \tdecay\ and demonstrates that SN 2016coi is on the upper end
of the \trise\ and \tdecay\ distribution. It is not, however, akin to SNe Ic
with very broad light curves \citep[e.g., SN 1997ef][]{Mazzali2000}. SN 2016coi is long in both rise and
decay compared to the mean and median values of both parameters for the
population, however the mean values are skewed to longer durations by the slowly
evolving SNe. \mej\ is considered in Section~\ref{sec:mod} but the results here
suggest that \mej\ of SN 2016coi is larger than average for He-poor SNe. 

\begin{figure}
	\centering
	\includegraphics[scale=0.43]{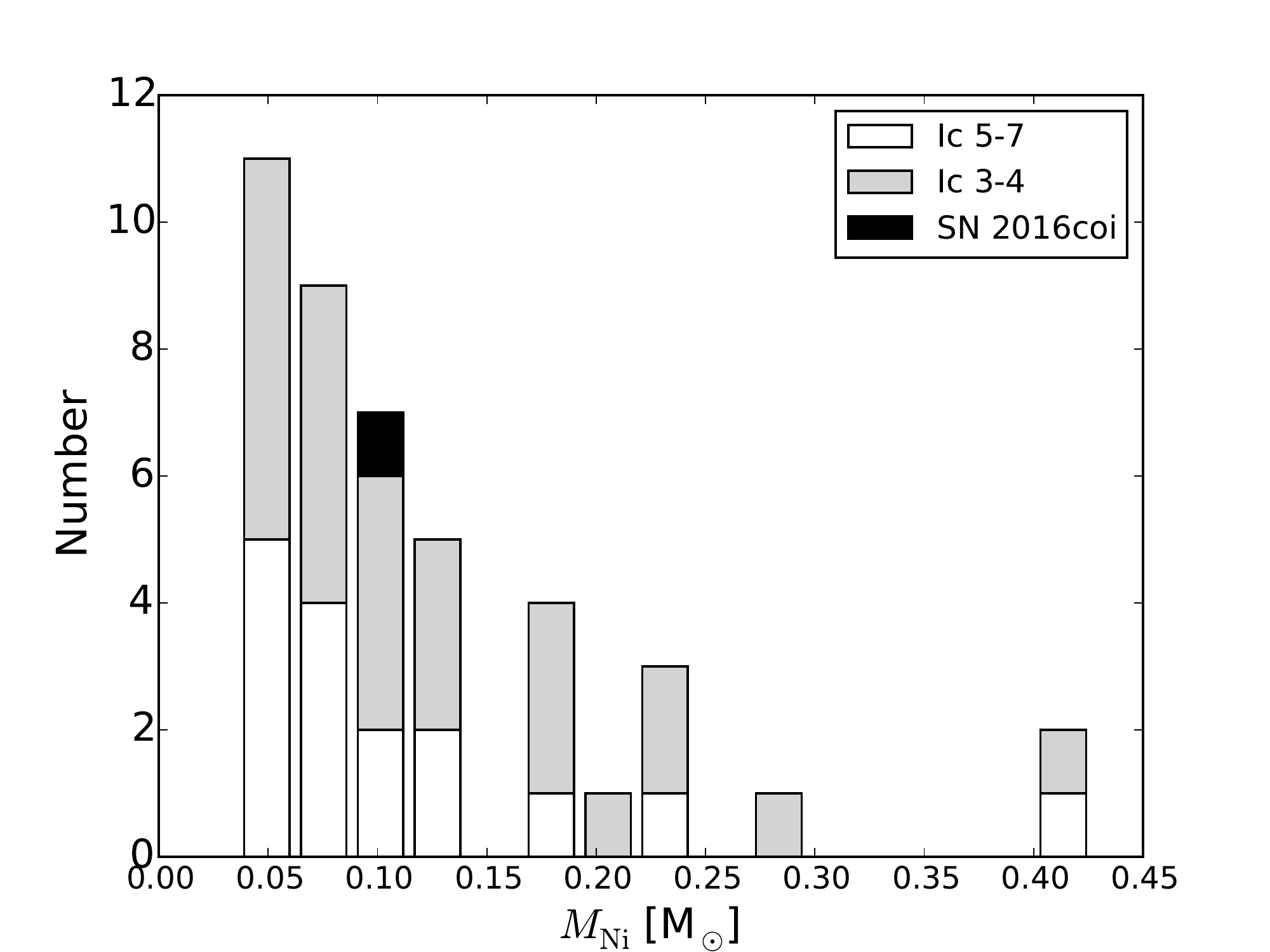}
	\caption{The nickel mass distribution as derived from the $4000 - 10000$ \AA\ pseudo-bolometric light curves for He-poor SNe \citep{Prentice2016}. To maximise the sample we have included SNe that have not been reclassified under the scheme given in \citet{Prentice2017} and assigned ``Ic-BL'' SNe to the Ic-3/4 distribution and all other SNe Ic to the Ic-5/6/7 group. SN 2016coi is at the median \mni\ for all the SNe and is below, but within one sigma, of the median for Ic-3/4 SNe.}
	\label{fig:mnihist}
\end{figure}

Figure~\ref{fig:mnihist} shows the distribution of \mni\ constructed from the $4000-10000$ \AA\ pseudo-bolometric light curves. Not all of the SNe have been reclassified under the scheme given in \citep{Prentice2017}, thus we group ``normal SNe Ic'' with Ic-5, 6 \& 7, and ``Ic-BL'' with Ic-3 and Ic-4. We do not include GRB-SNe in this plot. The median \mni\ is $0.09\pm_{0.03}^{0.08}$ \msun\ for all the SNe. In considering just the Ic-3/4 (broad-line) group, to which SN 2016coi belongs, we find the median \mni$=0.10\pm^{0.07}_{0.03}$ \msun. The distribution is clearly skewed, driven by a few luminous and long-rising SNe. Both measures suggest that \mni\ for SN 2016coi is consistent with the bulk of the population.

\section{Spectroscopy} \label{sec:spec}

Our spectroscopic coverage of SN 2016coi is dense, with 55 spectra in total.
These are presented as a select time series in Figure~\ref{fig:timeseries} and
fully in Figures~\ref{fig:s0}, \ref{fig:s1}, \ref{fig:s2}, and \ref{fig:s3} in
Appendix~\ref{sec:specplots}. Our spectral observations average one spectrum
every two days until two months after the date of classification. The spectral
sequence here is sufficiently dense to follow the evolution of features between
$4000$ \AA\ and $8000$ \AA\ during the photospheric phase in detail, when the SN
is defined by absorption features rather than emission lines. Late time
spectroscopy follows the SN from the photospheric phase and extends into the
early nebular phase.  The first indication of transition into the nebular phase
is seen at around $\sim +63$ d, shown in Figure~\ref{fig:timeseries}, as the
\ion{[O}{I]}$ \lambda \lambda$ 6300, 6363 emission line is clear to see around
$6300$ \AA. It is absent in the spectrum ten days previous. This line continues
to become more prominent against the fading continuum flux for the next month.
Comparison with the colour evolution of $g-r$ in Figure~\ref{fig:colours} shows
that as this feature gets stronger the colour curve turns from blue to red, the
reversal of the blue-ward evolution occurs on day $+110$ and it is around this
time that the SN can be considered to be in the nebular phase as emission lines
dominate the flux.

The journal of spectroscopic observations is presented in Table~\ref{tab:specobs}. We also discuss the presence of some static features in the pre-peak spectra in Appendix~\ref{sec:static}.

\begin{figure*}
	\centering
	\includegraphics[scale=0.73]{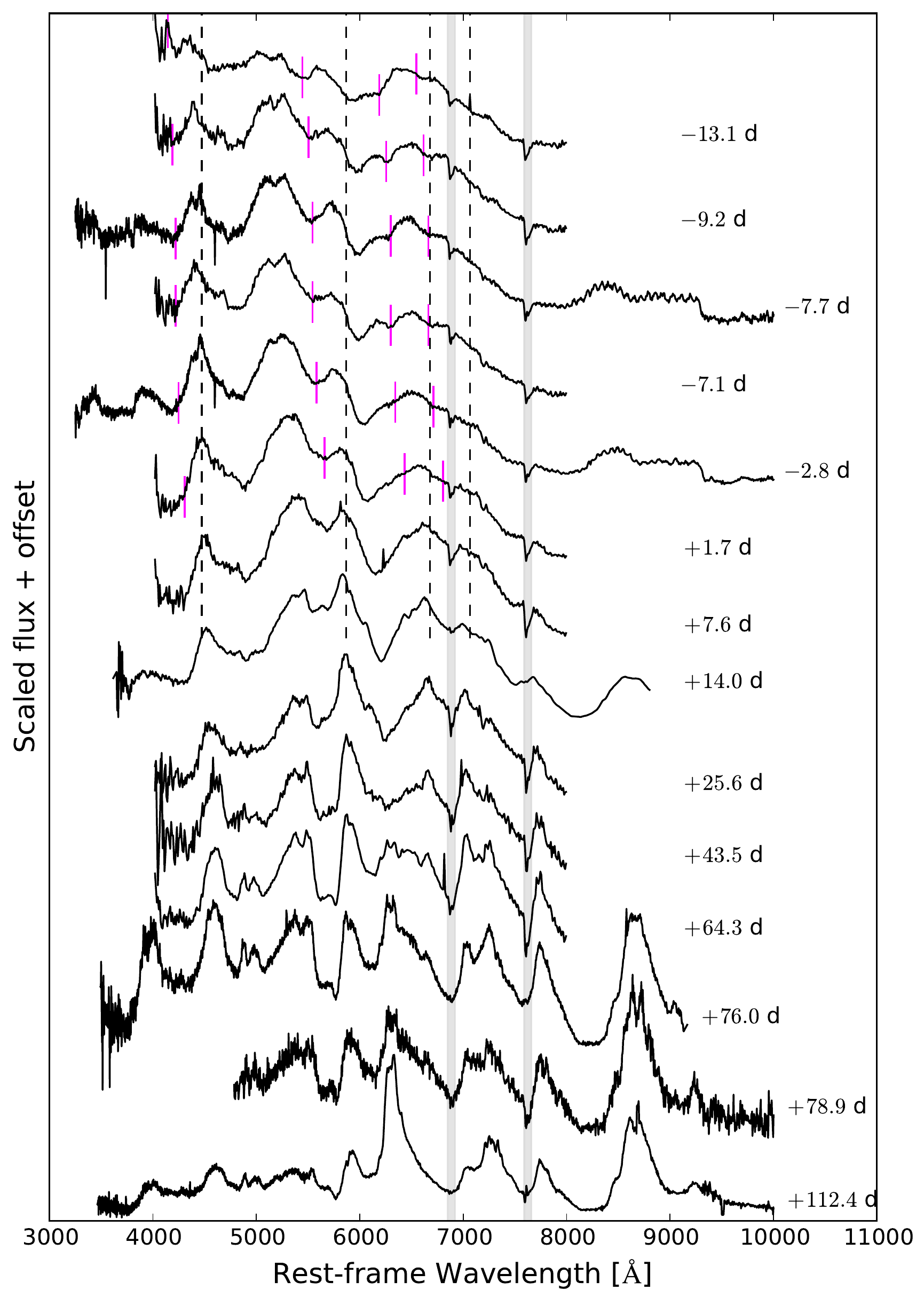}
	\caption{A select time series of spectra of SN 2016coi showing progression from $\sim 2$ d after explosion until the nebular phase.  Labelled versions of the spectra can be found in Section~\ref{sec:mod}. The light grey regions identify strong telluric features, the black dashed lines the rest wavelengths of \HeI\ \lam\lam 4472, 5876, 6678, 7065, and magenta the Doppler shifted position of these lines as given by the velocity from \HeI\ \lam 5876 and the minimum of the blue-ward absorption profile.} 
	\label{fig:timeseries}
\end{figure*}

\subsection{Preliminary classification}\label{sec:preliminaryclass}

\begin{figure}
	\centering
	\includegraphics[scale=0.43]{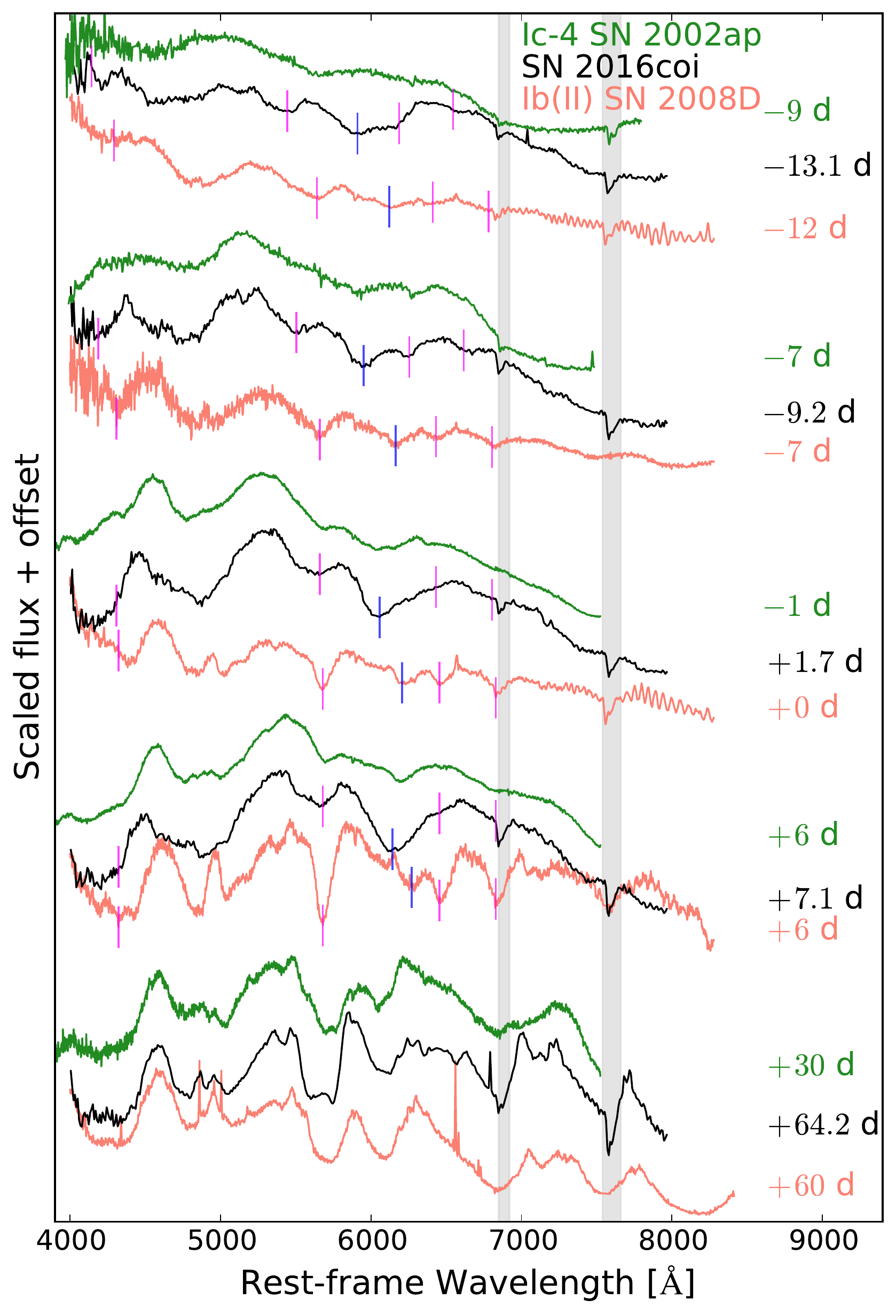}
	\caption{The spectra of SN 2016coi (black) in comparison with Ic-4 SN 2002ap (green) and He-rich SN 2008D (red) at various epochs. Magenta/blue lines show the position of Doppler shifted \HeI\ and \SiII\ lines The evolution of SN 2016coi is slower than that of SN 2002ap, which is to be expected as the time-scales are longer for SN 2016coi. It is noticeable that SN 2016coi has more features visible in the early spectra, especially with respect to the \NaI\ D line at $\sim 5500$ \AA. It can also be seen that the SN 2016coi line velocities are lower than SN 2002ap at very early times but are higher by peak. SN 2008D initially has broad lines that give way to a spectrum with strong narrow lines and dominated by He at peak.}
	\label{fig:comparespec}
\end{figure}

SN 2016coi was originally classified as a broad line SN Ic. Indeed, it has many
spectroscopic similarities to Ic-3/4 SNe. In Figure~\ref{fig:comparespec} we
show SN 2016coi in conjunction with Ic-4 SN 2002ap \citep{Mazzali2002} and SN
Ib(II) 2008D \citep{Mazzali2008,Modjaz2009}, the spectral epochs are scaled as close to SN 2016coi according to relative light curve width wherever possible. 
SN 2002ap is a typical He-poor SN with broad spectral features. SN
2008D, associated with X-ray flash 080109, showed broad lines in its early
spectra and was originally classified as a SN Ic. However, He lines gradually
appeared confirming its classification as a He-rich SN.  

The supernovae are similar in the early spectra, all showing broad absorption features. However, there are differences in velocity and strength of these features. In the case of SN 2008D ($\sim$ 12 days before \tmax) the first signs of broad He can be seen around $\sim 5600$ \AA\ and $\sim 6400$ \AA. SN 2016coi ($\sim$ 13 days before \tmax) shows more features than the SN 2002ap ($\sim 3$ 9 days before \tmax) but fewer than SN 2008D. In each case the \OI\ \lam\lam\ 7771,7774,7775 and \CaII\ NIR triplet remain blended. 
As the SNe move towards peak SN 2016coi retains a similar spectral shape to SN 2002ap, but with stronger features and a prominent $\sim 5500$ \AA\ absorption. In SN 2008D the lines become narrower and more prominent. Shortly after peak SN 2002ap and SN 2016coi show many of the same features with the key difference being the strength and velocity of these features. SN 2008D has progressed to look more like a standard He-rich SN. The later spectra demonstrate that there is a tendency to spectral similarity for He-rich and He-poor SNe.

\begin{figure}
	\centering
	\includegraphics[scale=0.43]{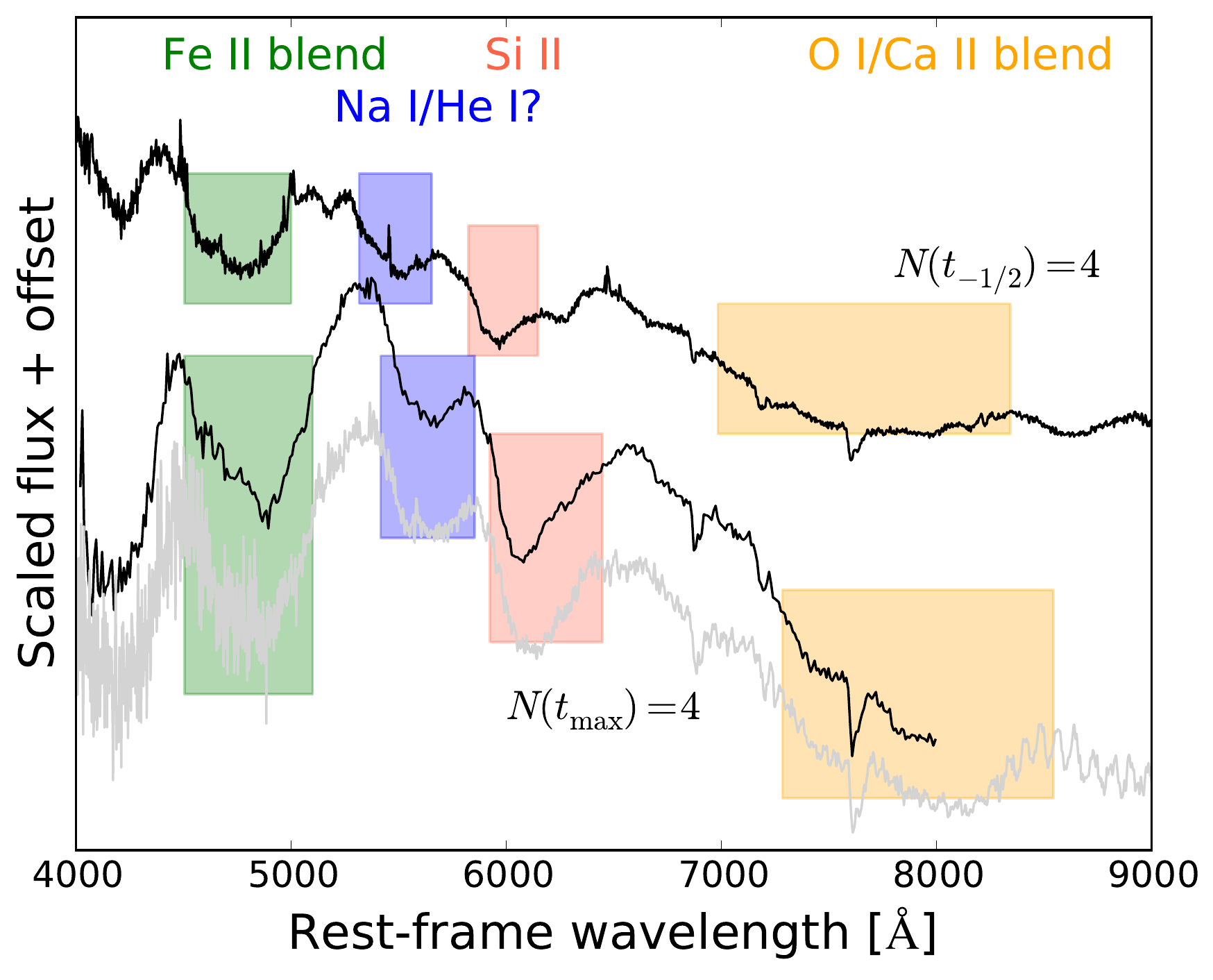}
	\caption{The classification spectra of SN 2016coi at \trise\ (top) and \tmax\ (bottom). The light grey spectrum is at $+2.6$ d and is included to show the blending of the \OI\ and \CaII\ lines around $7500-8000$ \AA. Highlighted are the line blends used to determine $N$, and consequently $\left<N\right>$. The SN has $N=4$ at both epochs which leads to the classification of a Ic-4 SN but matters are complicated if He is present in the ejecta as it would no longer fit into this classification scheme. }
	\label{fig:classification}
\end{figure}

If we consider the similarity of SN 2016coi and SN 2002ap at peak then, from the
classification scheme in \citet{Prentice2017}, typical Ic-4 SNe show three
blended lines at \trise\ and 5 at \tmax. SN 2016coi is a little different in
this regard in that it fulfils the criteria for 4 lines at \trise\ and 4 lines
at \tdecay, as shown in Figure~\ref{fig:classification}. Clearly the earliest
spectra are different as SN 2016coi shows more structure in its spectra that SN
2002ap, the most obvious difference is the appearance and strength of the
absorption features around 5500 \AA\ and 6000 \AA. In SNe Ic, the features are
normally attributed to \NaI\ D and \SiII\ \lam\ 6355 respectively but in light of
the comparison here, and the findings of \citet{Yamanaka2017}, could it be that
Helium contributes to, or is entirely responsible for, the former? In
Section~\ref{sec:mod} this possibility is investigated using a spectrum
synthesis code, here were consider more analytical methods.

\subsubsection{Testing for He via line profile}

\begin{figure}
	\centering
	\includegraphics[scale=0.43]{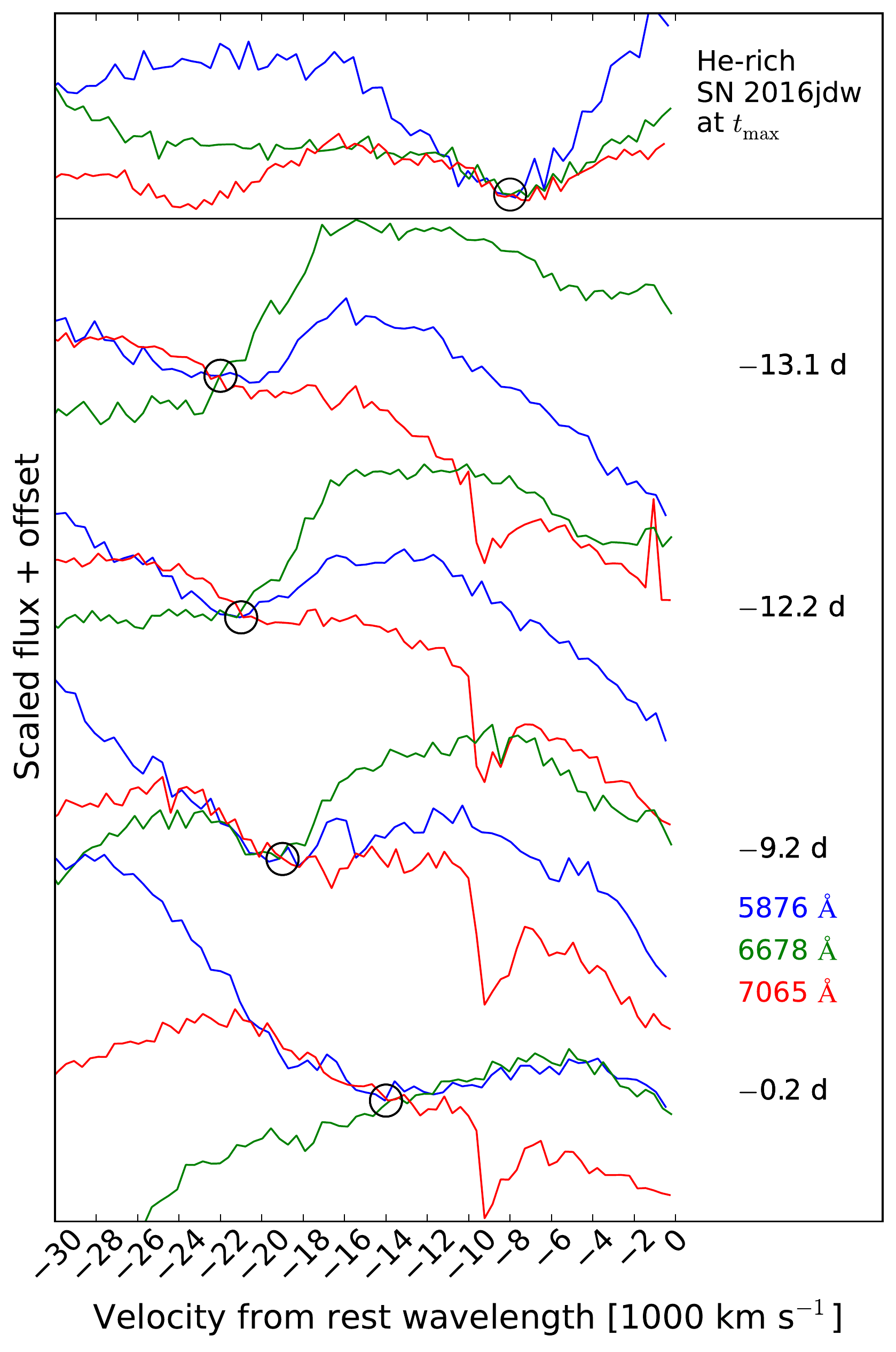}
	\caption{In order to test line profiles for a common origin we plot the pre-peak spectra of SN 2016coi, in velocity space relative to the rest wavelength of the He \lam\ 5876 (blue), \lam\ 6678 (green), and \lam\ 7065 (red) lines. The flux at a common velocity, as determined by the velocity measured from the 5876 \AA, relative to the rest wavelength is used to normalise each spectral region. For comparison the maximum light spectrum of the He-rich SN 2016jdw is included, which demonstrates how He forms within a common line-forming region. In SN 2016coi, the 5876 \AA\ line profile occasionally matches the shape of one of the other two in either the red or the blue, but at no point do the absorption features well match each other. }
	\label{fig:He}
\end{figure}

Figure~\ref{fig:He} demonstrates a test for common line forming regions using spectra in velocity-space.
Three prominent helium lines are plotted on top of each other at four separate epochs.
If the shape of the absorption profiles and the absorption minima (i.e. the velocities) are the same then the lines are formed in the same region, which is evidence for line transitions from the same element.

In the case of He-rich SN 2016jdw, the line profiles are very
similar (see top panel of Figure~\ref{fig:He}), indicating that the line forming
regions are the same and a result of absorption by He.  
However, with SN 2016coi the absorption profiles are dissimilar in both shape and absorption minima, The closest similarity is at $-9.2$ d.
The presence of broad lines complicates matters here as, aside from the width of the lines, multiple species occupy the same line forming region making it difficult to attribute the feature to one dominant transition. 
Thus, the identification of He cannot be confirmed as for the most part the absorption profiles are not similar at these epochs. 

\subsection{Photospheric phase}
The first spectrum is taken at $ -13.1$ d (Figure~\ref{fig:timeseries}), approximately 2 days after explosion. It is defined by several prominent absorption features.
In Section~\ref{sec:mod} we apply spectroscopic modelling to investigate these features further. In SN 2016coi, these lines, while broad, also appear well defined. This is unusual, as ``broad-lined'' SNe typically have fewer than 4 lines visible at this epoch (See Figure~\ref{fig:comparespec}). 
The blue-most absorption features at $\sim 5000$ \AA\ are normally attributed to blends of \FeII\ lines, while in the middle of the spectrum the two features at $\sim 5500$ \AA\ and $\sim 6000$ \AA\ are usually attributed to blends of \NaI\ D and \HeI\ \lam\ 5876, and \SiII\ \lam\ 6355 and \Ha\ respectively in He-rich SNe and just \NaI\ D and \SiII\ \lam\ 6355 in He-poor SNe. 7000 -- 8000 \AA\ is dominated by a blend of \OI\ \lam \lam\ 7771, 7774, 7775 and the \CaII\ NIR triplet. 
In the blue, the \FeII\ group around 5000 \AA\ appears to remain blended until a week after maximum, when there are weak indications of the three prominent \FeII\ lines (\lam\lam\ 4924, 5018, 5169). This behaviour is common in broad lined SNe.
Evolution of the 5500 and 6000 \AA\ features (Figure~\ref{fig:timeseries}) indicate that both are constructed of several components. The 5500 \AA\ feature appears to be formed from at least two components of similar strength. At around $0$ to $+7$ d (Figure~\ref{fig:timeseries}) the red component briefly becomes the stronger of the two after which the blue component dominates. At no point do these lines fully de-blend. The $\sim 6000$ \AA\ line de-blends into what is normally considered to be \SiII\ \lam 6355 and \CII\ \lam\ 6580 \AA, the latter of which could be \HeI\ \lam 6678. The 6000 \AA\ feature remains very strong and well defined until $\sim$ 3 weeks after maximum. The emission peak immediately blue-ward, associated with \NaI\ D, becomes sharp from $\sim +7$ d and can be traced all the way to the nebular phase. The \OI\ and \CaII\ NIR blend remains in place until $\sim +6$ days, at which point the \OI\ absorption becomes distinct.

\subsubsection{Line velocities} \label{sec:vels}

\begin{table}
	\caption{Lines used to define velocity }
	\begin{tabular}{ll}
    \hline
	Ion & $\lambda$/ [\AA ] \\
    \hline
	\ion{Fe}{II} & 4924\\
	\ion{Fe}{II} & 5018\\
	\ion{Fe}{II} & 5169\\
    \HeI & 5876\\
	\ion{Na}{I} & 5891\\	
	\ion{Si}{II} & 6355\\
	\ion{O}{I} & 7774\\
	\ion{Ca}{II} & NIR triplet\\
    \hline
	\end{tabular}
	\label{tab:lines}
\end{table}
We calculate the line velocities for various features, which we attribute to the lines given in Table~\ref{tab:lines}. It is important to note however that our labelling is not meant to be a conclusive line identification (see Section~\ref{sec:mod}) and that some features are blends of several lines. This is especially important with the \FeII\ region around 5000 \AA\ where the \FeII\ \lam\lam 4924, 5018, 5169 lines blend with each other and with other lines in the region.

Measurements are taken from the minimum of the absorption feature. The uncertainty on this value is then the range of velocities in that region that returns a similar flux to that of the minimum. For example, a narrow line will result in a small range of velocities as the flux rises rapidly around the minimum. For broader lines the absorption is shallower, resulting in a larger range of possible velocities. Thus, the uncertainty is related to the degree of line blending. Also, in highly blended lines it is extremely unlikely that the measured minimum is caused by a single line (see Section~\ref{sec:02ap}), hence velocity measurements are highly uncertain and this is reflected in the range of values. At later times, as lines de-blend, it becomes easier to associate a particular feature with a particular line. 
 
\begin{figure}
	\centering
	\includegraphics[scale=0.43]{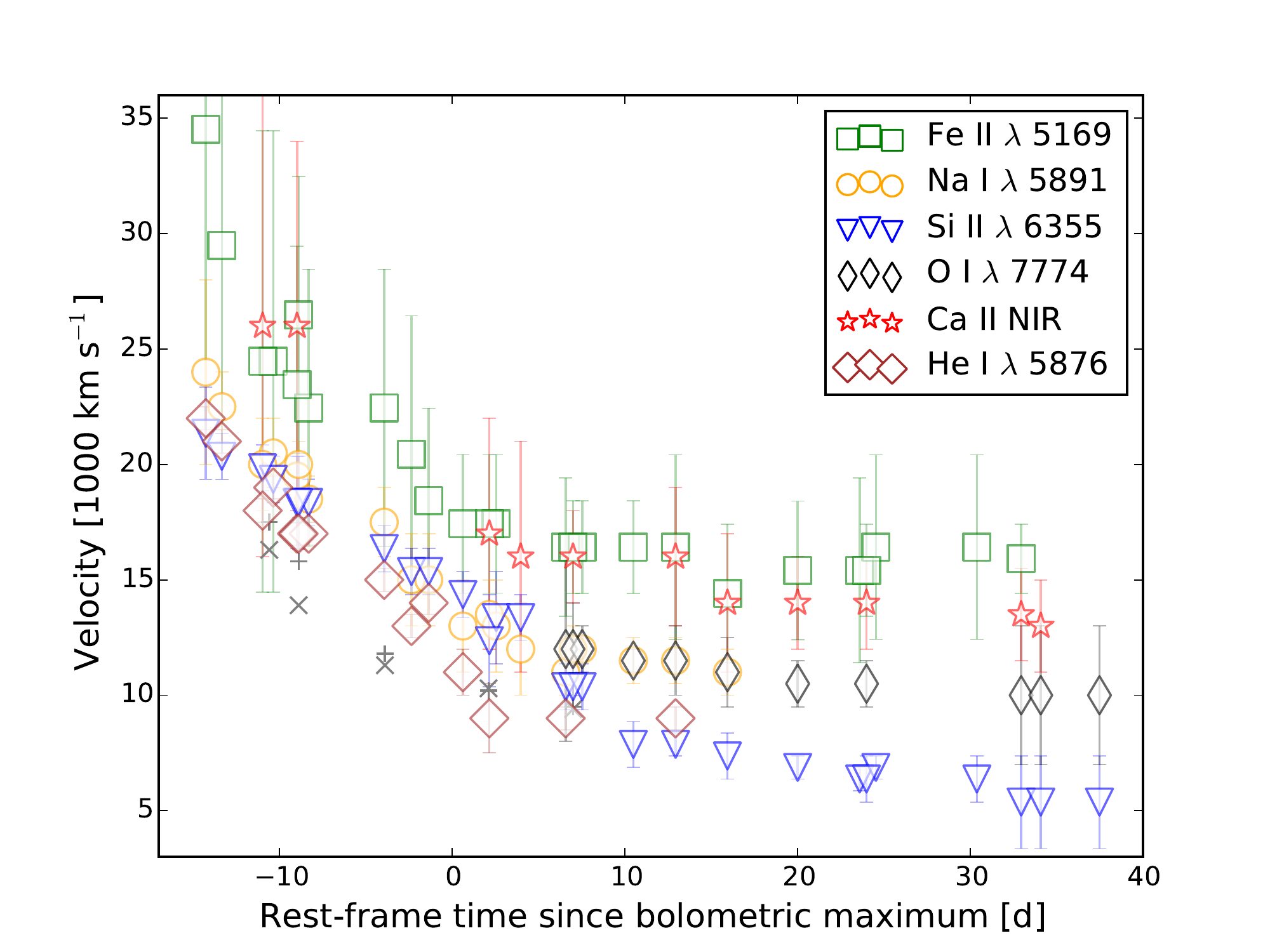}
	\caption{The line velocities of SN 2016coi as measured from absorption minima. Error bars represent the valid spread of velocities that the line could take and, as discussed in the text, we caution against taking line velocities from blended lines. Before peak it appears that the line forming region for \CaII\ and \FeII\ occurs at a similar velocity coordinate, and the same can be said for \SiII\ and \NaI/\HeI. As the photosphere recedes the apparent ejecta stratification becomes clear with \FeII\ and \CaII\ retaining high velocities, \NaI/\HeI\ and \OI\ occupying the same shell, and \SiII\ splitting further. \HeI\ closely traces the photospheric velocity \vph, as derived in Section~\ref{sec:mod}. In those models however, the $\sim 5400$ \AA\ feature is dominated by \NaI\ by maximum light. Thus, the detection of He, and subsequent velocities, are tenuous at this time. } 
	\label{fig:vels}
\end{figure}

Figure~\ref{fig:vels} plots the line velocities derived the absorption minima of the features with their associated labels. It is clear that the \FeII\ velocity is higher than any other measured line at the very earliest epoch ($35,000\pm{10000}$ \kms, modelling suggests 26000 \kms at $-13.1$ d) and, throughout the $\sim 55$ d period over which we make measurements, remains the highest velocity line with the possible exception of \CaII, where the earliest measurement of this feature indicates similar line velocities. 
The projection of Fe, a heavy ion, to high velocities seems counter-intuitive as the lighter elements could be expected to be in the outer layers of the ejecta. In the case of GRB-SNe iron-group elements can be ejected to high velocities as part of a jet \citep{Ashall2017}. Alternatively, because a small abundance of Fe is required to provide opacity, it could be a consequence of dredge-up of primordial material. 
Past maximum it appears that \CaII\ and \FeII\ diverge but note that for the most part both features still show some significant broadening, as indicated by the error bars. As the \CaII\ NIR triplet is a series of lines, and it is hard to attribute the minimum to exactly one line, then it is likely that \FeII\ and \CaII\ form lines at similar velocities. This was noticed in \citet{Prentice2017} for other He-poor SNe. The \FeII\ line finds a plateau at $\sim 16,000$ \kms\ around \tmax\ while for \CaII\ this value is $\sim 14,000$ \kms.

The velocity of \HeI\ \lam\ 5876 is measured from the same absorption feature as \NaI\ D. However, there is an additional constraint on the range of valid velocities for helium as the \HeI\ \lam\lam\ 6678, 7065 lines must also match features in the spectra.  
This means that the position in the absorption feature used to measure velocity varies between \HeI\ \lam 5876 and \NaI\ D, hence differing velocity evolutions, as can be seen in Figure~\ref{fig:vels}.

The \NaI\ D/\HeI\ and \SiII\ lines follow a very similar evolution until near \tmax. If the $\sim 5500$ \AA\ is assumed to be \NaI\ then these remain similar until around a week after maximum at which point \SiII\ appears to level off at around $6,000$ \kms\ while \NaI\ levels off at $\sim 12,000$ \kms. It appears that there is a shell of material that is mixed Na and Si. When the photosphere recedes far enough, the base of the Na layer is revealed and the two velocities decouple.
If the feature is \HeI\ then the \HeI\ velocities remain around or just below \SiII\ and \HeI\ has a sharper velocity gradient than \SiII. \HeI\ then levels off at $\sim 9000$ \kms\ around \tmax. In He-rich SNe \HeI, \CaII, and \FeII\ are typically found at similar velocities, above that of \SiII\ and \OI.

A week after \tmax\ is also the first opportunity to estimate \OI\ as it has de-blended from the \CaII\ NIR triplet, and it can be seen that the line velocity matches that of \NaI. Given that the \OI\ velocity is below that of the earlier \NaI\ and \SiII\ velocities, we can suggest that there is a shell in the ejecta which contains all three elements, this may also include He.

\begin{figure}
	\centering
	\includegraphics[scale=0.5]{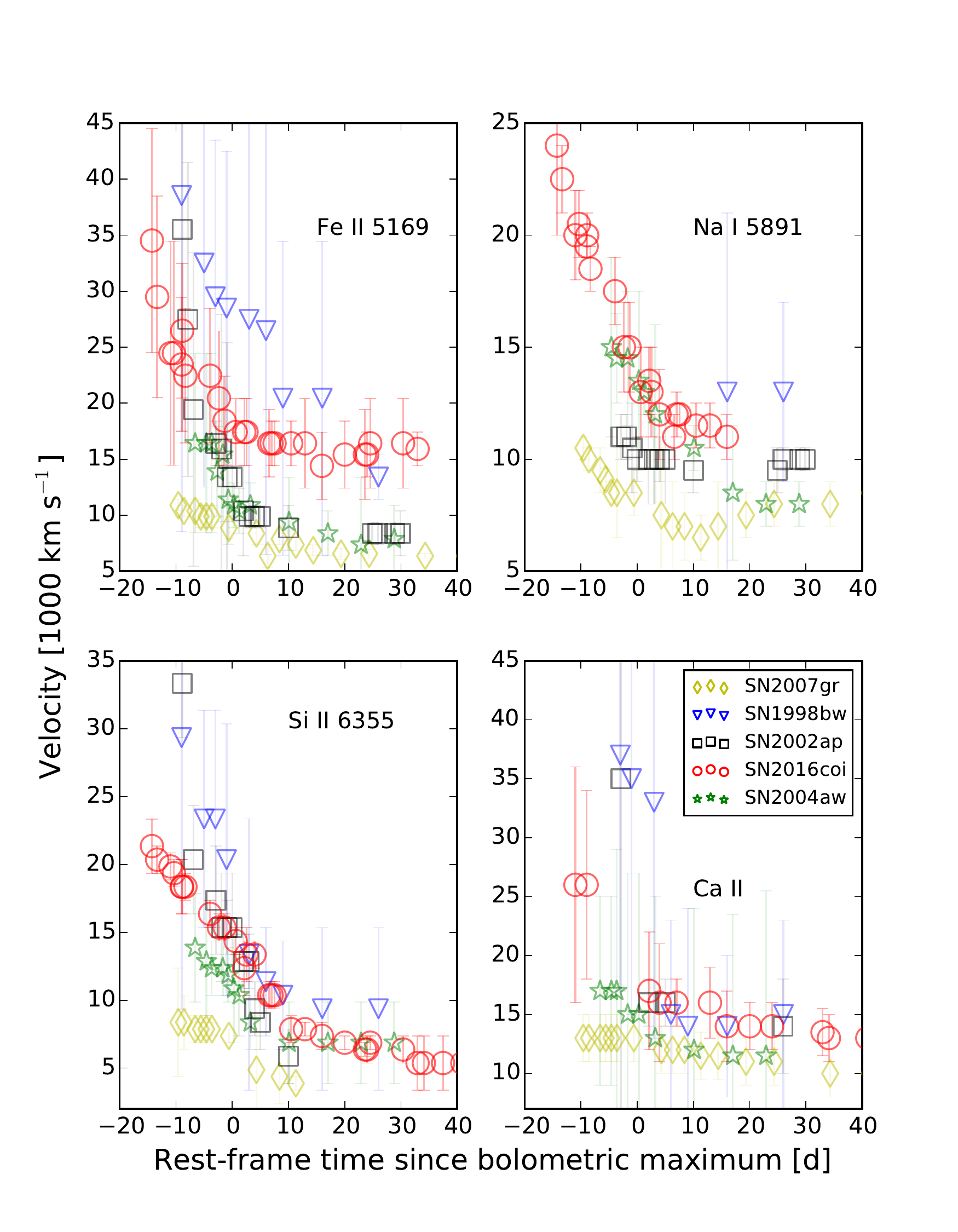}
	\caption{Comparative line velocities between different He-poor SNe types, measured as per Figure\ref{fig:vels}. Included are Ic-3 GRB SN 1998bw (blue), Ic-4 SN 2002ap (black), Ic-6 SN 2004aw (green), Ic-7 SN 2007gr (yellow), and SN 2016coi (red). Difficulties with measuring velocities of highly blended lines are discussed in the text; this is why the line velocities of SN 1998bw can appear discontinuous.}
	\label{fig:velscomp}
\end{figure}

Figure~\ref{fig:velscomp} presents comparison of the line velocities with He-poor SNe, marked according to classification. The velocities of SN 2016coi are comparable to other Ic-3/4 SNe.

\section{Modelling} \label{sec:mod}

\subsection{The photospheric phase}

\begin{figure}
	\centering
	\includegraphics[scale=0.8]{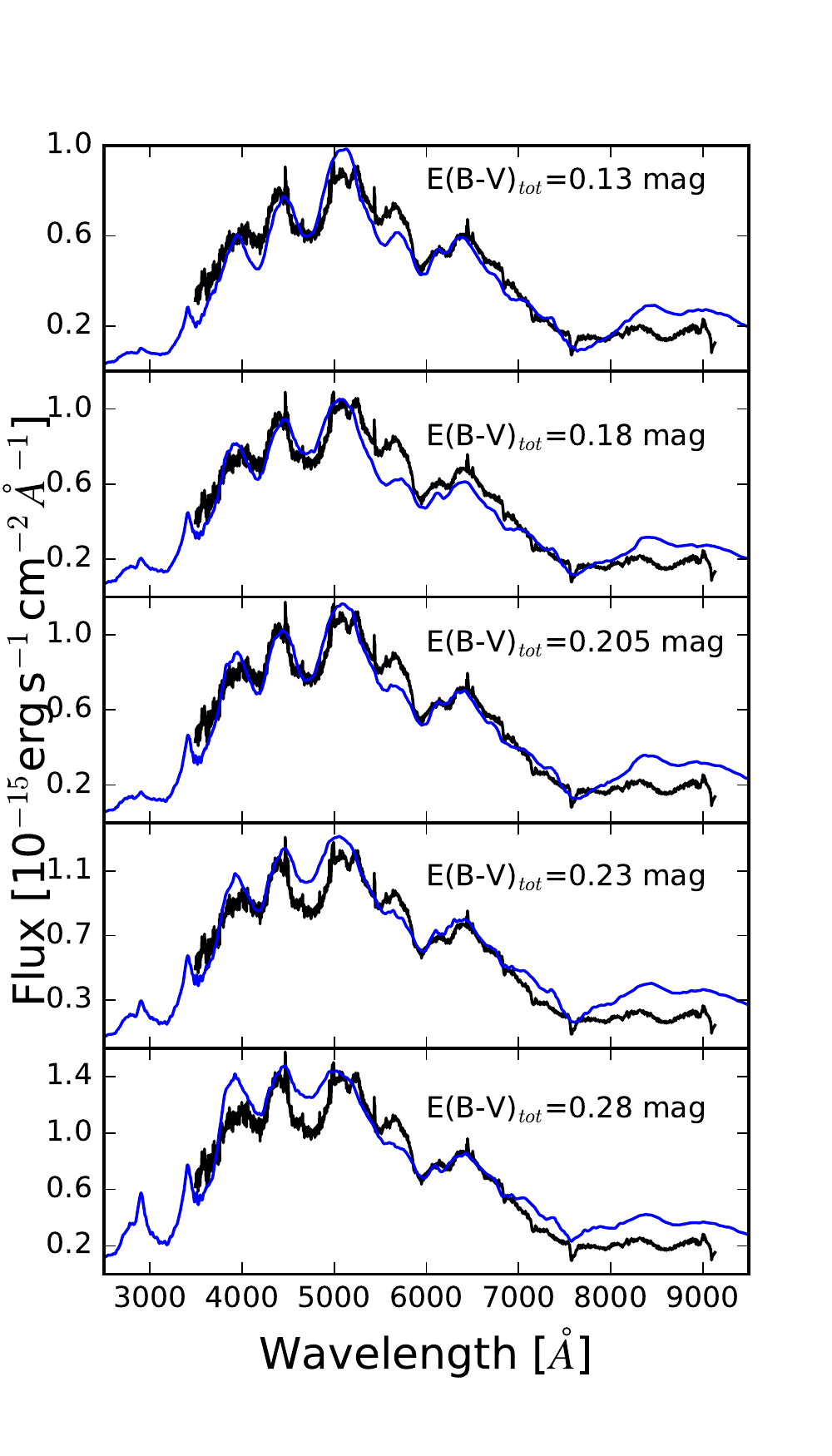}
	\caption{A set of models (blue) produced at $t=5.4$\,d after explosion compared with the $-10.9$ d spectrum (black). The observations have been corrected for different amounts of host galaxy extinction in addition to \Emw\ $=0.08$ mag, and a model produced for each spectrum. The best fits suggest \Eh\ is $0.1-0.15$ mag, greater than this and the luminosity requires a higher temperature which changes the ionization regime of the spectra.}
	\label{fig:reddmodel}
\end{figure}

\begin{figure*}
	\centering
	\includegraphics[scale=0.5]{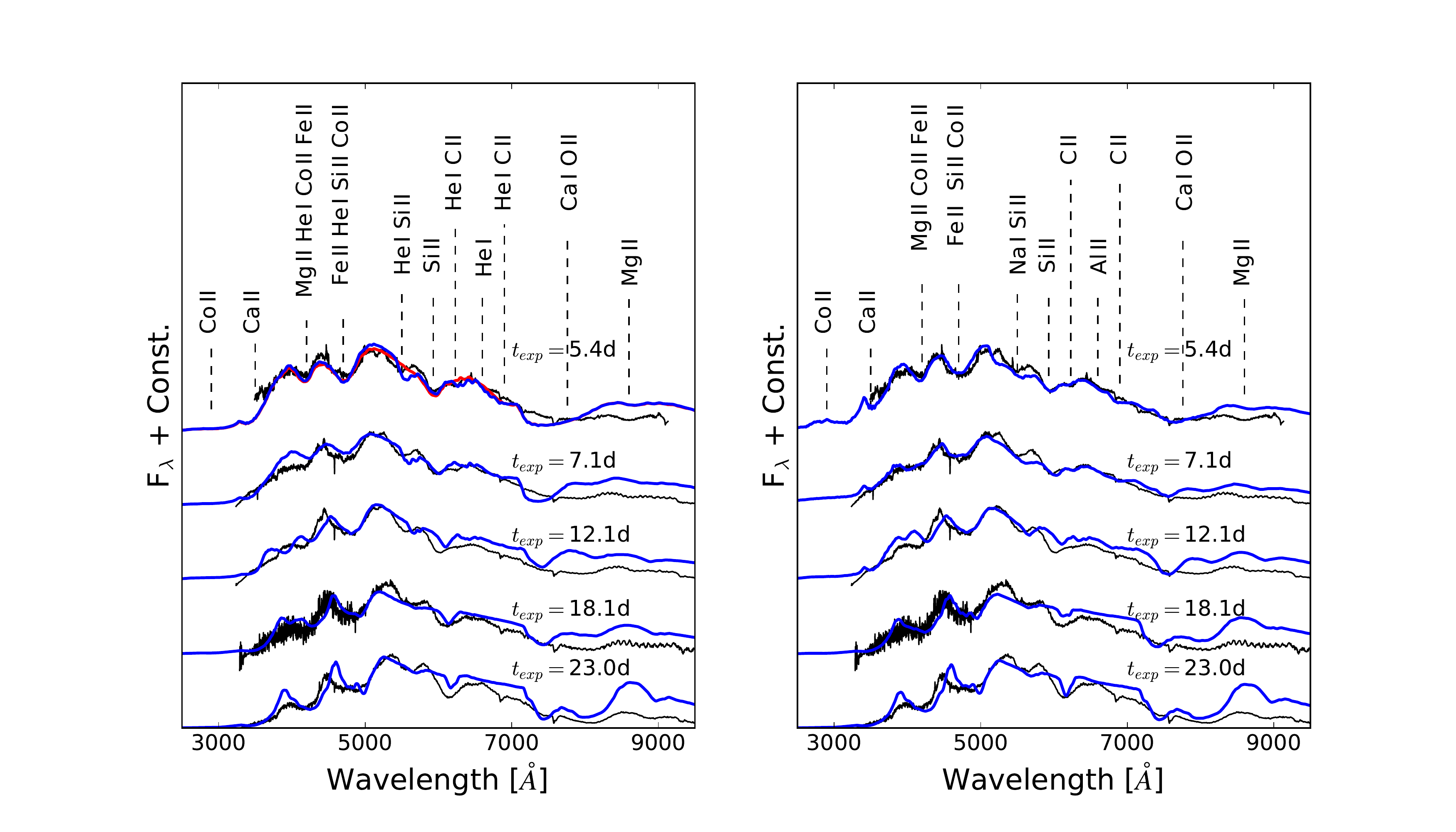}
	\caption{A time series of spectral models (blue) and observations (black) at five different epochs. {\it Left:} The models were produced with He enhanced with departure coefficients. The red model at $t=5.4$\,d is the same as the blue model but with no departure coefficient applied, demonstrating that the \HeI\ lines need to be non-thermally excited. {\it Right:} spectral models with no He abundance.}
	\label{fig:model}
\end{figure*}

\begin{table*}
	\centering
 \caption{Basic parameters of models form Figure \ref{fig:model}.}
	\begin{tabular}{llllll}
      \hline
    &	&	  No He     	&	&	with He	&\\
   $t_\mathrm{exp}$& $t-$\tmax & UVOIR \loglp & \vph & UVOIR \loglp & \vph \\
   
   [days] & [days] &  & [\kms] &   & [\kms]  \\
     \hline
	5.4		&	-10.6	&42.08&	16300	&42.08& 17500	\\
    7.1		&	-8.9	&42.26&	13900	&42.31&	15800	\\
    12.1	&	-3.9	&42.48&	11300&42.49&	11800\\
    18.1	&	+2.1	& 	42.58&	10300&42.54&	10200	\\
    23.0	&	+7.0	&42.52	&	9400	&42.53&9500	\\
     \hline
	\end{tabular}
	\label{tab:inputmodel}
\end{table*}

To determine the elements which
make up the spectra, and to examine the range of possible values for \Eh,
we turn to spectral modelling.
This technique utilizes the fact that SNe are in homologous expansion within a day of explosion, such that $r=v_\mathrm{ph}\times t_\mathrm{exp}$ where $r$ is the radial distance, \vph\ is the photospheric velocity, and $t_\mathrm{exp}$ is the time of explosion. We use a 1D Monte Carlo spectra synthesis code \citep[See][]{Abbott1985,Mazzali1993,Lucy1999,Mazzali2000b}, which follows the propagation of photon packets through a SN ejecta.

The code makes used of the Schuster-Schwarzschild approximation, which assumes that the radiative energy is emitted from an inner boundary blackbody. This approximation is useful as it does not require in-depth knowledge of the radiation transport below the photosphere, while it still produces good results. The code works best at early times, when most of the \Nifs\ is located below the photosphere. At late times significant gamma-ray trapping occurs above the photosphere, and the assumption of the model begins to break down.  This code has been used to model SNe Ia \citep[e.g., SNe 2014J and 1986G][]{Ashall2014,Ashall2016}, SE-SNe \citep[e.g., SNe 1994I and 2008D][]{Sauer2006,Mazzali2008}, and GRB-SNe \citep[e.g., SNe 2003dh and 2016jca][]{Mazzali2003, Ashall2017}.

The photon packets have two possible fates; they either re-enter the photosphere, through a process known as back scattering or escape the ejecta and are ``observed''. 
Packets can undergo Thomson scattering and line absorption. If a photon packet is absorbed it is re-emitted following a photon branching scheme which allows both fluorescence (blue to red) and reverse fluorescence (red to blue) to take place. A modified nebular approximation is used to treat the ionization/excitation state of the gas, to account for non-local thermodynamic equilibrium (NLTE) effects. The radiation field and state of the gas are iterated until convergence is reached.  The final spectrum is obtained by computing the formal integral. 
Non-thermal effects can be simulated in the code in a parametrised way 
through the use of departure coefficients, which modify the populations of the 
excited levels of the relevant ions \citep[e.g.,][]{Mazzali2009}.
To treat \HeI\ we use departure coefficients of $10^4$ \citep{Lucy1991, Mazzali1998,Hachinger2012}.
The purpose of the code is to produce optimally fitting synthetic spectra, by varying the elemental abundance, photospheric velocity and bolometric luminosity.
The code requires an input density profile, which can be scaled in time to the epoch of the spectrum, due to the homologous expansion of the ejecta.

As SN 2016coi shows spectroscopic similarity to Ic-4 SN 2002ap, but with a LC width
which is $\sim40\%$ larger, we use the same density profile that was used
to model SN 2002ap \citep{Mazzali2002}, but scaled up in  \mej\ by 40$\%$.  The
calculated \mej\ is $2.5-4$ \msun\ with a \ke\ $=(4.5-7) \times10^{51}$
erg, with a specific kinetic energy \eom\ of $\sim 1.6$ [$10^{51}$ erg/\msun] throughout (see \cite{Mazzali2017} for a discussion on uncertainties in spectral modelling).

\subsubsection{Investigating \Eh}

As discussed in Section~\ref{sec:host}, \Eh\ is very uncertain. Therefore, we produced a set of four 
models at  
$-10.9$\,d relative to bolometric maximum, and $ 5.4$ d from explosion, see Figure~\ref{fig:reddmodel}.
The models have varying \Eh\ $= 0.05, 0.10, 0.15$ and $0.2$\,mag. 
It is apparent that most of the models produce reasonable fits,
but the model with \Eh\ $=0.2$\,mag is too hot and has a slightly worse fit (i.e., there is not enough absorption in the features at 4200 and 4700 \AA).
Also the model with  \Eh\ $=0.05$\,mag, similar to that derived from the interstellar \NaI\ D absorption, produces acceptable fits. However, other 
properties  (i.e. $g-r$ colour curve and Ni mass to ejecta mass ratio)
of SN~2016coi are in tension with this value of the extinction. 
Therefore, we choose to take a value 
of \Eh\ of 0.125\,mag, which is in-between the two best fits (\Eh\ = 0.1 \& 0.15\,mag).
This value of \Eh\ also provides good spectral fits, see the middle panel of  Figure~\ref{fig:reddmodel}.

\subsubsection{Modelling results}
With the distance, extinction and density profile determined, we  produced spectral models at five different epochs
($t_\mathrm{exp}=5.4, 7.1, 12.1, 18.1$ and 23.0\,d) to determine which
ions contribute to the formation of the spectra, and abundances.
We do this both with and without He, and discuss the
faults and merits with both sets of models. The basic input parameters of the models 
are presented in Table~\ref{tab:inputmodel}. When determining the properties of a SN spectrum it is important to have the correct flux level in the UV/blue, as there is re-processing 
of flux from the blue to the red due to the Doppler overlapping of the 
spectral lines. 
This process is known as line blanketing, a process by which photon packets
only escape the SN ejecta in a ``line free'' window, which is always
red-ward from where they are emitted.

\subsection{Models without He}
The right-hand panel in Figure~\ref{fig:model} presents the spectral models of SN\,2016coi
without He. The main ions are labelled at the top of the panel, and the same lines tend to form the spectra at all epochs.  The photospheric velocity covers a range of 
9400 to 16300 \kms, and the 
bolometric luminosities are roughly consistent 
with those derived in Section~\ref{sec:bol}.
The blue part of the spectrum consists of \MgII\ resonance lines 
(\lam \lam\ 2803,2796), the 
\CaII\ ground state lines (\lam \lam\ 3968, 3934), as well as
blends of metals including \CoII\ lines, 
the strongest of which are \lam \lam\ 3502, 3446, 3387.
The feature at $\sim$4200 \AA, is dominated by
\MgII\ (\lam \lam\ 4481.13 4481.32),
\CoII\ (\lam \lam\ 4569, 4497, 4517) and \FeII\ (\lam\ 4549).
The feature at $\sim$4700 \AA\ consists of a blend of \FeII\ 
(\lam \lam\ 5169 5198),
\SiII\ (\lam \lam\ 5056 5041) and \CoII\ (\lam \lam\ 5017, 5126, 5121, 4980). 
 At $\sim$5600 \AA\ absorption is caused by 
 \NaI\ (\lam \lam\ 5896 5890), and 
 the small absorption on the red-ward side of the feature is
 produced by \SiII\ (\lam \lam\ 5958,5979).
 \SiII\ (\lam \lam\ 6347, 6371) forms the spectra at, $\sim$6000 \AA, 
 and the smaller feature at $\sim$6300 \AA\ is produced by \CII\ 
 (\lam \lam\ 6578, 6583). 
 The notch at $\sim$6700 \AA\ is produced by 
 \AlII\ (\lam \lam\ 7042, 7056.7, 7063.7), 
 there is also \CII\ (\lam\ 7236, 7231) absorption in the 
 same wavelength range as the telluric feature at $\sim$6900 \AA.
 The broad feature at $\sim$7500 \AA\ is a blend 
 of \CaII\ (\lam \lam\ 8498, 8542, 8662) and
 \OI\ (\lam \lam\ 7771, 7774, 7775). 
Finally, the feature at $\sim$8700 \AA\ is dominated 
by \MgII\ (\lam \lam\ 9218, 9244).
Although these models produce a good fits, 
arguably better than those with He, the abundances we 
require for Al and Na are unusual (see Section~\ref{sec:abundance}) 
and are an argument against these line identification.

\subsection{Models with He}

The left panel in Figure \ref{fig:model} contains the spectral models  of
SN~2016coi with He. The main ions which contribute to each feature are labelled
at the top of the plot, and the basic input parameters can be found  in
Table~\ref{tab:inputmodel}. 

The models are similar to those without helium except now the feature at $\sim$4200 \AA, contains \HeI\ (\lam \lam\ 4471.47, 4471.68, 4471.48, 4388), the one at
$\sim$4700 \AA\ has \HeI\
(\lam \lam\ 4923 5016), and the feature at $\sim$5600 \AA\ consist of only \HeI\ (\lam \lam\
5875.61, 5875.64, 5875.96, 5875.63) at $t_\mathrm{exp}$=5.4\,d,
although the small absorption on the red-ward side of the feature is \SiII\
(\lam \lam\ 5958,5979). 

The notch at $\sim$6700 \AA\ is produced by \HeI\ (\lam \lam\ 7065.17, 7065.21
7065.70) absorption, and there is \HeI\ (\lam\ 7281) and \CII\ (\lam \lam\ 7236,
7231)  absorption in the same wavelength range as the telluric feature at $\sim$
6900 \AA. 

For these models the line identification was made at $t_\mathrm{exp}=5.4$\,d.
However, it should be noted that by $t_\mathrm{exp}=18.1$\,d ($\sim 2$ d after
maximum light) the dominant ion in the $\sim$ 5600 \AA\ feature is \NaI, and the
dominant ion in the  $\sim$ 6300 \AA\ feature is \CII. This could however change
if the departure coefficients were different. For example \cite{Hachinger2012}
determined that at 22.1\,d past maximum light, the departure coefficients in the
deep atmosphere layers were $\sim 10^3$, but in the outer atmosphere layers they
were $\sim 10^7$. So it could  be the case that He absorption could produce these
features at later times if the departure coefficients are increased. However, in
the observations these 'He' features do disappear over time, unlike in SNe Ib,
and in other SNe Ic (such as 2002ap) \NaI\ appears $\sim 3$ days before peak,
some $7-10$ days after explosion.

He excitation usually increases with time as more non-thermal particles
penetrate through the SN ejecta. In this case, He lines are possibly present
early on but do not grow in strength over time, rather the opposite. This
suggest that there is only a very small amount of He in the outer layers of the
exploding star, and that there may be some \Nifs\ mixed out to high velocities
to non-thermally excite the lines at early times. At later times, as density
decreases, the opacity in the outer layers is too small for the deposition of
fast particles, even when locally produced.

\subsection{Abundances} \label{sec:abundance}
\subsubsection{The He free models}
The bottom panel of Figure~\ref{fig:abuplot} shows the abundances as a function of velocity for the models that do not include He. The abundances here are generally consistent with the explosion of a C/O core of a massive star, with a few exceptions. The Na abundance is $\sim6$ percent,
given that Na is not produced in a SE-SN explosion, this 
Na must come from the progenitor system. However, 
the metallicity of UGC~11868 is approximately one third solar, 
and the solar abundance of Na is $\sim1\times 10^{-6}$ 
\citep{Asplund2009}. Therefore, it seems physically unlikely that there could be this much Na in this object. 
Furthermore, the average Al abundance is $\sim0.6$ percent, 
This is much larger than the solar value of $\sim2\times 10^{-6}$, demonstrating that 
these line identifications are unlikely. 

\subsubsection{The models with He}
\begin{figure*}
	\centering
	\includegraphics[scale=0.8]{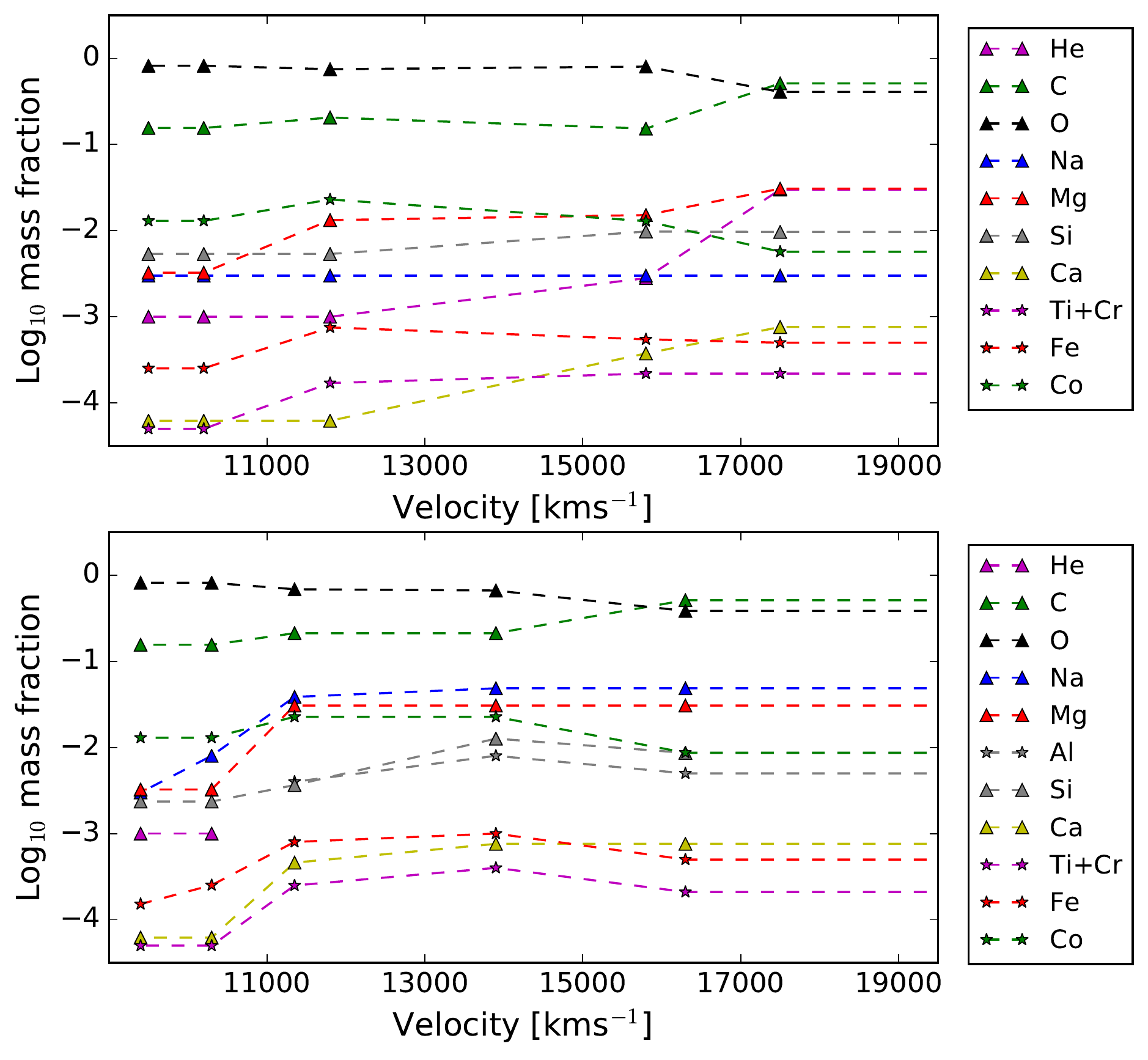}
	\caption{Abundance as a function of velocity for models presented in the previous Figure. The top panel corresponds to the models with He, and the bottom to the models without He.  }
	\label{fig:abuplot}
\end{figure*}

The top panel in Figure~\ref{fig:abuplot} shows the abundances in velocity space for the models that include He.
These abundances are consistent with what could be expected from a core of a massive C/O star. Carbon dominates at the highest velocity, and oxygen at lower velocities. The Na abundance in this model is about 
$\sim3\times 10^{-5}$, which is more consistent with 
solar values, and no Al is required for this model. 
The abundance of He decreases as a function of velocity,
this coincides with the decreasing strength of the He features in the models.
The outer layers (at \vph\ $=17500$ \kms) have a He abundance of 3 percent, whereas 
when the photosphere has receded to 9400 \kms\ the He abundance is 0.1 percent. 
As time passes it should become easier for non-thermal electrons, energised by gamma-rays from \Nifs\ decay, to reach the He layer and excite He into the states required to produce lines in the optical.
However, the small abundance of He in the outer layers, which gets more diffuse as time increases, means that at later times there would be no indication of He in the spectra because the He opacity is insignificant.

In conclusion, the  models indicate that the debated lines could be produced by Na and Al or from He, but 
the abundances from our models suggests that the He identification is more likely.

\subsection{Models of the nebular phase}\label{sec:neb}
\begin{figure}
	\centering
	\includegraphics[scale=0.43]{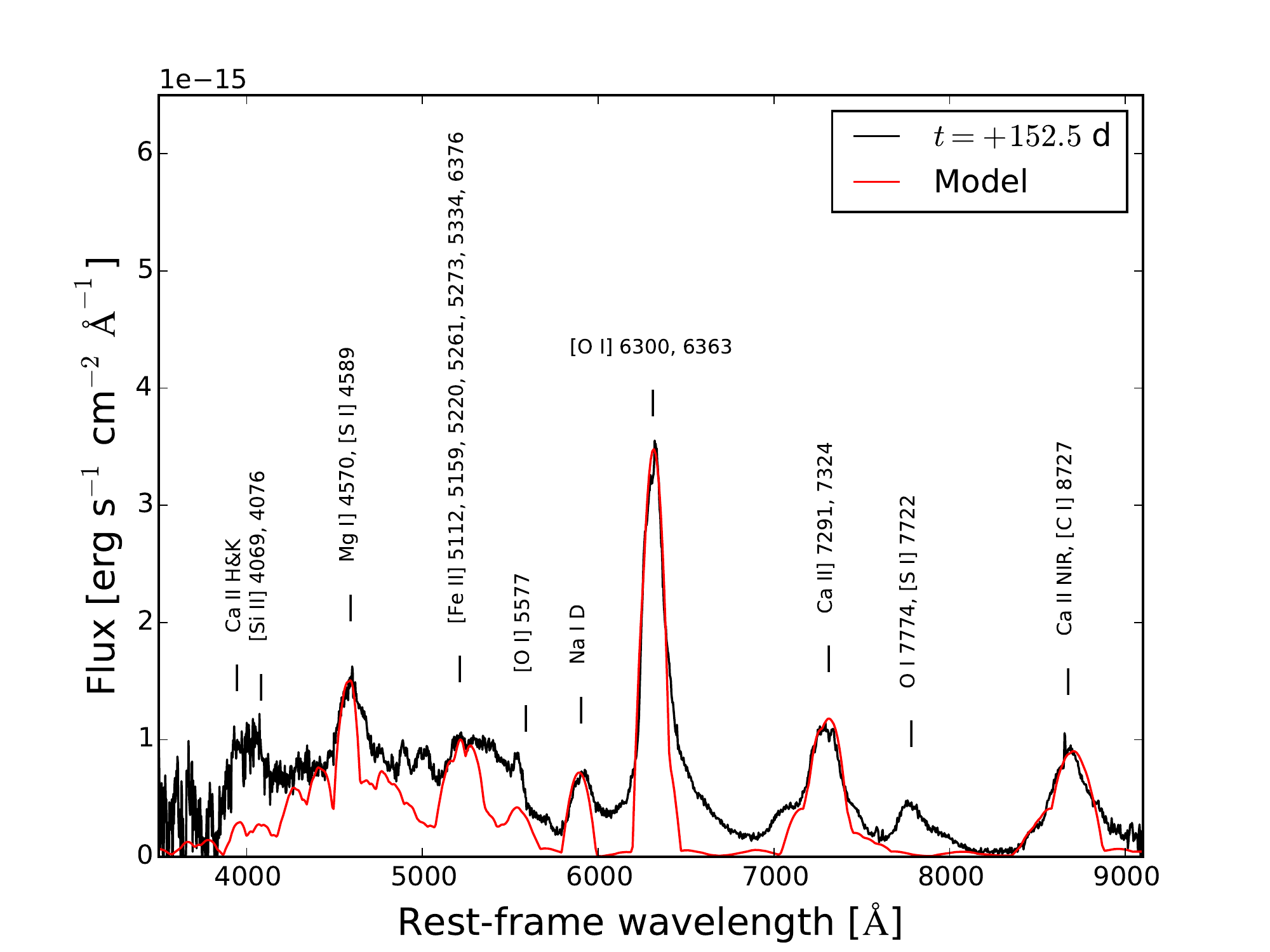}
	\caption{The nebular model of SN 2016coi (red) with the $+152.5$ d spectrum (black)}
	\label{fig:neb}
\end{figure}

We modelled the nebular-epoch spectrum of SN\,2016coi using our non-local
thermodynamic equilibrium (NLTE) code. This code has been described extensively
in \cite{Mazzali2007} and used for both SNe\,Ia and Ib/c \citep[e.g.][]{Mazzali2017}.
It computes the energy produces in gamma-rays and positrons by radioactive decay
of \Nifs\ into \Cofs\ and then \Fefs, it then follows the propagation of the gamma-rays and positrons in the SN ejecta, computes the heating caused by the collisions that characterise the thermalization of these particles, and balances it with cooling via line emission. Both permitted and forbidden transitions can be sources of cooling. An emerging spectrum is computed based on line emissivity and a geometry which can either be stratified in abundance and density or a single zone, which is then bounded by an outer velocity. 

In the case of SN\,2016coi we simply use the one-zone model, as we aim at getting an approximate value of the \Nifs\ mass and the emitting gas. In SNe\,Ic the emitting mass is mostly oxygen, and the range can be from less than 1 \msun\ \citep[SN\,1993J][]{Sauer2006} to about 10 \msun\ in GRB-SNe like 1998bw \citep{Mazzali2001} to several dozen \msun\ in PISN candidates like SN\,2007bi \citep{GalYam2007}. 

The optical spectrum of SN\,2016coi can be reproduced by
an emitting nebula with outer boundary velocity 5000 \kms\ and mass $\sim 1.50 $
\msun within that velocity, Figure~\ref{fig:neb}.  The \Nifs\ mass required to
heat the gas is $0.14$ \msun, which makes SN\,2016coi as luminous as most SNe
Ic, but significantly less luminous than GRB/SNe.  The \Nifs\ mass is verified
also by the flux in the forbidden [\FeII] lines near 5200\,\AA.  The oxygen
mass, as determined by the emission at \lam \lam\ 6300, 6363\,\AA, is 0.75
\msun. Some unburned carbon is also present. \ion{C}{I} \lam\ 8727 contributes to
the red wing of the Ca-dominated emission near 8600\,\AA, and the mass required
is 0.12 \msun. Calcium produces strong emission lines, esp. the semiforbidden
\CaII] \lam\lam\ 7290, 7324, but also H\&K \lam \lam\ 3933, 3968 and the IR
triplet near 8500\,\AA. These are intrinsically strong lines, so that a small
mass of Ca, $\sim 0.005$ \msun, is sufficient. Other strong lines are due to
magnesium ([\MgI\ \lam\ 4570) and sodium (\NaI\ D \lam\ 5890), but the masses of
these elements are quite small ($\sim 2\times10^{-3}$ and $2\times10^{-4}$
\msun, respectively) Finally, the mass of the most abundant intermediate-mass
elements, silicon and sulphur, can be determined indirectly.  For Si, the model
should not exceed the observed emission near 6500\,\AA, which could be \SiI]
\lam\ 6527, which sets a limit of $\sim 1.50$ \msun\ and leads to a strong
emission near 1.6 $\mu$m, while for sulphur the strongest optical line is \SII] \lam\ 
4069, which is well reproduced for a mass of 0.12 \msun, leading to strong NIR
emission near 1.1 $\mu$m.  It is unusual for the Si/S ratio to be as large as 10
or more, so if the upper limit for sulphur is established silicon should
probably not exceed $\sim 0.5$ \msun. The availability of NIR data would improve
the constraints on these elements. The model does not use clumping. If clumping
is used, a slightly worse ratio of the permitted and semi-forbidden \CaII] lines
is produced, suggesting that the density is too high in that case.

On the other hand, photospheric phase modelling indicates that the mass enclosed
within 5000\,\kms\ should be $\sim 1$ \msun. As oxygen and \Nifs\ alone already
reach almost that value, this suggests that the mass of silicon and sulphur
should be small. A reasonable match to the late-time spectrum can be found for a
\Nifs\ mass of 0.14 \msun, an oxygen mass of 0.75 \msun, carbon 0.11 \msun,
calcium 0.005 \msun. This model has a very weak NIR flux. If the bluest part of
the spectrum is not well calibrated the poor apparent fit to the \CaII\ H\&K and
\SI] emissions does not constitute a problem. The model does not use clumping.

\section{Discussion}\label{sec:discussion}
\subsection{Helium}
\citealt{Yamanaka2017} claimed detection of He in the early spectra of SN 2016coi initially. Here we consider the evidence based on our own analysis. In the first instance we consider the arguments for He:

\begin{itemize}
	\item{Several absorption features line up with prominent He lines in the early spectra}
    \item{Using a sensible departure coefficient, our spectral models can include realistic quantities of He}
    \item{Abundance estimates for other likely ions, in particular Na and Al, disfavour these elements having a strong contribution at early times}
    \item{The very early spectra show more absorption features than other ''broad-lined'' SNe, suggesting the nature of this object is a little different}
\end{itemize}

However, there are also reasons to doubt this identification. 

\begin{itemize}
	\item{The He lines do not behave as they do in He-rich SNe as the lines decay in strength over time rather than increase in strength \citep[e.g.,][]{Liu2016,Prentice2017}. }
    \item{The features attributed to \HeI\ do not share the same shape, suggesting that they are not tracing the same line-forming region. In He-rich SNe the shape of the \HeI\ \lam \lam\ 6678, 7065 lines are extremely similar}
    \item{The velocity of He in relation to other elements is unlike that in He-rich SNe where He is found to have the highest ejecta velocity (or second highest if H is present). We instead find that the prominent $\sim 5400$ \AA\ feature attributed to \HeI\ \lam 5876 shows a velocity similar to \SiII}
    \item{If the $\sim 5400$ \AA\ feature remains He dominated after maximum then it levels off to a velocity lower than that of \FeII\ and \CaII}
\end{itemize}

Thus, we have a model which suggests that there may be a small amount of He in the ejecta,
but its behaviour is unlike that in any He-rich SN. NIR spectra would aid in the
identification of helium as \HeI\ \lam\ 2.058 $\mu$m is a strong line that is somewhat
isolated from other prominent features. If He is present, then the progenitor
appears to have been almost completely stripped of it prior to explosion,
leaving some residual amount to imprint on the spectra.

\subsection{Classification}\label{sec:class}

The classification of SN 2016coi is complicated by the uncertain presence of He
in the ejecta, \cite{Yamanaka2017} took the view that this was a broad-lined
type Ib. In the previous section however, it was noted that the identification
of He is promising, but not conclusive.  In \citet{Prentice2017} two methods for
classifying SE-SNe were presented for He-rich and He-poor SNe.  In that work no SE-SNe was found to have just a small
quantity of He present, it was either clearly there or absent. There are more
than a dozen examples of SNe classified as type Ic with spectroscopy earlier
than a week before maximum light, when features akin to those in SN 2016coi
would be visible.  It is unexplained how the envelope stripping process could
leave a clear distinction between SNe with enough He to form strong lines and no
He. If SN 2016coi was to have He in the ejecta then would this represent a
unique intermediate case? This seems unlikely, as our results suggest that if
there is He present in the ejecta it is not in the form of a He shell, but a
result of some mixing into the C/O layer below this (as per the line
velocities). It may be the case here that residual He is non-thermally excited
by \Nifs\ mixed into the outer layers but the abundance is so small that it
rapidly becomes too diffuse to further influence line formation. Indeed, the
velocity of \HeI\ lines closely follows that of the photosphere. This may also suggest
that because \Nifs\ mixing is expected to occur in SNe~Ic, especially those with
broad lines, and because these features are not seen then they are genuinely
free of He. 

SN 2016coi is spectroscopically similar to Ic-4 SNe around peak, regardless of the arguments for and against the presence of He. Thus, based upon the method for SN classification presented in \citet{Prentice2017}, we classify SN 2016coi as Ic-4(14/21) where 4 is the mean number of features visible in the spectra at \trise\ and \tmax, 14 is the velocity of the \SiII\ line at \tmax\ in units of 1000 \kms\ (see Section~\ref{sec:vels}), and 21 is \tdecay\ in units of rest-frame days. This classification system provides a quick reference to compare He-poor SNe line blending and physical properties.

The typical specific kinetic energy of Ic-4 SNe is $\sim 2$, but they show a
wide range of \lp, \ek, and \mej, as well as temporal parameters.  SN 2002ap is
relatively dim, had low \mej\ $\sim 2.5$ \msun\ and \ek\ $\sim 4\times10^{51}$
erg whereas Ic-4 SN 1997ef had \mej\ $\sim 8 M_\odot$ and \ek\ $\sim
16\times10^{51}$ erg. SN 2016coi sits between the two with of \mej\ $=4-7$
\msun, \ek$=(4.5-7)\times10^{51}$ erg and \eom\ $\sim 1.6$ [$10^{51}$ erg/\msun].

A final point on the classification is that as more examples of this type of SN are discovered, and if the presence of He is confirmed, then the taxonomy should be adjusted to account for these objects and their physical parameters. We suggest that the scheme from \cite{Prentice2017} could be adapted to account for this by referring to such SNe as Ic(b)-\mN.

\section{Conclusions}\label{sec:conclusions}

We have presented the photometric and spectroscopic evolution of type Ic-4(14/21) SN 2016coi for around 460 days, from shortly after explosion until the nebular phase. 
Our dense sample of 56 spectra has allowed us to trace the evolution of features within the SN and the transition to the nebular phase in unprecedented detail. 
It has been found that the SN has quite typical light curve parameters in comparison with other non-GRB associated type Ic events with a $griz$ pseudo-bolometric \lp$=(1.9\pm{0.1})\times10^{42}$ \ergs and $ugriz$ \lp$=(2.4\pm{0.1})\times10^{42}$, and an estimated total bolometric UVOIR luminosity of \lp$\sim 3\times10^{42}$. 
The $griz$ temporal parameters, with \trise\ $=12.4\pm{0.5}$ d and \tdecay\ $23\pm{1}$ d, are a little more extreme than normal SNe Ic but not as extreme as the broadest SNe Ic. The SN follows a late time decay rate (non-host subtracted) of 0.015 mag d$^{-1}$.
The SN synthesised $\sim 0.14$ \msun\ of \Nifs\ based upon the estimated UVOIR light curve peak and the results of spectral modelling (contrasting with $\sim$0.09 \msun\ from the $griz$ \lp\ alone).
We note that there is significant uncertainty in the distance to the SN and the value of \Eh. We have attempted to constrain these properties where possible but accept that \mni\ and \lp\ are greatly affected by changes to these values.

Spectroscopically the SN shows broad absorption features common the Ic-3/4 SNe,
but a strong absorption around $\sim 5400$ \AA\ in the early spectra is unusual.
This feature, in conjunction with weaker features at $\sim 6100$ \AA\ and $\sim
6500$ \AA, suggests the presence of helium in the ejecta. However, spectral
models and observational evidence indicates that if this is He then it is in
very low quantities and located at low velocities in a C/O dominated shell. It
becomes diffuse quickly as the lines weaken with strength towards maximum light,
contrary to the behaviour of the lines in He-rich SNe. By $\sim+2$ d the feature
is dominated by Na. We conclude that the SN is a ''broad-lined '' type Ic that
likely retains trace He mixed in the a C/O dominated shell, which is
non-thermally excited at early times by \Nifs\ projected into the outer layers.
It does not represent a case of a SN Ib with weak He lines.

Spectral models indicate that SN had \ek\ $=(4.5-7)\times10^{51}$ erg and \mej\ $= 2.5-4$ \msun, with $\sim 1.5$ \msun\ below $5000$ \kms. This makes SN 2016coi one of the more massive He-poor SE-SNe and similar to Ic-6 SN 2004aw \citep{Mazzali2017}. 
The host is LMC-like, and the SN exploded away from star-forming regions. Additionally we suggest that the metallicity of the host is similar to that of the LMC. 
From this and \mej\ we estimate that the progenitor, likely a Wolf-Rayet star at the time of explosion, had $M_\mathrm{ZAMS}=23-28$ \msun.

\section*{Acknowledgements}
SJP and CA are funded by a Science and Technology Facilities Council (STFC) grant. 
CA acknowledges the support provided by the National Science Foundation under Grant No. AST-1613472.
The Liverpool Telescope is operated on the island of La Palma by Liverpool John Moores University in the Spanish Observatorio del Roque de los Muchachos of the Instituto de Astrofisica de Canarias with financial support from the STFC. This paper makes use of data from Las Cumbres Observatory and the Supernova Key Project. DAH, CM, and GH are supported by NSF grant number 1313484.  Research by DJS and L.T. is supported by NSF grant AST-1412504 and AST-1517649. Research at Lick Observatory is partially supported by a generous gift from Google.
Funding for the LJT has been provided by the Chinese Academy of Sciences (CAS) and the People's Government of Yunnan Province. The LJT is jointly operated and administrated by Yunnan Observatories and Center for Astronomical Mega-Science, CAS. 
J.-J. Zhang is supported by National Science Foundation of China (NSFC, grants 11403096 and 11773067), the Youth Innovation Promotion Association of the CAS, the Western Light Youth Project, and the Key Research Program of the CAS (Grant NO. KJZD-EW-M06). 
Finally, we thank the anonymous referee for their time, consideration, and patience. 
%




\bibliographystyle{mnras}
\bibliography{allbib} 

\begin{thebibliography}{}
\makeatletter
\relax
\def\mn@urlcharsother{\let\do\@makeother \do\$\do\&\do\#\do\^\do\_\do\%\do\~}
\def\mn@doi{\begingroup\mn@urlcharsother \@ifnextchar [ {\mn@doi@}
  {\mn@doi@[]}}
\def\mn@doi@[#1]#2{\def\@tempa{#1}\ifx\@tempa\@empty \href
  {http://dx.doi.org/#2} {doi:#2}\else \href {http://dx.doi.org/#2} {#1}\fi
  \endgroup}
\def\mn@eprint#1#2{\mn@eprint@#1:#2::\@nil}
\def\mn@eprint@arXiv#1{\href {http://arxiv.org/abs/#1} {{\tt arXiv:#1}}}
\def\mn@eprint@dblp#1{\href {http://dblp.uni-trier.de/rec/bibtex/#1.xml}
  {dblp:#1}}
\def\mn@eprint@#1:#2:#3:#4\@nil{\def\@tempa {#1}\def\@tempb {#2}\def\@tempc
  {#3}\ifx \@tempc \@empty \let \@tempc \@tempb \let \@tempb \@tempa \fi \ifx
  \@tempb \@empty \def\@tempb {arXiv}\fi \@ifundefined
  {mn@eprint@\@tempb}{\@tempb:\@tempc}{\expandafter \expandafter \csname
  mn@eprint@\@tempb\endcsname \expandafter{\@tempc}}}

\bibitem[\protect\citeauthoryear{{Abbott} \& {Lucy}}{{Abbott} \&
  {Lucy}}{1985}]{Abbott1985}
{Abbott} D.~C.,  {Lucy} L.~B.,  1985, \mn@doi [\apj] {10.1086/162834}, \href
  {http://adsabs.harvard.edu/abs/1985ApJ...288..679A} {288, 679}

\bibitem[\protect\citeauthoryear{{Arcavi} et~al.,}{{Arcavi}
  et~al.}{2011}]{Arcavi2011}
{Arcavi} I.,  et~al., 2011, \mn@doi [\apjl] {10.1088/2041-8205/742/2/L18},
  \href {http://adsabs.harvard.edu/abs/2011ApJ...742L..18A} {742, L18}

\bibitem[\protect\citeauthoryear{{Arnett}}{{Arnett}}{1982}]{Arnett1982}
{Arnett} W.~D.,  1982, \mn@doi [\apj] {10.1086/159681}, \href
  {http://adsabs.harvard.edu/abs/1982ApJ...253..785A} {253, 785}

\bibitem[\protect\citeauthoryear{{Ashall}, {Mazzali}, {Bersier}, {Hachinger},
  {Phillips}, {Percival}, {James}  \& {Maguire}}{{Ashall}
  et~al.}{2014}]{Ashall2014}
{Ashall} C.,  {Mazzali} P.,  {Bersier} D.,  {Hachinger} S.,  {Phillips} M.,
  {Percival} S.,  {James} P.,   {Maguire} K.,  2014, \mn@doi [\mnras]
  {10.1093/mnras/stu1995}, \href
  {http://adsabs.harvard.edu/abs/2014MNRAS.445.4427A} {445, 4427}

\bibitem[\protect\citeauthoryear{{Ashall}, {Mazzali}, {Pian}  \&
  {James}}{{Ashall} et~al.}{2016}]{Ashall2016}
{Ashall} C.,  {Mazzali} P.~A.,  {Pian} E.,   {James} P.~A.,  2016, \mn@doi
  [\mnras] {10.1093/mnras/stw2114}, \href
  {http://adsabs.harvard.edu/abs/2016MNRAS.463.1891A} {463, 1891}

\bibitem[\protect\citeauthoryear{{Ashall} et~al.,}{{Ashall}
  et~al.}{2017}]{Ashall2017}
{Ashall} C.,  et~al., 2017, preprint, \href
  {http://adsabs.harvard.edu/abs/2017arXiv170204339A} {} (\mn@eprint {arXiv}
  {1702.04339})

\bibitem[\protect\citeauthoryear{{Asplund}, {Grevesse}, {Sauval}  \&
  {Scott}}{{Asplund} et~al.}{2009}]{Asplund2009}
{Asplund} M.,  {Grevesse} N.,  {Sauval} A.~J.,   {Scott} P.,  2009, \mn@doi
  [\araa] {10.1146/annurev.astro.46.060407.145222}, \href
  {http://adsabs.harvard.edu/abs/2009ARA%26A..47..481A} {47, 481}

\bibitem[\protect\citeauthoryear{{Barnsley}, {Smith}  \& {Steele}}{{Barnsley}
  et~al.}{2012}]{Barnsley2012}
{Barnsley} R.~M.,  {Smith} R.~J.,   {Steele} I.~A.,  2012, \mn@doi
  [Astronomische Nachrichten] {10.1002/asna.201111634}, \href
  {http://adsabs.harvard.edu/abs/2012AN....333..101B} {333, 101}

\bibitem[\protect\citeauthoryear{{Beasor} \& {Davies}}{{Beasor} \&
  {Davies}}{2016}]{Beasor2016}
{Beasor} E.~R.,  {Davies} B.,  2016, \mn@doi [\mnras] {10.1093/mnras/stw2054},
  \href {http://adsabs.harvard.edu/abs/2016MNRAS.463.1269B} {463, 1269}

\bibitem[\protect\citeauthoryear{{Becker}}{{Becker}}{2015}]{Becker2015}
{Becker} A.,  2015, {HOTPANTS: High Order Transform of PSF ANd Template
  Subtraction}, Astrophysics Source Code Library (\mn@eprint {ascl} {1504.004})

\bibitem[\protect\citeauthoryear{{Berg} et~al.,}{{Berg}
  et~al.}{2012}]{Berg2012}
{Berg} D.~A.,  et~al., 2012, \mn@doi [\apj] {10.1088/0004-637X/754/2/98}, \href
  {http://adsabs.harvard.edu/abs/2012ApJ...754...98B} {754, 98}

\bibitem[\protect\citeauthoryear{{Bianco} et~al.,}{{Bianco}
  et~al.}{2014}]{Bianco2014}
{Bianco} F.~B.,  et~al., 2014, \mn@doi [\apjs] {10.1088/0067-0049/213/2/19},
  \href {http://adsabs.harvard.edu/abs/2014ApJS..213...19B} {213, 19}

\bibitem[\protect\citeauthoryear{{Brown} et~al.,}{{Brown}
  et~al.}{2013}]{Brown2013}
{Brown} T.~M.,  et~al., 2013, \mn@doi [\pasp] {10.1086/673168}, \href
  {http://adsabs.harvard.edu/abs/2013PASP..125.1031B} {125, 1031}

\bibitem[\protect\citeauthoryear{{Cardelli}, {Clayton}  \& {Mathis}}{{Cardelli}
  et~al.}{1989}]{CCM}
{Cardelli} J.~A.,  {Clayton} G.~C.,   {Mathis} J.~S.,  1989, \mn@doi [\apj]
  {10.1086/167900}, \href {http://adsabs.harvard.edu/abs/1989ApJ...345..245C}
  {345, 245}

\bibitem[\protect\citeauthoryear{{Cioni}}{{Cioni}}{2009}]{Cioni2009}
{Cioni} M.-R.~L.,  2009, \mn@doi [\aap] {10.1051/0004-6361/200912138}, \href
  {http://adsabs.harvard.edu/abs/2009A%26A...506.1137C} {506, 1137}

\bibitem[\protect\citeauthoryear{{Clocchiatti} \& {Wheeler}}{{Clocchiatti} \&
  {Wheeler}}{1997}]{Clocchiatti1997}
{Clocchiatti} A.,  {Wheeler} J.~C.,  1997, \apj, \href
  {http://adsabs.harvard.edu/abs/1997ApJ...491..375C} {491, 375}

\bibitem[\protect\citeauthoryear{{Corsi} et~al.,}{{Corsi}
  et~al.}{2011}]{Corsi2011}
{Corsi} A.,  et~al., 2011, \mn@doi [\apj] {10.1088/0004-637X/741/2/76}, \href
  {http://adsabs.harvard.edu/abs/2011ApJ...741...76C} {741, 76}

\bibitem[\protect\citeauthoryear{{Corsi} et~al.,}{{Corsi}
  et~al.}{2012}]{Corsi2012}
{Corsi} A.,  et~al., 2012, \mn@doi [\apjl] {10.1088/2041-8205/747/1/L5}, \href
  {http://adsabs.harvard.edu/abs/2012ApJ...747L...5C} {747, L5}

\bibitem[\protect\citeauthoryear{{Drout} et~al.,}{{Drout}
  et~al.}{2011}]{Drout2011}
{Drout} M.~R.,  et~al., 2011, \mn@doi [\apj] {10.1088/0004-637X/741/2/97},
  \href {http://adsabs.harvard.edu/abs/2011ApJ...741...97D} {741, 97}

\bibitem[\protect\citeauthoryear{{Drout} et~al.,}{{Drout}
  et~al.}{2016}]{Drout2016}
{Drout} M.~R.,  et~al., 2016, \mn@doi [\apj] {10.3847/0004-637X/821/1/57},
  \href {http://adsabs.harvard.edu/abs/2016ApJ...821...57D} {821, 57}

\bibitem[\protect\citeauthoryear{{Eldridge} \& {Maund}}{{Eldridge} \&
  {Maund}}{2016}]{Eldridge2016}
{Eldridge} J.~J.,  {Maund} J.~R.,  2016, \mn@doi [\mnras]
  {10.1093/mnrasl/slw099}, \href
  {http://adsabs.harvard.edu/abs/2016MNRAS.461L.117E} {461, L117}

\bibitem[\protect\citeauthoryear{{Eldridge}, {Fraser}, {Maund}  \&
  {Smartt}}{{Eldridge} et~al.}{2015}]{Eldridge2015}
{Eldridge} J.~J.,  {Fraser} M.,  {Maund} J.~R.,   {Smartt} S.~J.,  2015,
  \mn@doi [\mnras] {10.1093/mnras/stu2197}, \href
  {http://adsabs.harvard.edu/abs/2015MNRAS.446.2689E} {446, 2689}

\bibitem[\protect\citeauthoryear{{Elias-Rosa} et~al.,}{{Elias-Rosa}
  et~al.}{2016}]{EliasRosa2016}
{Elias-Rosa} N.,  et~al., 2016, \mn@doi [\mnras] {10.1093/mnras/stw2253}, \href
  {http://adsabs.harvard.edu/abs/2016MNRAS.463.3894E} {463, 3894}

\bibitem[\protect\citeauthoryear{{Faber} et~al.,}{{Faber}
  et~al.}{2003}]{Faber2003}
{Faber} S.~M.,  et~al., 2003, in {Iye} M.,  {Moorwood} A.~F.~M.,  eds,
  \procspie Vol. 4841, Instrument Design and Performance for Optical/Infrared
  Ground-based Telescopes. pp 1657--1669, \mn@doi{10.1117/12.460346}

\bibitem[\protect\citeauthoryear{{Fan}, {Bai}, {Zhang}, {Wang}, {Chang}, {Xin}
  \& {Zhang}}{{Fan} et~al.}{2015}]{Fan15}
{Fan} Y.-F.,  {Bai} J.-M.,  {Zhang} J.-J.,  {Wang} C.-J.,  {Chang} L.,  {Xin}
  Y.-X.,   {Zhang} R.-L.,  2015, \mn@doi [Research in Astronomy and
  Astrophysics] {10.1088/1674-4527/15/6/014}, \href
  {http://adsabs.harvard.edu/abs/2015RAA....15..918F} {15, 918}

\bibitem[\protect\citeauthoryear{{Filippenko} et~al.,}{{Filippenko}
  et~al.}{1995}]{Filippenko1995}
{Filippenko} A.~V.,  et~al., 1995, \mn@doi [\apjl] {10.1086/309659}, \href
  {http://adsabs.harvard.edu/abs/1995ApJ...450L..11F} {450, L11}

\bibitem[\protect\citeauthoryear{{Folatelli} et~al.,}{{Folatelli}
  et~al.}{2014}]{Folatelli2014}
{Folatelli} G.,  et~al., 2014, \mn@doi [\apj] {10.1088/0004-637X/792/1/7},
  \href {http://adsabs.harvard.edu/abs/2014ApJ...792....7F} {792, 7}

\bibitem[\protect\citeauthoryear{{Foley}, {Berger}, {Fox}, {Levesque},
  {Challis}, {Ivans}, {Rhoads}  \& {Soderberg}}{{Foley}
  et~al.}{2011}]{Foley2011}
{Foley} R.~J.,  {Berger} E.,  {Fox} O.,  {Levesque} E.~M.,  {Challis} P.~J.,
  {Ivans} I.~I.,  {Rhoads} J.~E.,   {Soderberg} A.~M.,  2011, \mn@doi [\apj]
  {10.1088/0004-637X/732/1/32}, \href
  {http://adsabs.harvard.edu/abs/2011ApJ...732...32F} {732, 32}

\bibitem[\protect\citeauthoryear{{Fox} et~al.,}{{Fox} et~al.}{2014}]{Fox2014}
{Fox} O.~D.,  et~al., 2014, \mn@doi [\apj] {10.1088/0004-637X/790/1/17}, \href
  {http://adsabs.harvard.edu/abs/2014ApJ...790...17F} {790, 17}

\bibitem[\protect\citeauthoryear{{Fukugita}, {Ichikawa}, {Gunn}, {Doi},
  {Shimasaku}  \& {Schneider}}{{Fukugita} et~al.}{1996}]{Fukugita1996}
{Fukugita} M.,  {Ichikawa} T.,  {Gunn} J.~E.,  {Doi} M.,  {Shimasaku} K.,
  {Schneider} D.~P.,  1996, \mn@doi [\aj] {10.1086/117915}, \href
  {http://adsabs.harvard.edu/abs/1996AJ....111.1748F} {111, 1748}

\bibitem[\protect\citeauthoryear{{Gal-Yam} et~al.,}{{Gal-Yam}
  et~al.}{2009}]{GalYam2007}
{Gal-Yam} A.,  et~al., 2009, \mn@doi [\nat] {10.1038/nature08579}, \href
  {http://adsabs.harvard.edu/abs/2009Natur.462..624G} {462, 624}

\bibitem[\protect\citeauthoryear{{Georgy}, {Ekstr{\"o}m}, {Meynet}, {Massey},
  {Levesque}, {Hirschi}, {Eggenberger}  \& {Maeder}}{{Georgy}
  et~al.}{2012}]{Georgy2012}
{Georgy} C.,  {Ekstr{\"o}m} S.,  {Meynet} G.,  {Massey} P.,  {Levesque} E.~M.,
  {Hirschi} R.,  {Eggenberger} P.,   {Maeder} A.,  2012, \mn@doi [\aap]
  {10.1051/0004-6361/201118340}, \href
  {http://adsabs.harvard.edu/abs/2012A%26A...542A..29G} {542, A29}

\bibitem[\protect\citeauthoryear{{Hachinger}, {Mazzali}, {Taubenberger},
  {Hillebrandt}, {Nomoto}  \& {Sauer}}{{Hachinger}
  et~al.}{2012}]{Hachinger2012}
{Hachinger} S.,  {Mazzali} P.~A.,  {Taubenberger} S.,  {Hillebrandt} W.,
  {Nomoto} K.,   {Sauer} D.~N.,  2012, \mn@doi [\mnras]
  {10.1111/j.1365-2966.2012.20464.x}, \href
  {http://adsabs.harvard.edu/abs/2012MNRAS.422...70H} {422, 70}

\bibitem[\protect\citeauthoryear{{Helmboldt}, {Walterbos}, {Bothun}, {O'Neil}
  \& {de Blok}}{{Helmboldt} et~al.}{2004}]{Helmboldt2004}
{Helmboldt} J.~F.,  {Walterbos} R.~A.~M.,  {Bothun} G.~D.,  {O'Neil} K.,   {de
  Blok} W.~J.~G.,  2004, \mn@doi [\apj] {10.1086/423126}, \href
  {http://adsabs.harvard.edu/abs/2004ApJ...613..914H} {613, 914}

\bibitem[\protect\citeauthoryear{{Hoeflich} et~al.,}{{Hoeflich}
  et~al.}{2017}]{Hoeflich2017}
{Hoeflich} P.,  et~al., 2017, \mn@doi [\apj] {10.3847/1538-4357/aa84b2}, \href
  {http://adsabs.harvard.edu/abs/2017ApJ...846...58H} {846, 58}

\bibitem[\protect\citeauthoryear{{Holoien} et~al.,}{{Holoien}
  et~al.}{2016}]{Holoien2016Atel}
{Holoien} T.~W.-S.,  et~al., 2016, The Astronomer's Telegram, \href
  {http://adsabs.harvard.edu/abs/2016ATel.9086....1H} {9086}

\bibitem[\protect\citeauthoryear{{Huang}, {Li}, {Wang}, {Shang}, {Zhang}, {Hu},
  {Qiu}  \& {Jiang}}{{Huang} et~al.}{2012}]{Huang12}
{Huang} F.,  {Li} J.-Z.,  {Wang} X.-F.,  {Shang} R.-C.,  {Zhang} T.-M.,  {Hu}
  J.-Y.,  {Qiu} Y.-L.,   {Jiang} X.-J.,  2012, \mn@doi [Research in Astronomy
  and Astrophysics] {10.1088/1674-4527/12/11/012}, \href
  {http://adsabs.harvard.edu/abs/2012RAA....12.1585H} {12, 1585}

\bibitem[\protect\citeauthoryear{{Iwamoto} et~al.,}{{Iwamoto}
  et~al.}{1998}]{Iwamoto1998}
{Iwamoto} K.,  et~al., 1998, \mn@doi [\nat] {10.1038/27155}, \href
  {http://adsabs.harvard.edu/abs/1998Natur.395..672I} {395, 672}

\bibitem[\protect\citeauthoryear{{James} et~al.,}{{James}
  et~al.}{2004}]{James2004}
{James} P.~A.,  et~al., 2004, \mn@doi [\aap] {10.1051/0004-6361:20031568},
  \href {http://adsabs.harvard.edu/abs/2004A%26A...414...23J} {414, 23}

\bibitem[\protect\citeauthoryear{{Jordi}, {Grebel}  \& {Ammon}}{{Jordi}
  et~al.}{2006}]{Jordi2006}
{Jordi} K.,  {Grebel} E.~K.,   {Ammon} K.,  2006, \mn@doi [\aap]
  {10.1051/0004-6361:20066082}, \href
  {http://adsabs.harvard.edu/abs/2006A%26A...460..339J} {460, 339}

\bibitem[\protect\citeauthoryear{{Kilpatrick} et~al.,}{{Kilpatrick}
  et~al.}{2017}]{Kilpatrick2017}
{Kilpatrick} C.~D.,  et~al., 2017, \mn@doi [\mnras] {10.1093/mnras/stw3082},
  \href {http://adsabs.harvard.edu/abs/2017MNRAS.465.4650K} {465, 4650}

\bibitem[\protect\citeauthoryear{{Langer}}{{Langer}}{2012}]{Langer2012}
{Langer} N.,  2012, \mn@doi [\araa] {10.1146/annurev-astro-081811-125534},
  \href {http://adsabs.harvard.edu/abs/2012ARA%26A..50..107L} {50, 107}

\bibitem[\protect\citeauthoryear{{Liu}, {Modjaz}, {Bianco}  \& {Graur}}{{Liu}
  et~al.}{2016}]{Liu2016}
{Liu} Y.-Q.,  {Modjaz} M.,  {Bianco} F.~B.,   {Graur} O.,  2016, \mn@doi [\apj]
  {10.3847/0004-637X/827/2/90}, \href
  {http://adsabs.harvard.edu/abs/2016ApJ...827...90L} {827, 90}

\bibitem[\protect\citeauthoryear{{Lucy}}{{Lucy}}{1991}]{Lucy1991}
{Lucy} L.~B.,  1991, \mn@doi [\apj] {10.1086/170787}, \href
  {http://adsabs.harvard.edu/abs/1991ApJ...383..308L} {383, 308}

\bibitem[\protect\citeauthoryear{{Lucy}}{{Lucy}}{1999}]{Lucy1999}
{Lucy} L.~B.,  1999, \aap, \href
  {http://adsabs.harvard.edu/abs/1999A%26A...345..211L} {345, 211}

\bibitem[\protect\citeauthoryear{{Maeda} et~al.,}{{Maeda}
  et~al.}{2008}]{Maeda2008}
{Maeda} K.,  et~al., 2008, \mn@doi [Science] {10.1126/science.1149437}, \href
  {http://adsabs.harvard.edu/abs/2008Sci...319.1220M} {319, 1220}

\bibitem[\protect\citeauthoryear{{Maeda} et~al.,}{{Maeda}
  et~al.}{2015}]{Maeda2015}
{Maeda} K.,  et~al., 2015, \mn@doi [\apj] {10.1088/0004-637X/807/1/35}, \href
  {http://adsabs.harvard.edu/abs/2015ApJ...807...35M} {807, 35}

\bibitem[\protect\citeauthoryear{{Maund}, {Smartt}, {Kudritzki},
  {Podsiadlowski}  \& {Gilmore}}{{Maund} et~al.}{2004}]{Maund2004}
{Maund} J.~R.,  {Smartt} S.~J.,  {Kudritzki} R.~P.,  {Podsiadlowski} P.,
  {Gilmore} G.~F.,  2004, \mn@doi [\nat] {10.1038/nature02161}, \href
  {http://adsabs.harvard.edu/abs/2004Natur.427..129M} {427, 129}

\bibitem[\protect\citeauthoryear{{Maund} et~al.,}{{Maund}
  et~al.}{2015}]{Maund2015}
{Maund} J.~R.,  et~al., 2015, \mn@doi [\mnras] {10.1093/mnras/stv2098}, \href
  {http://adsabs.harvard.edu/abs/2015MNRAS.454.2580M} {454, 2580}

\bibitem[\protect\citeauthoryear{{Mazzali}}{{Mazzali}}{2000}]{Mazzali2000b}
{Mazzali} P.~A.,  2000, \aap, \href
  {http://adsabs.harvard.edu/abs/2000A%26A...363..705M} {363, 705}

\bibitem[\protect\citeauthoryear{{Mazzali} \& {Lucy}}{{Mazzali} \&
  {Lucy}}{1993}]{Mazzali1993}
{Mazzali} P.~A.,  {Lucy} L.~B.,  1993, \aap, \href
  {http://adsabs.harvard.edu/abs/1993A%26A...279..447M} {279, 447}

\bibitem[\protect\citeauthoryear{{Mazzali} \& {Lucy}}{{Mazzali} \&
  {Lucy}}{1998}]{Mazzali1998}
{Mazzali} P.~A.,  {Lucy} L.~B.,  1998, \mn@doi [\mnras]
  {10.1046/j.1365-8711.1998.01323.x}, \href
  {http://adsabs.harvard.edu/abs/1998MNRAS.295..428M} {295, 428}

\bibitem[\protect\citeauthoryear{{Mazzali}, {Iwamoto}  \& {Nomoto}}{{Mazzali}
  et~al.}{2000}]{Mazzali2000}
{Mazzali} P.~A.,  {Iwamoto} K.,   {Nomoto} K.,  2000, \mn@doi [\apj]
  {10.1086/317808}, \href {http://adsabs.harvard.edu/abs/2000ApJ...545..407M}
  {545, 407}

\bibitem[\protect\citeauthoryear{{Mazzali}, {Nomoto}, {Patat}  \&
  {Maeda}}{{Mazzali} et~al.}{2001}]{Mazzali2001}
{Mazzali} P.~A.,  {Nomoto} K.,  {Patat} F.,   {Maeda} K.,  2001, \mn@doi [\apj]
  {10.1086/322420}, \href {http://adsabs.harvard.edu/abs/2001ApJ...559.1047M}
  {559, 1047}

\bibitem[\protect\citeauthoryear{{Mazzali} et~al.,}{{Mazzali}
  et~al.}{2002}]{Mazzali2002}
{Mazzali} P.~A.,  et~al., 2002, \mn@doi [\apjl] {10.1086/341504}, \href
  {http://adsabs.harvard.edu/abs/2002ApJ...572L..61M} {572, L61}

\bibitem[\protect\citeauthoryear{{Mazzali} et~al.,}{{Mazzali}
  et~al.}{2003}]{Mazzali2003}
{Mazzali} P.~A.,  et~al., 2003, \mn@doi [\apjl] {10.1086/381259}, \href
  {http://adsabs.harvard.edu/abs/2003ApJ...599L..95M} {599, L95}

\bibitem[\protect\citeauthoryear{{Mazzali} et~al.,}{{Mazzali}
  et~al.}{2006}]{Mazzali2006b}
{Mazzali} P.~A.,  et~al., 2006, \mn@doi [\apj] {10.1086/504415}, \href
  {http://adsabs.harvard.edu/abs/2006ApJ...645.1323M} {645, 1323}

\bibitem[\protect\citeauthoryear{{Mazzali} et~al.,}{{Mazzali}
  et~al.}{2007}]{Mazzali2007}
{Mazzali} P.~A.,  et~al., 2007, \mn@doi [\apj] {10.1086/521873}, \href
  {http://adsabs.harvard.edu/abs/2007ApJ...670..592M} {670, 592}

\bibitem[\protect\citeauthoryear{{Mazzali} et~al.,}{{Mazzali}
  et~al.}{2008}]{Mazzali2008}
{Mazzali} P.~A.,  et~al., 2008, \mn@doi [Science] {10.1126/science.1158088},
  \href {http://adsabs.harvard.edu/abs/2008Sci...321.1185M} {321, 1185}

\bibitem[\protect\citeauthoryear{{Mazzali}, {Deng}, {Hamuy}  \&
  {Nomoto}}{{Mazzali} et~al.}{2009}]{Mazzali2009}
{Mazzali} P.~A.,  {Deng} J.,  {Hamuy} M.,   {Nomoto} K.,  2009, \mn@doi [\apj]
  {10.1088/0004-637X/703/2/1624}, \href
  {http://adsabs.harvard.edu/abs/2009ApJ...703.1624M} {703, 1624}

\bibitem[\protect\citeauthoryear{{Mazzali}, {Maurer}, {Valenti}, {Kotak}  \&
  {Hunter}}{{Mazzali} et~al.}{2010}]{Mazzali2010}
{Mazzali} P.~A.,  {Maurer} I.,  {Valenti} S.,  {Kotak} R.,   {Hunter} D.,
  2010, \mn@doi [\mnras] {10.1111/j.1365-2966.2010.17133.x}, \href
  {http://adsabs.harvard.edu/abs/2010MNRAS.408...87M} {408, 87}

\bibitem[\protect\citeauthoryear{{Mazzali}, {Walker}, {Pian}, {Tanaka},
  {Corsi}, {Hattori}  \& {Gal-Yam}}{{Mazzali} et~al.}{2013}]{Mazzali2013}
{Mazzali} P.~A.,  {Walker} E.~S.,  {Pian} E.,  {Tanaka} M.,  {Corsi} A.,
  {Hattori} T.,   {Gal-Yam} A.,  2013, \mn@doi [\mnras] {10.1093/mnras/stt605},
  \href {http://adsabs.harvard.edu/abs/2013MNRAS.432.2463M} {432, 2463}

\bibitem[\protect\citeauthoryear{{Mazzali} et~al.,}{{Mazzali}
  et~al.}{2014}]{Mazzali2014}
{Mazzali} P.~A.,  et~al., 2014, \mn@doi [\mnras] {10.1093/mnras/stu077}, \href
  {http://adsabs.harvard.edu/abs/2014MNRAS.439.1959M} {439, 1959}

\bibitem[\protect\citeauthoryear{{Mazzali}, {Sauer}, {Pian}, {Deng},
  {Prentice}, {Ben Ami}, {Taubenberger}  \& {Nomoto}}{{Mazzali}
  et~al.}{2017}]{Mazzali2017}
{Mazzali} P.~A.,  {Sauer} D.~N.,  {Pian} E.,  {Deng} J.,  {Prentice} S.,  {Ben
  Ami} S.,  {Taubenberger} S.,   {Nomoto} K.,  2017, \mn@doi [\mnras]
  {10.1093/mnras/stx992}, \href
  {http://adsabs.harvard.edu/abs/2017MNRAS.469.2498M} {469, 2498}

\bibitem[\protect\citeauthoryear{{Metzger}, {Margalit}, {Kasen}  \&
  {Quataert}}{{Metzger} et~al.}{2015}]{Metzger2015}
{Metzger} B.~D.,  {Margalit} B.,  {Kasen} D.,   {Quataert} E.,  2015, \mn@doi
  [\mnras] {10.1093/mnras/stv2224}, \href
  {http://adsabs.harvard.edu/abs/2015MNRAS.454.3311M} {454, 3311}

\bibitem[\protect\citeauthoryear{{Milisavljevic} et~al.,}{{Milisavljevic}
  et~al.}{2015}]{Mili2015}
{Milisavljevic} D.,  et~al., 2015, \mn@doi [\apj] {10.1088/0004-637X/799/1/51},
  \href {http://adsabs.harvard.edu/abs/2015ApJ...799...51M} {799, 51}

\bibitem[\protect\citeauthoryear{{Modjaz}, {Kirshner}, {Blondin}, {Challis}  \&
  {Matheson}}{{Modjaz} et~al.}{2008}]{Modjaz2008}
{Modjaz} M.,  {Kirshner} R.~P.,  {Blondin} S.,  {Challis} P.,   {Matheson} T.,
  2008, \mn@doi [\apjl] {10.1086/593135}, \href
  {http://adsabs.harvard.edu/abs/2008ApJ...687L...9M} {687, L9}

\bibitem[\protect\citeauthoryear{{Modjaz} et~al.,}{{Modjaz}
  et~al.}{2009}]{Modjaz2009}
{Modjaz} M.,  et~al., 2009, \mn@doi [\apj] {10.1088/0004-637X/702/1/226}, \href
  {http://adsabs.harvard.edu/abs/2009ApJ...702..226M} {702, 226}

\bibitem[\protect\citeauthoryear{{Modjaz}, {Kewley}, {Bloom}, {Filippenko},
  {Perley}  \& {Silverman}}{{Modjaz} et~al.}{2011}]{Modjaz2011}
{Modjaz} M.,  {Kewley} L.,  {Bloom} J.~S.,  {Filippenko} A.~V.,  {Perley} D.,
  {Silverman} J.~M.,  2011, \mn@doi [\apjl] {10.1088/2041-8205/731/1/L4}, \href
  {http://adsabs.harvard.edu/abs/2011ApJ...731L...4M} {731, L4}

\bibitem[\protect\citeauthoryear{{Modjaz} et~al.,}{{Modjaz}
  et~al.}{2014}]{Modjaz2014}
{Modjaz} M.,  et~al., 2014, \mn@doi [\aj] {10.1088/0004-6256/147/5/99}, \href
  {http://adsabs.harvard.edu/abs/2014AJ....147...99M} {147, 99}

\bibitem[\protect\citeauthoryear{{Modjaz}, {Liu}, {Bianco}  \&
  {Graur}}{{Modjaz} et~al.}{2016}]{Modjaz2016}
{Modjaz} M.,  {Liu} Y.~Q.,  {Bianco} F.~B.,   {Graur} O.,  2016, \mn@doi [\apj]
  {10.3847/0004-637X/832/2/108}, \href
  {http://adsabs.harvard.edu/abs/2016ApJ...832..108M} {832, 108}

\bibitem[\protect\citeauthoryear{{Nomoto}, {Yamaoka}, {Pols}, {van den Heuvel},
  {Iwamoto}, {Kumagai}  \& {Shigeyama}}{{Nomoto} et~al.}{1994}]{Nomoto1994}
{Nomoto} K.,  {Yamaoka} H.,  {Pols} O.~R.,  {van den Heuvel} E.~P.~J.,
  {Iwamoto} K.,  {Kumagai} S.,   {Shigeyama} T.,  1994, \mn@doi [\nat]
  {10.1038/371227a0}, \href {http://adsabs.harvard.edu/abs/1994Natur.371..227N}
  {371, 227}

\bibitem[\protect\citeauthoryear{{Piascik}, {Steele}, {Bates}, {Mottram},
  {Smith}, {Barnsley}  \& {Bolton}}{{Piascik} et~al.}{2014}]{Piascik2014}
{Piascik} A.~S.,  {Steele} I.~A.,  {Bates} S.~D.,  {Mottram} C.~J.,  {Smith}
  R.~J.,  {Barnsley} R.~M.,   {Bolton} B.,  2014, in Ground-based and Airborne
  Instrumentation for Astronomy V. p. 91478H, \mn@doi{10.1117/12.2055117}

\bibitem[\protect\citeauthoryear{{Pignata} et~al.,}{{Pignata}
  et~al.}{2011}]{Pignata2011}
{Pignata} G.,  et~al., 2011, \mn@doi [\apj] {10.1088/0004-637X/728/1/14}, \href
  {http://adsabs.harvard.edu/abs/2011ApJ...728...14P} {728, 14}

\bibitem[\protect\citeauthoryear{{Podsiadlowski}, {Joss}  \&
  {Hsu}}{{Podsiadlowski} et~al.}{1992}]{Pod1992}
{Podsiadlowski} P.,  {Joss} P.~C.,   {Hsu} J.~J.~L.,  1992, \mn@doi [\apj]
  {10.1086/171341}, \href {http://adsabs.harvard.edu/abs/1992ApJ...391..246P}
  {391, 246}

\bibitem[\protect\citeauthoryear{{Poznanski}, {Ganeshalingam}, {Silverman}  \&
  {Filippenko}}{{Poznanski} et~al.}{2011}]{Poznanski2011}
{Poznanski} D.,  {Ganeshalingam} M.,  {Silverman} J.~M.,   {Filippenko} A.~V.,
  2011, \mn@doi [\mnras] {10.1111/j.1745-3933.2011.01084.x}, \href
  {http://adsabs.harvard.edu/abs/2011MNRAS.415L..81P} {415, L81}

\bibitem[\protect\citeauthoryear{{Poznanski}, {Prochaska}  \&
  {Bloom}}{{Poznanski} et~al.}{2012}]{Poznanski2012}
{Poznanski} D.,  {Prochaska} J.~X.,   {Bloom} J.~S.,  2012, \mn@doi [\mnras]
  {10.1111/j.1365-2966.2012.21796.x}, \href
  {http://adsabs.harvard.edu/abs/2012MNRAS.426.1465P} {426, 1465}

\bibitem[\protect\citeauthoryear{{Prentice} \& {Mazzali}}{{Prentice} \&
  {Mazzali}}{2017}]{Prentice2017}
{Prentice} S.~J.,  {Mazzali} P.~A.,  2017, \mn@doi [\mnras]
  {10.1093/mnras/stx980}, \href
  {http://adsabs.harvard.edu/abs/2017MNRAS.469.2672P} {469, 2672}

\bibitem[\protect\citeauthoryear{{Prentice} et~al.,}{{Prentice}
  et~al.}{2016}]{Prentice2016}
{Prentice} S.~J.,  et~al., 2016, \mn@doi [\mnras] {10.1093/mnras/stw299}, \href
  {http://adsabs.harvard.edu/abs/2016MNRAS.458.2973P} {458, 2973}

\bibitem[\protect\citeauthoryear{{Ryder} et~al.,}{{Ryder}
  et~al.}{2018}]{Ryder2018}
{Ryder} S.~D.,  et~al., 2018, preprint, \href
  {http://adsabs.harvard.edu/abs/2018arXiv180105125R} {} (\mn@eprint {arXiv}
  {1801.05125})

\bibitem[\protect\citeauthoryear{{Sanders} et~al.,}{{Sanders}
  et~al.}{2012}]{Sanders2012}
{Sanders} N.~E.,  et~al., 2012, \mn@doi [\apj] {10.1088/0004-637X/758/2/132},
  \href {http://adsabs.harvard.edu/abs/2012ApJ...758..132S} {758, 132}

\bibitem[\protect\citeauthoryear{{Sauer}, {Mazzali}, {Deng}, {Valenti},
  {Nomoto}  \& {Filippenko}}{{Sauer} et~al.}{2006}]{Sauer2006}
{Sauer} D.~N.,  {Mazzali} P.~A.,  {Deng} J.,  {Valenti} S.,  {Nomoto} K.,
  {Filippenko} A.~V.,  2006, \mn@doi [\mnras]
  {10.1111/j.1365-2966.2006.10438.x}, \href
  {http://adsabs.harvard.edu/abs/2006MNRAS.369.1939S} {369, 1939}

\bibitem[\protect\citeauthoryear{{Schlafly} \& {Finkbeiner}}{{Schlafly} \&
  {Finkbeiner}}{2011}]{Schlafly2011}
{Schlafly} E.~F.,  {Finkbeiner} D.~P.,  2011, \mn@doi [\apj]
  {10.1088/0004-637X/737/2/103}, \href
  {http://adsabs.harvard.edu/abs/2011ApJ...737..103S} {737, 103}

\bibitem[\protect\citeauthoryear{{Shappee} et~al.,}{{Shappee}
  et~al.}{2014}]{Shappee2014}
{Shappee} B.,  et~al., 2014, in American Astronomical Society Meeting Abstracts
  \#223. p. 236.03

\bibitem[\protect\citeauthoryear{{Shivvers} et~al.,}{{Shivvers}
  et~al.}{2017}]{Shivvers2017}
{Shivvers} I.,  et~al., 2017, \mn@doi [\pasp] {10.1088/1538-3873/aa54a6}, \href
  {http://adsabs.harvard.edu/abs/2017PASP..129e4201S} {129, 054201}

\bibitem[\protect\citeauthoryear{{Smartt}}{{Smartt}}{2009}]{Smartt2009}
{Smartt} S.~J.,  2009, \mn@doi [\araa] {10.1146/annurev-astro-082708-101737},
  \href {http://adsabs.harvard.edu/abs/2009ARA%26A..47...63S} {47, 63}

\bibitem[\protect\citeauthoryear{{Smith}, {Gehrz}, {Hinz}, {Hoffmann}, {Hora},
  {Mamajek}  \& {Meyer}}{{Smith} et~al.}{2003}]{Smith2003}
{Smith} N.,  {Gehrz} R.~D.,  {Hinz} P.~M.,  {Hoffmann} W.~F.,  {Hora} J.~L.,
  {Mamajek} E.~E.,   {Meyer} M.~R.,  2003, \mn@doi [\aj] {10.1086/346278},
  \href {http://adsabs.harvard.edu/abs/2003AJ....125.1458S} {125, 1458}

\bibitem[\protect\citeauthoryear{{Steele} et~al.,}{{Steele}
  et~al.}{2004}]{Steele2004}
{Steele} I.~A.,  et~al., 2004, in {Oschmann} Jr. J.~M.,  ed.,  \procspie Vol.
  5489, Ground-based Telescopes. pp 679--692, \mn@doi{10.1117/12.551456}

\bibitem[\protect\citeauthoryear{{Stevance} et~al.,}{{Stevance}
  et~al.}{2017}]{Stevance2017}
{Stevance} H.~F.,  et~al., 2017, \mn@doi [\mnras] {10.1093/mnras/stx970}, \href
  {http://adsabs.harvard.edu/abs/2017MNRAS.469.1897S} {469, 1897}

\bibitem[\protect\citeauthoryear{{Stritzinger} \& {Leibundgut}}{{Stritzinger}
  \& {Leibundgut}}{2005}]{Stritzinger2005}
{Stritzinger} M.,  {Leibundgut} B.,  2005, \mn@doi [\aap]
  {10.1051/0004-6361:20041630}, \href
  {http://adsabs.harvard.edu/abs/2005A%26A...431..423S} {431, 423}

\bibitem[\protect\citeauthoryear{{Stritzinger} et~al.,}{{Stritzinger}
  et~al.}{2018}]{Stritzinger2018}
{Stritzinger} M.~D.,  et~al., 2018, \mn@doi [\aap]
  {10.1051/0004-6361/201730843}, \href
  {http://adsabs.harvard.edu/abs/2018A%26A...609A.135S} {609, A135}

\bibitem[\protect\citeauthoryear{{Taddia} et~al.,}{{Taddia}
  et~al.}{2015}]{Taddia2015}
{Taddia} F.,  et~al., 2015, \mn@doi [\aap] {10.1051/0004-6361/201423915}, \href
  {http://adsabs.harvard.edu/abs/2015A%26A...574A..60T} {574, A60}

\bibitem[\protect\citeauthoryear{{Taddia} et~al.,}{{Taddia}
  et~al.}{2018}]{Taddia2018}
{Taddia} F.,  et~al., 2018, \mn@doi [\aap] {10.1051/0004-6361/201730844}, \href
  {http://adsabs.harvard.edu/abs/2018A%26A...609A.136T} {609, A136}

\bibitem[\protect\citeauthoryear{{Tartaglia} et~al.,}{{Tartaglia}
  et~al.}{2017}]{Tartaglia2017}
{Tartaglia} L.,  et~al., 2017, \mn@doi [\apjl] {10.3847/2041-8213/aa5c7f},
  \href {http://adsabs.harvard.edu/abs/2017ApJ...836L..12T} {836, L12}

\bibitem[\protect\citeauthoryear{{Taubenberger} et~al.,}{{Taubenberger}
  et~al.}{2006}]{Taubenberger2006}
{Taubenberger} S.,  et~al., 2006, \mn@doi [\mnras]
  {10.1111/j.1365-2966.2006.10776.x}, \href
  {http://adsabs.harvard.edu/abs/2006MNRAS.371.1459T} {371, 1459}

\bibitem[\protect\citeauthoryear{{Tonry} et~al.,}{{Tonry}
  et~al.}{2012}]{Tonry12}
{Tonry} J.~L.,  et~al., 2012, \mn@doi [\apj] {10.1088/0004-637X/750/2/99},
  \href {http://adsabs.harvard.edu/abs/2012ApJ...750...99T} {750, 99}

\bibitem[\protect\citeauthoryear{{Valenti} et~al.,}{{Valenti}
  et~al.}{2016}]{Valenti2016}
{Valenti} S.,  et~al., 2016, \mn@doi [\mnras] {10.1093/mnras/stw870}, \href
  {http://adsabs.harvard.edu/abs/2016MNRAS.459.3939V} {459, 3939}

\bibitem[\protect\citeauthoryear{{Van Dyk} et~al.,}{{Van Dyk}
  et~al.}{2014}]{VanDyk2014}
{Van Dyk} S.~D.,  et~al., 2014, \mn@doi [\aj] {10.1088/0004-6256/147/2/37},
  \href {http://adsabs.harvard.edu/abs/2014AJ....147...37V} {147, 37}

\bibitem[\protect\citeauthoryear{{Wang} et~al.,}{{Wang} et~al.}{2008}]{Wang08}
{Wang} X.,  et~al., 2008, \mn@doi [\apj] {10.1086/526413}, \href
  {http://adsabs.harvard.edu/abs/2008ApJ...675..626W} {675, 626}

\bibitem[\protect\citeauthoryear{{Woosley}, {Eastman}, {Weaver}  \&
  {Pinto}}{{Woosley} et~al.}{1994}]{Woosley1994}
{Woosley} S.~E.,  {Eastman} R.~G.,  {Weaver} T.~A.,   {Pinto} P.~A.,  1994,
  \mn@doi [\apj] {10.1086/174319}, \href
  {http://adsabs.harvard.edu/abs/1994ApJ...429..300W} {429, 300}

\bibitem[\protect\citeauthoryear{{Yamanaka} et~al.,}{{Yamanaka}
  et~al.}{2017}]{Yamanaka2017}
{Yamanaka} M.,  et~al., 2017, \mn@doi [\apj] {10.3847/1538-4357/aa5f57}, \href
  {http://adsabs.harvard.edu/abs/2017ApJ...837....1Y} {837, 1}

\bibitem[\protect\citeauthoryear{{Yoon}, {Gr{\"a}fener}, {Vink}, {Kozyreva}  \&
  {Izzard}}{{Yoon} et~al.}{2012}]{Yoon2012}
{Yoon} S.-C.,  {Gr{\"a}fener} G.,  {Vink} J.~S.,  {Kozyreva} A.,   {Izzard}
  R.~G.,  2012, \mn@doi [\aap] {10.1051/0004-6361/201219790}, \href
  {http://adsabs.harvard.edu/abs/2012A%26A...544L..11Y} {544, L11}

\bibitem[\protect\citeauthoryear{{Zhang}, {Wang}, {Bai}, {Zhang}, {Wang},
  {Liu}, {Zhao}  \& {Chen}}{{Zhang} et~al.}{2014}]{Zhang14}
{Zhang} J.-J.,  {Wang} X.-F.,  {Bai} J.-M.,  {Zhang} T.-M.,  {Wang} B.,  {Liu}
  Z.-W.,  {Zhao} X.-L.,   {Chen} J.-C.,  2014, \mn@doi [\aj]
  {10.1088/0004-6256/148/1/1}, \href
  {http://adsabs.harvard.edu/abs/2014AJ....148....1Z} {148, 1}

\bibitem[\protect\citeauthoryear{{de Vaucouleurs}, {de Vaucouleurs}, {Corwin},
  {Buta}, {Paturel}  \& {Fouqu{\'e}}}{{de Vaucouleurs}
  et~al.}{1991}]{devac1991}
{de Vaucouleurs} G.,  {de Vaucouleurs} A.,  {Corwin} Jr. H.~G.,  {Buta} R.~J.,
  {Paturel} G.,   {Fouqu{\'e}} P.,  1991, {Third Reference Catalogue of Bright
  Galaxies. Volume I: Explanations and references. Volume II: Data for galaxies
  between 0$^{h}$ and 12$^{h}$. Volume III: Data for galaxies between 12$^{h}$
  and 24$^{h}$.}

\makeatother
\end{thebibliography}



%
\appendix

\section{Static features in the spectra}\label{sec:static}
\begin{figure}
	\centering
	\includegraphics[scale=0.55]{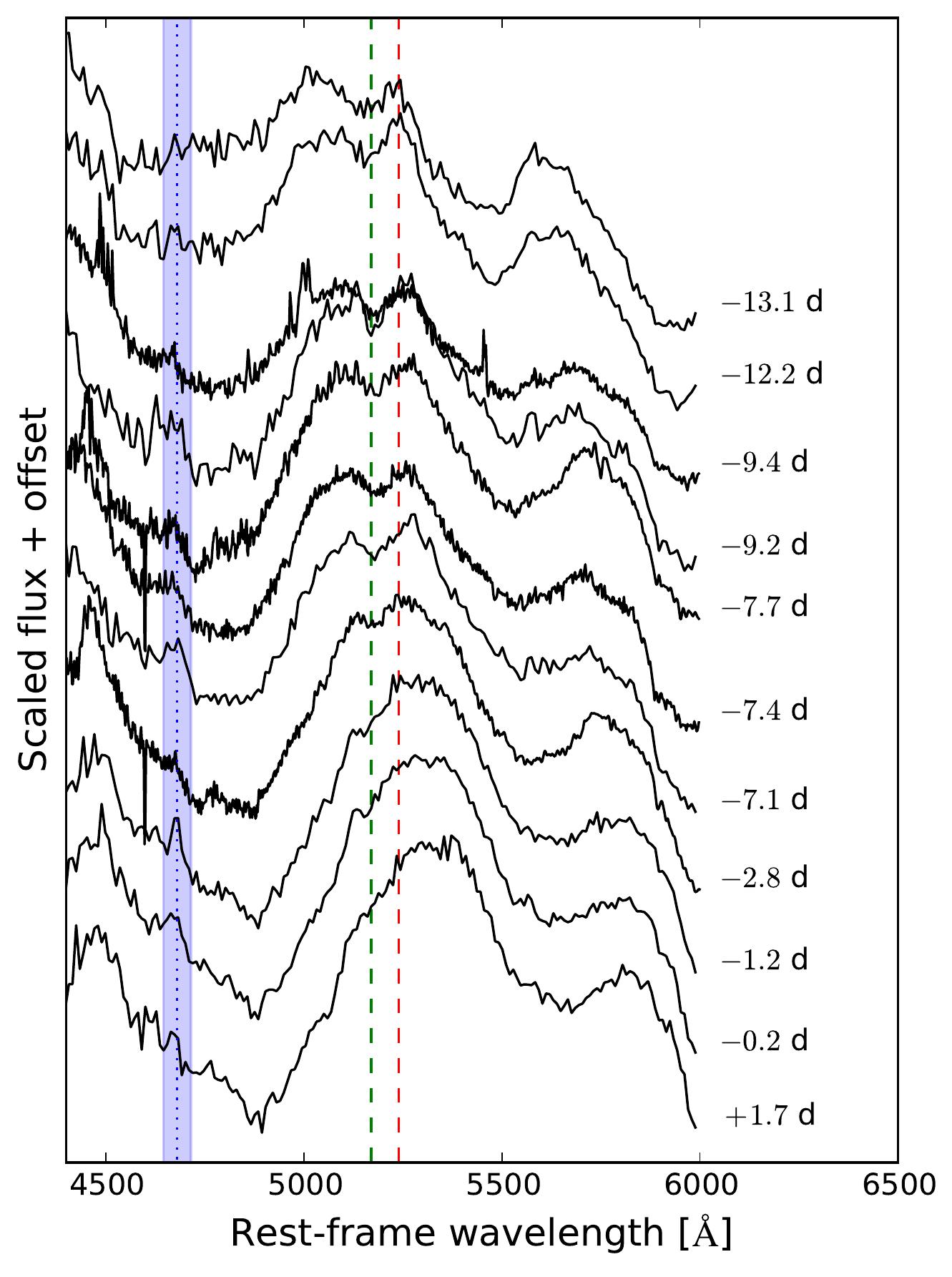}
	\caption{The pre-peak spectra of SN 2016coi showing static features. The blue dotted line and region surrounds a line that may be in emission, this does not move as the absorption feature moves in velocity underneath suggesting that it is external to the SN ejecta. This feature is present in spectra from three different instruments located in China, La Palma, and Hawaii, which suggests it is not noise. The green dashed line and the red dashed line show a feature that is in absorption (green) and possibly emission (dashed). The feature disappears as the pseudo-emission peak moves red-ward.}
	\label{fig:csm}
\end{figure}

The early spectra of SN 2016coi show two regions where features appear to be static despite the rapid evolution in line velocity and the degree of blending, they are shown in Figure~\ref{fig:csm}. Analysis of these features is beyond the scope of this work but we briefly discuss their effect on the early time spectra.

Aside from the abnormal strength of the absorption $\sim 5500$ \AA\ the early spectra also show an line absorption around $\sim 5170$ \AA\ and a possible re-emission peak at 5240 \AA. The absorption is rather weak, and by peak it has disappeared into the absorption caused by the blend of lines blue-ward of its position. Strangely, the velocity evolution of this line, if any, is also weak, especially compared with how the absorption profiles blue-ward and red-ward are decreasing in velocity. At $t = -13.1$ d the absorption profile looks to be significantly broader than at later times, this however may be deceptive as the S/N of the spectrum masks the true position of the base and the re-emission peak that the absorption profile is sitting on is being narrowed and shifted red-ward over time but the competing absorption features either side of it.

We can find no explanation for this in the ions normally found in SN ejecta. If the ``emission'' at 5240 \AA\ is the rest wavelength of the absorption then the line velocity is extremely small $\sim 4000$ \AA\ when compared with other line velocities at this epoch.

There also appears to be a weak ``emission feature'' around 4670 \AA, which is static against the red-ward movement of the broad absorption feature behind it. The feature does not appear to be systematic as it is present in spectra from SPRAT, Floyds, and LJT. Assuming this feature is some emission line at the rest wavelength then estimate of the half width at half maximum ($\sim 15-25$ \AA) gives a velocity of $\sim 1000$ \kms. The line disappears as the pseudo-emission peak immediately blue-ward moves into the same wavelength regime and is not seen again. 
If the line was the result of recombination from ionised circumstellar medium (CSM) it could be attributable to \ion{He}{II} \lam\ 4686 offset by $1000$ \kms, however there appears to be no evidence for other emission lines and so we remain sceptical. 
Note that in the post-peak spectra there is a small pseudo-emission peak in the \FeII\ blend that is attributable to the de-blending of \FeII\ \lam\ 4924 and \lam\ 5018 lines from the \FeII\ \lam\ 5169 line. This is commonly seen in the evolution of SE-SNe.

Recognising the behaviour of the 4670 \AA\ feature is important because it could be mistaken for a pseudo-emission peak related to the Doppler shifted \FeII\ 5018 \AA\ line. As classification is dependent upon the blending of the three strong \FeII\ lines in this region (see Section~\ref{sec:class}) then the behaviour of this feature must be catalogued. As the line is static we can rule out its being a consequence of \FeII.

\section{Line velocity measurements of blended lines}\label{sec:02ap}
\begin{figure}
	\centering
	\includegraphics[scale=0.43]{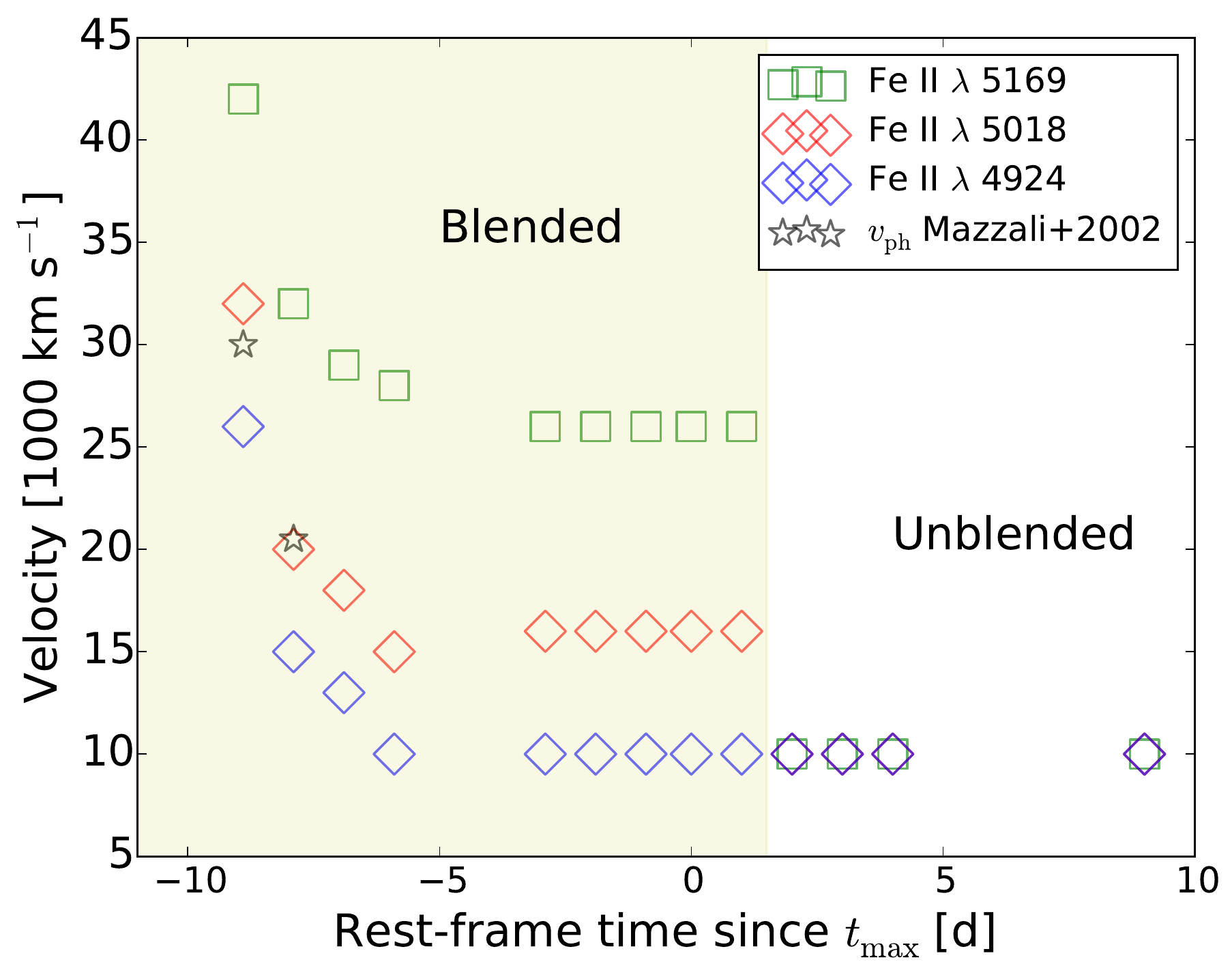}
	\caption{The line velocity of the absorption feature at $\sim 4800$ \AA\ in SN 2002ap as measured using three iron lines; \FeII\ 4924, 5018, 5169. The periods when the lines are blended and clearly unblended are marked. We also include the photospheric velocity measurements determined through spectral modelling \citep{Mazzali2002}. It can be see that \FeII\ 5169 is a poor proxy for the line velocity at peak because when the lines de-blend the velocity measurements are suddenly some $\sim 20,000$ \kms\ lower. It may be tempting to use \FeII\ 4924 as it appears to match the line velocity at the de-blending stage, however at early times this line returns velocities below that of the photospheric velocity, a situation that not physical. }
	\label{fig:02ap}
\end{figure}

Blended lines cause a problem for line velocity measurements as the minima of the absorption features do not necessarily represent a single atomic transition. To demonstrate, we use the spectra of SN 2002ap and measure the velocity of the feature in the region around $4500-5000$ \AA\ from the absorption minimum and using the three strong iron transitions in this region; \FeII\ 4924, 5018, 5169. The results are shown in Figure~\ref{fig:02ap} where it can be seen that it is not possible to define the velocity throughout by the use of a single line, we also show \vph\ as found by \cite{Mazzali2002}.

This plot shows that the \FeII\ \lam\ 5169 line cannot be used to accurately define the velocity of \FeII\ in blended lines, and if one is to use this line in this respect then appropriate steps must be taken to quantify the uncertainties.
The inclusion of \vph\ also demonstrates that \FeII\ \lam\ 5169 is a poor proxy for the photospheric velocity.

\section{Spectral plots}\label{sec:specplots}
\begin{figure*}
	\centering
	\includegraphics[scale=0.8]{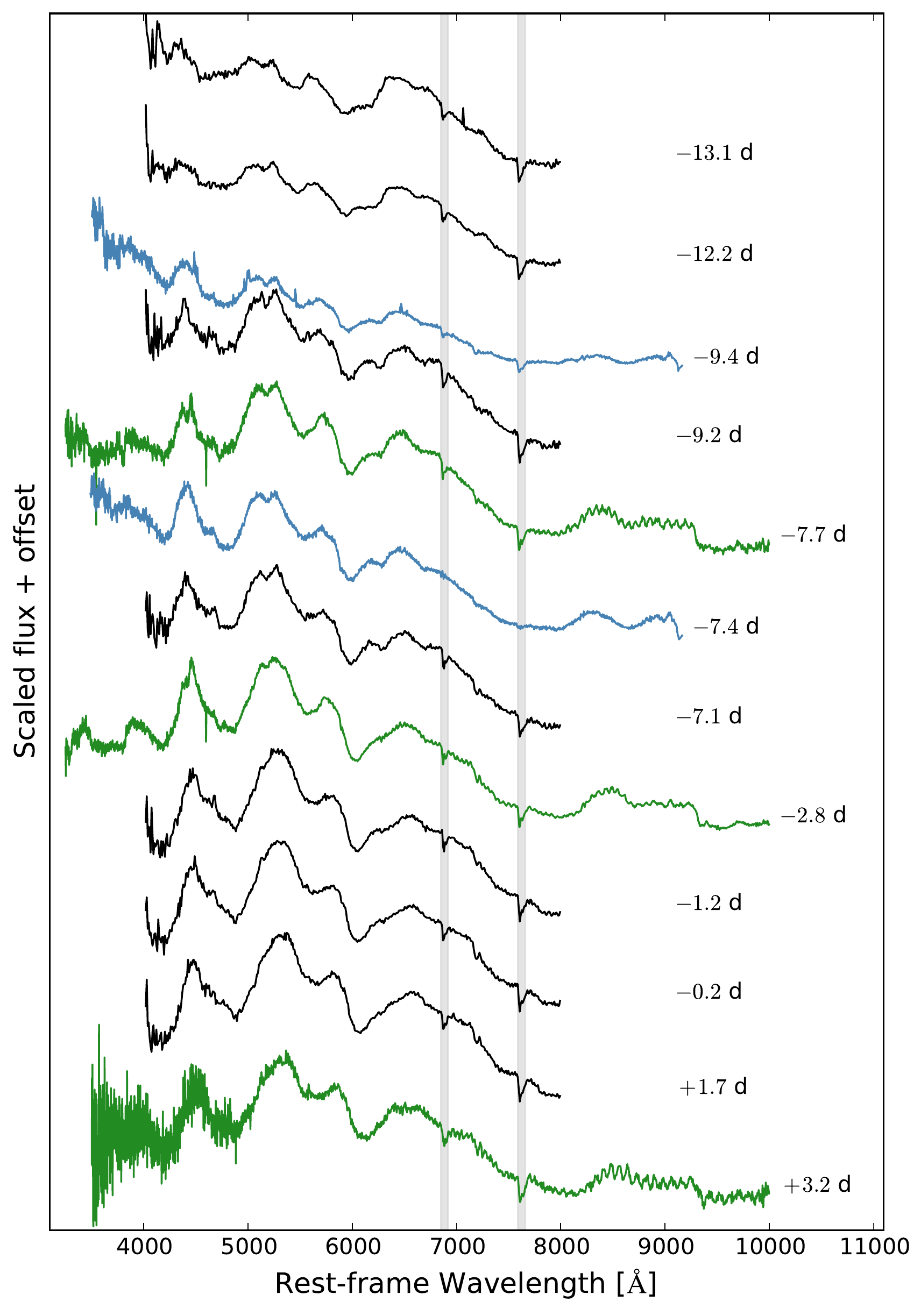}
	\caption{Extinction corrected spectra of SN 2016coi. The grey shaded regions represent prominent telluric features. Spectra are from LT (black), LJT (blue), LCO (green). In certain cases spectra have been truncated to remove excessive noise.}
	\label{fig:s0}
\end{figure*}

\begin{figure*}
	\centering
	\includegraphics[scale=0.8]{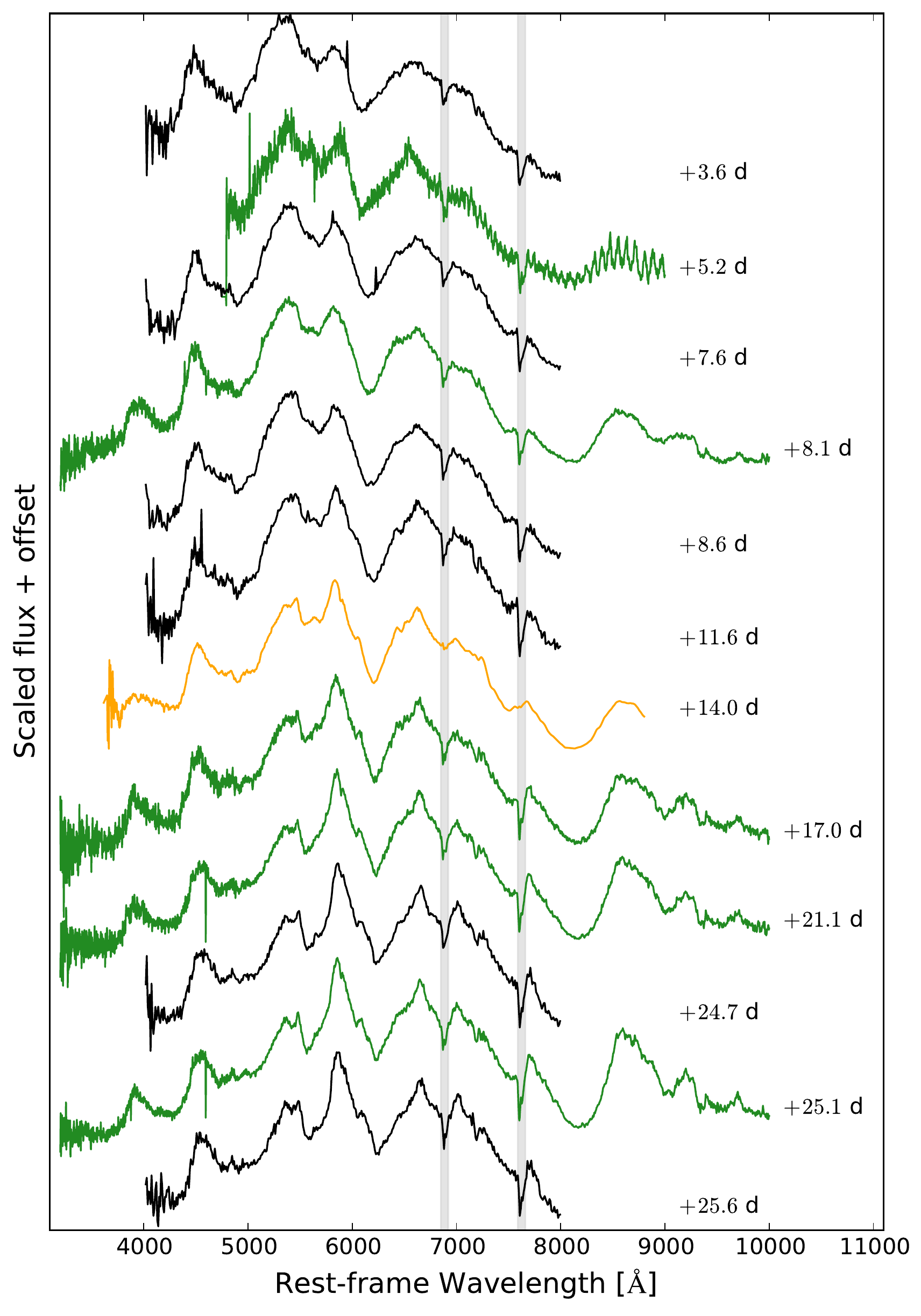}
	\caption{As Figure~\ref{fig:s0}. Additional spectra from Xianglong (orange). }
	\label{fig:s1}
\end{figure*}

\begin{figure*}
	\centering
	\includegraphics[scale=0.8]{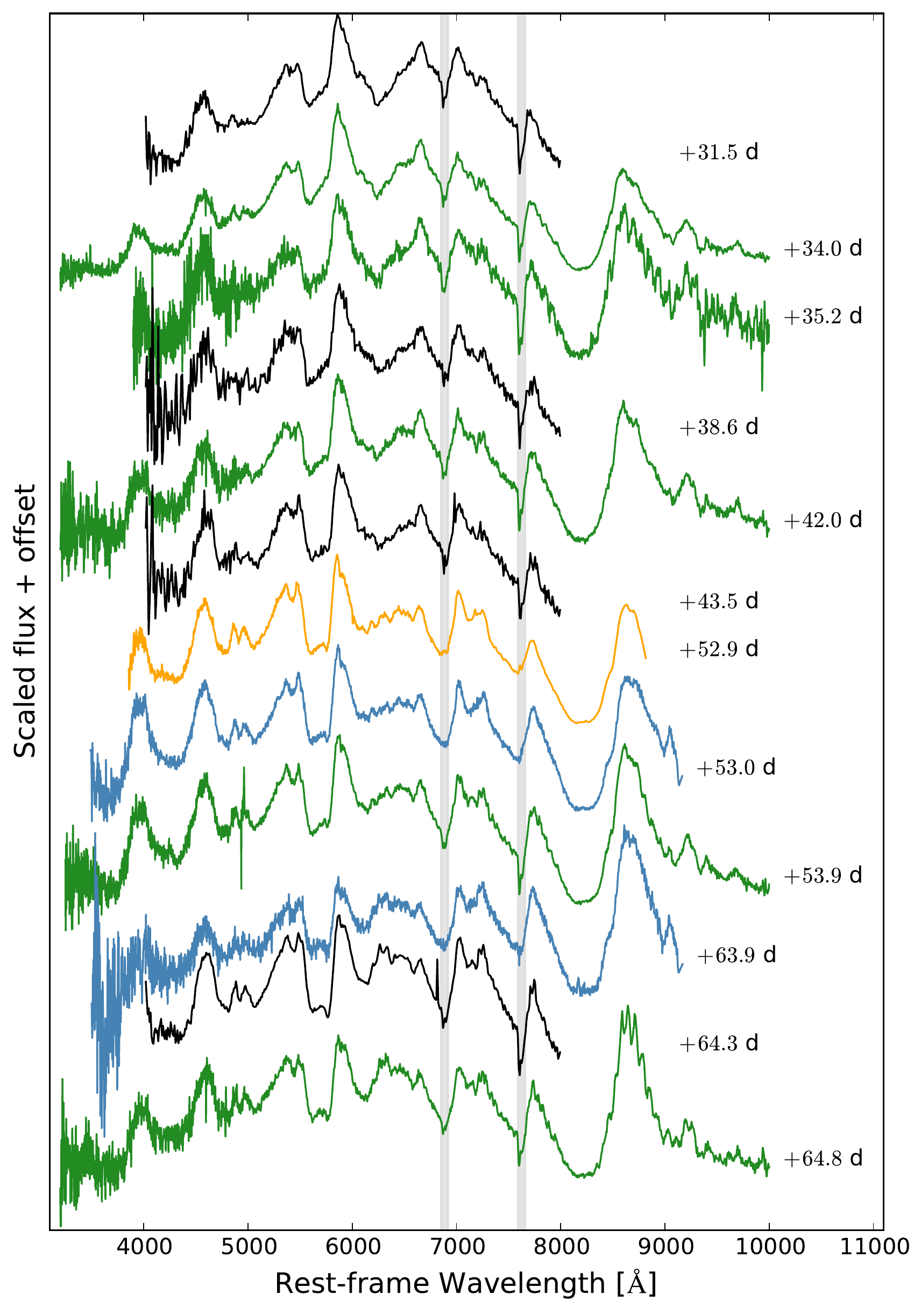}
	\caption{As Figure~\ref{fig:s0}}
	\label{fig:s2}
\end{figure*}

\begin{figure*}
	\centering
	\includegraphics[scale=0.8]{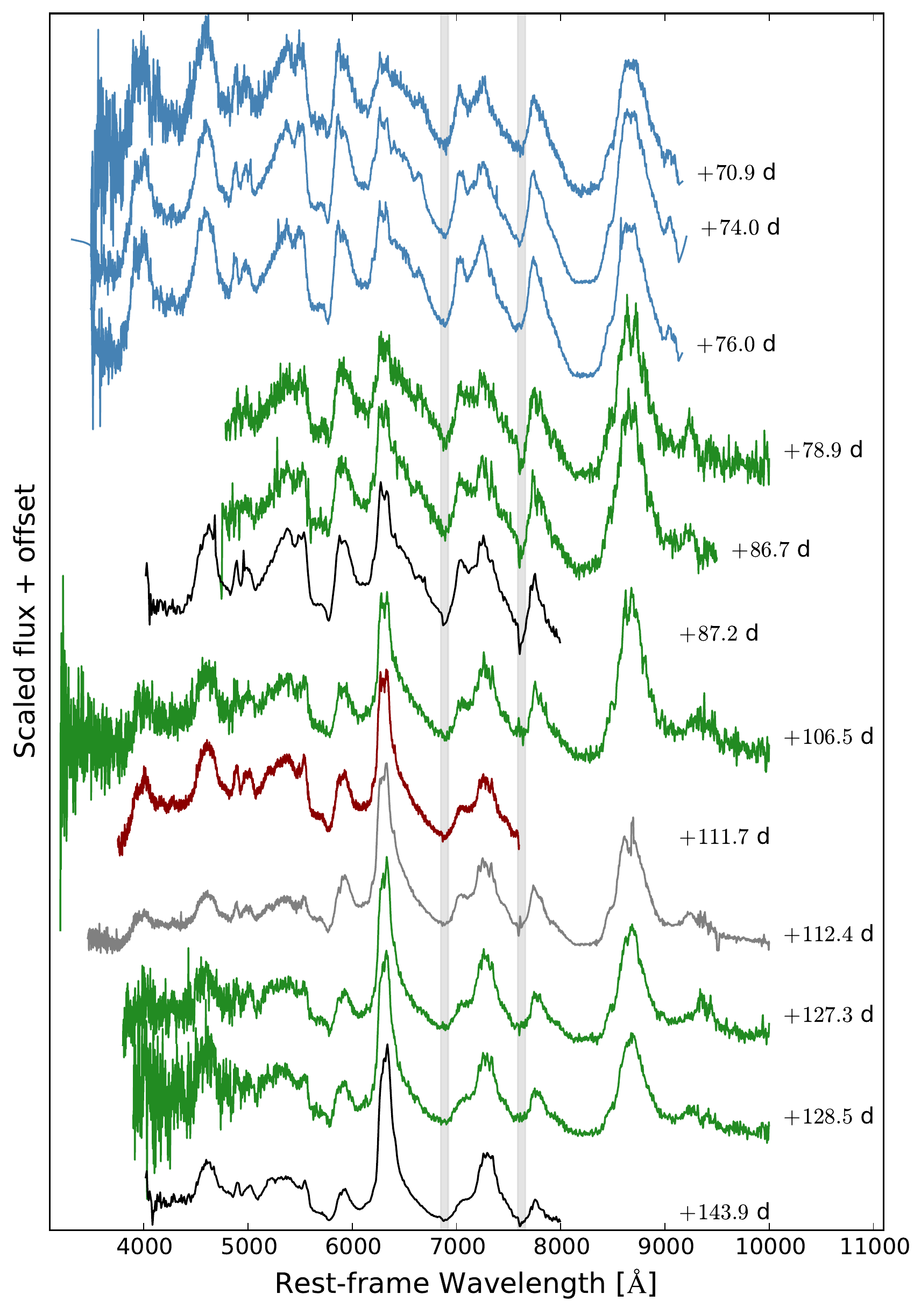}
	\caption{As Figure~\ref{fig:s0}. Additional spectra from the INT (dark red) and Lick (grey)}
	\label{fig:s3}
\end{figure*}

\begin{figure*}
	\centering
	\includegraphics[scale=0.8]{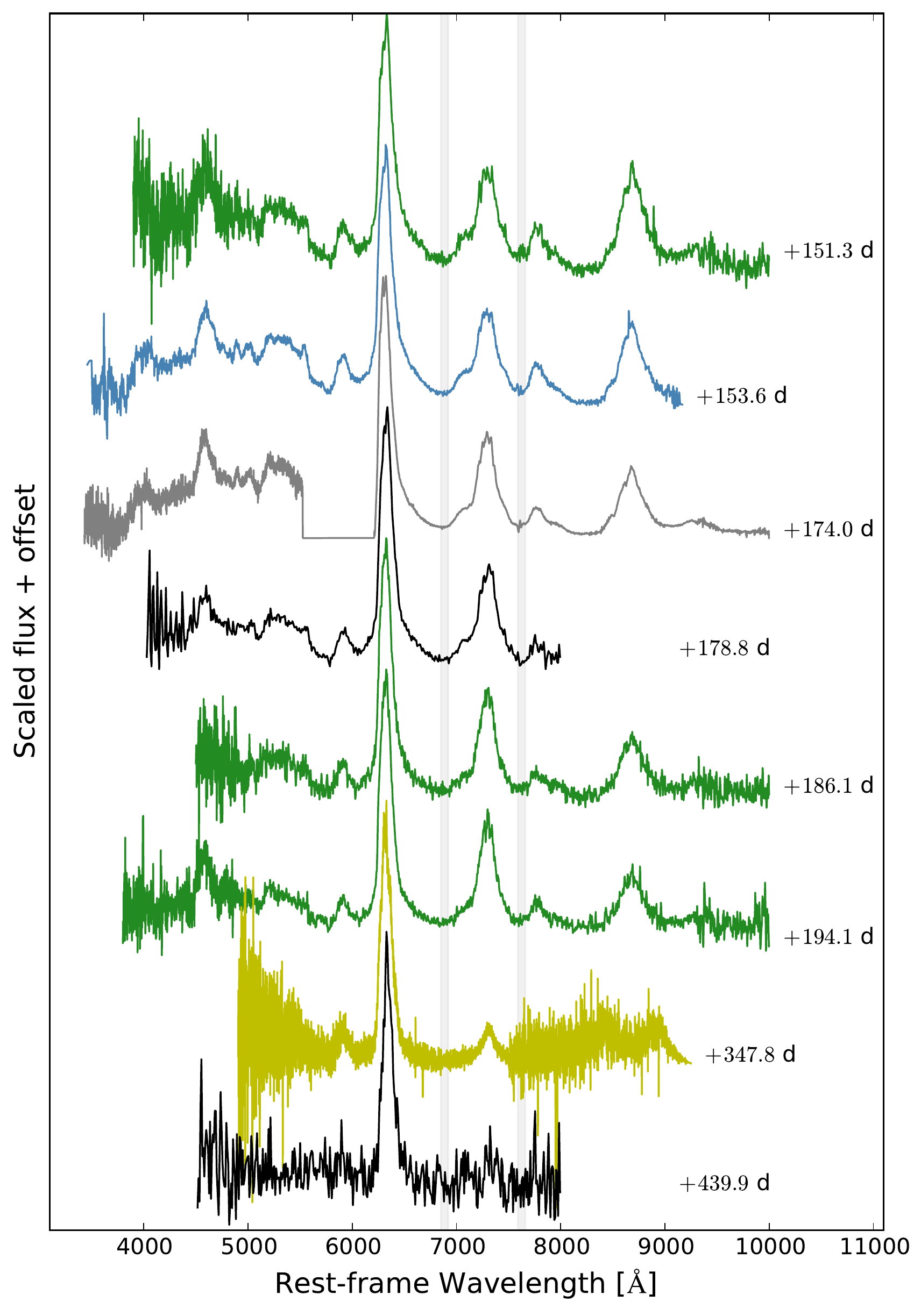}
	\caption{As Figure~\ref{fig:s0}. The DEIMOS spectrum is in yellow.}
	\label{fig:s4}
\end{figure*}

\section{Tables}\label{sec:phottabs}

\begin{table*}
	\centering
	\caption{Journal of spectroscopic observations}
	\begin{tabular}{lcccc}
    \hline
	MJD & $t-$\tmax & Telescope & Wavelength range* & Exposure time \\
    	&  [Rest-frame d]	& & [\AA] & [s] \\
        \hline
57536.19&-13.1&LT&4000 -- 8000&200 \\
57537.13&-12.2&LT&4000 -- 8000&100 \\
57539.88&-9.5&LJT&3500 -- 9200&1800 \\
57540.11&-9.2&LT&4000 -- 8000&100 \\
57541.59&-7.8&LCO (Haleakala)&3200 -- 10000&1200 \\
57541.87&-7.5&LJT&3500 -- 9200&1800 \\
57542.17&-7.2&LT&4000 -- 8000&100 \\
57546.56&-2.8&LCO (Haleakala)&3200 -- 10000&1200\\
57548.14&-1.2&LT&4000 -- 8000&100 \\
57549.15&-0.2&LT&4000 -- 8000&100 \\
57551.14&1.6&LT&4000 -- 8000&100\\
57552.67&3.2&LCO (SSO)&3200 -- 10000&1200 \\
57553.07&3.6&LT&4000 -- 8000&100 \\
57554.67&5.2&LCO (SSO)*&3200 -- 10000&400 \\
57557.11&7.6&LT&4000 -- 8000&100 \\
57557.53&8.0&LCO (Haleakala)&3200 -- 10000&1200 \\
57558.07&8.6&LT&4000 -- 8000&100 \\
57561.04&11.5&LT&4000 -- 8000&100 \\
57563.74&14.2&XLT&3600 -- 8800&2400 \\
57566.51&17.0&LCO (Haleakala)&3200 -- 10000&1200 \\
57570.59&21.0&LCO (Haleakala)&3200 -- 10000&1200 \\
57574.20&24.6&LT&4000 -- 8000&150 \\
57574.59&25.0&LCO (Haleakala)&3200 -- 10000&1200 \\
57575.14&25.6&LT&4000 -- 8000&150 \\
57581.01&31.4&LT&4000 -- 8000&150\\
57583.59&34.0&LCO (Haleakala)&3200 -- 10000&1800 \\
57584.72&35.1&LCO (SSO)&3200 -- 10000&1800 \\
57588.14&38.5&LT&4000 -- 8000&200 \\
57591.58&41.9&LCO (Haleakala)&3200 -- 10000&1800 \\
57593.12&43.5&LT&4000 -- 8000&300 \\
57602.65&53.0&XLT&3600 -- 8800&2700 \\
57602.68&53.0&LJT&3500 -- 9200&1800 \\
57603.53&53.9&LCO (Haleakala)&3200 -- 10000&1800 \\
57613.59&63.9&LJT&3500 -- 9200&2700 \\
57613.93&64.2&LT&4000 -- 8000&1000 \\
57614.47&64.7&LCO (Haleakala)&3200 -- 10000&1800 \\
57620.57&70.8&LJT&3500 -- 9200&2700 \\
57623.69&73.9&LJT&3500 -- 9200&2700 \\
57625.66&75.9&LJT&3500 -- 9200&2700 \\
57628.62&78.8&LCO (SSO)*&3200 -- 10000&1500 \\
57636.50&86.7&LCO (Haleakala)*&3200 -- 10000&1500 \\
57636.99&87.2&LT&4000 -- 8000&700 \\
57656.36&106.5&LCO (Haleakala)&3200 -- 10000&1800 \\
57661.5&111.6&INT&3700 -- 7600&3600 \\
57662.20&112.3&Lick&3400 -- 10000&2700 \\
57677.20&127.3&LCO (Haleakala)&3200 -- 10000&1800 \\
57678.41&128.5&LCO (Haleakala)&3200 -- 10000&1800 \\
57693.84&143.8&LT&4000 -- 8000&1500 \\
57701.32&151.3&LCO (Haleakala)&3200 -- 10000&1800 \\
57703.57&153.5&LJT&3500 -- 9200&2400 \\
57724.11&174.0&Lick&3400 -- 10000&1200 \\
57728.84&178.7&LT&4000 -- 8000&1500 \\
57736.23&186.1&LCO (Haleakala)&3200 -- 10000&1800 \\
57744.21&194.0&LCO (Haleakala)&3200 -- 10000&1800 \\
57898.5 &347.8& Keck & 4000 -- 10500& 900\\ 
57990.94&439.9&LT&4000 -- 8000&2400 \\
	\hline
    \multicolumn{5}{l}{*Denotes LCO data obtained independently of SNEx}\\
	\end{tabular}
	\label{tab:specobs}
\end{table*}

\begin{table*}
	\caption{$ugriz$ observations of SN 2016coi}
	\begin{tabular}{cccccccccccc}
MJD & $u$ & $\delta u$& $g$ & $\delta g$& $r$ & $\delta r$& $i$ & $\delta i$& $z$ & $\delta z$ & Telescope \\  
\hline
57537.07&-&-&15.554&0.012&14.956&0.006&15.258&0.01&-&-&LCO (SAAO) \\
57537.11&-&-&15.555&0.007&14.962&0.004&15.266&0.002&-&-&LT \\
57538.12&-&-&15.255&0.002&14.695&0.002&15.011&0.008&14.933&0.001&LT \\
57539.12&-&-&14.988&0.007&14.479&0.009&14.839&0.018&-&-&LCO (SAAO) \\
57539.83&15.869&0.013&14.899&0.001&14.383&0.001&14.755&0.002&14.591&0.001&LJT \\
57540.14&15.731&0.013&14.804&0.003&14.281&0.002&14.669&0.001&14.541&0.002&LT \\
57541.86&-&-&14.649&0.012&14.152&0.004&14.551&0.003&14.342&0.010&LJT \\
57542.12&-&-&14.511&0.006&14.028&0.002&14.418&0.006&-&-&LCO (SAAO) \\
57542.16&15.633&0.005&14.511&0.001&14.028&0.001&14.421&0.001&14.25&0.001&LT \\
57544.50&-&-&14.235&0.005&13.814&0.002&14.206&0.003&-&-&LCO (Haleakala)* \\
57545.11&-&-&14.251&0.011&13.747&0.005&14.162&0.008&-&-&LCO (SAAO) \\
57548.17&15.767&0.006&14.171&0.001&13.749&0.001&14.037&0.001&13.78&0.001&LT \\
57549.16&15.747&0.013&14.176&0.001&13.609&0.002&14.019&0.001&13.729&0.002&LT \\
57551.13&15.923&0.018&14.186&0.007&13.571&0.002&13.971&0.002&13.666&0.003&LT \\
57552.07&-&-&14.015&0.002&13.432&0.005&-&-&-&-&LCO (SAAO) \\
57552.56&-&-&14.192&0.002&13.541&0.001&13.940&0.001&-&-&LCO (Haleakala)* \\
57553.08&16.186&0.014&14.251&0.010&13.594&0.003&13.947&0.002&13.622&0.006&LT \\
57557.11&16.647&0.032&14.485&0.014&13.633&0.004&13.957&0.001&13.629&0.002&LT \\
57557.16&-&-&14.468&0.012&13.618&0.009&-&-&-&-&LCO (SAAO) \\
57558.08&16.729&0.058&14.555&0.002&13.667&0.003&13.972&0.001&13.633&0.001&LT \\
57561.03&-&-&14.750&0.012&13.790&0.005&14.041&0.005&13.713&0.011&LT \\
57561.14&-&-&14.764&0.011&13.749&0.012&13.977&0.007&-&-&LCO (SAAO) \\
57564.16&-&-&14.999&0.080&13.887&0.011&14.093&0.007&-&-&LCO (SAAO) \\
57567.60&-&-&15.337&0.005&14.037&0.001&14.196&0.002&-&-&LCO (Haleakala)* \\
57568.11&-&-&15.285&0.012&14.092&0.004&14.253&0.005&13.874&0.005&LT \\
57568.15&-&-&15.269&0.008&14.066&0.011&14.212&0.005&-&-&LCO (SAAO) \\
57570.40&-&-&15.482&0.004&14.171&0.001&14.272&0.002&-&-&LCO (Haleakala)* \\
57574.13&-&-&15.616&0.006&14.34&0.005&-&-&-&-&LCO (SAAO) \\
57574.19&-&-&15.654&0.005&14.390&0.001&14.467&0.001&14.055&0.001&LT \\
57576.38&-&-&15.77&0.006&14.444&0.001&14.493&0.002&-&-&LCO (Haleakala)* \\
57576.41&-&-&15.723&0.003&14.423&0.003&14.487&0.004&-&-&LCO (CTIO) \\
57579.07&-&-&15.8&0.007&14.587&0.002&14.619&0.002&14.159&0.003&LT \\
57579.60&-&-&-&-&14.641&0.005&14.658&0.011&-&-&LCO (SSO)* \\
57581.10&-&-&15.875&0.007&14.634&0.004&14.654&0.004&-&-&LCO (SAAO) \\
57582.36&-&-&16.041&0.012&14.705&0.003&14.689&0.003&-&-&LCO (Haleakala)* \\
57586.13&-&-&16.022&0.002&14.850&0.002&14.828&0.004&14.317&0.001&LT \\
57587.09&-&-&16.049&0.004&14.883&0.002&14.865&0.001&14.337&0.001&LT \\
57587.11&-&-&16.061&0.014&14.885&0.007&14.844&0.007&-&-&LCO (SAAO) \\
57591.55&-&-&-&-&15.026&0.007&14.952&0.006&-&-&LCO (Haleakala)* \\
57593.14&-&-&16.169&0.004&15.052&0.002&15.014&0.001&14.458&0.001&LT \\
57594.73&-&-&16.193&0.008&15.087&0.005&15.034&0.006&-&-&LCO (SSO) \\
57596.44&-&-&16.27&0.009&15.107&0.002&15.069&0.003&-&-&LCO (Haleakala)* \\
57601.32&-&-&16.284&0.005&15.225&0.004&15.141&0.005&-&-&LCO (CTIO) \\
57602.67&-&-&16.375&0.001&15.333&0.001&15.266&0.001&14.637&0.001&LJT \\
57605.54&-&-&16.344&0.008&15.324&0.003&15.224&0.004&-&-&LCO (Haleakala)* \\
57606.52&-&-&16.357&0.007&15.34&0.003&15.255&0.004&-&-&LCO (Haleakala)* \\
57609.01&-&-&16.419&0.009&15.417&0.002&15.352&0.001&14.692&0.002&LT \\
57611.65&-&-&16.413&0.010&15.427&0.003&15.369&0.003&-&-&LCO (SSO)* \\
57613.58&-&-&16.491&0.006&15.543&0.001&15.481&0.001&14.787&0.001&LJT \\
57617.75&-&-&-&-&15.720&0.012&15.706&0.026&-&-&LJT \\
57620.03&-&-&16.552&0.010&15.631&0.003&15.536&0.001&14.864&0.004&LT \\
57620.56&-&-&-&-&15.617&0.003&15.54&0.002&-&-&LJT \\
57621.02&-&-&16.588&0.029&15.659&0.02&15.577&0.018&-&-&LCO (SAAO) \\
57622.64&-&-&16.673&0.005&15.733&0.002&15.679&0.001&14.977&0.004&LJT \\
57623.68&-&-&16.692&0.010&15.719&0.001&-&-&14.951&0.002&LJT \\
57625.02&-&-&16.621&0.001&15.706&0.001&15.619&0.001&14.925&0.001&LT \\
57620.19&-&-&16.531&0.0&15.637&0.0&15.546&0.0&-&-&LT \\
57620.56&-&-&-&-&15.617&0.003&15.54&0.002&-&-&LJT \\
57621.02&-&-&16.588&0.029&15.659&0.02&15.577&0.018&-&-&LCO (SAAO) \\

\hline
\multicolumn{12}{l}{*Denotes LCO data obtained independently of SNEx}\\
    \end{tabular}
    \label{tab:ugriztable}
\end{table*}

\begin{table*}
	\contcaption{$ugriz$ observations of SN 2016coi}
	\begin{tabular}{cccccccccccc}
MJD & $u$ & $\delta u$& $g$ & $\delta g$& $r$ & $\delta r$& $i$ & $\delta i$& $z$ & $\delta z$ & Telescope \\    
\hline
57622.64&-&-&16.673&0.005&15.733&0.002&15.679&0.001&14.977&0.004&LJT \\
57623.68&-&-&16.692&0.010&15.719&0.001&-&-&14.951&0.002&LJT \\
57625.65&-&-&16.654&0.003&15.746&0.001&-&-&-&-&LJT \\
57626.58&-&-&16.657&0.014&15.692&0.005&-&-&-&-&LCO (Haleakala)* \\
57627.97&-&-&16.609&0.012&15.74&0.008&15.638&0.009&-&-&LCO (SAAO) \\
57634.45&-&-&16.686&0.013&15.786&0.006&15.814&0.006&-&-&LCO (SSO)* \\
57635.00&-&-&16.739&0.001&15.866&0.001&15.799&0.001&15.087&0.001&LT \\
57635.98&-&-&16.737&0.014&15.872&0.012&15.825&0.014&-&-&LCO (SAAO) \\
57636.00&-&-&16.755&0.003&15.874&0.001&15.817&0.001&15.114&0.002&LT \\
57638.98&-&-&16.798&0.001&15.928&0.001&15.878&0.001&15.134&0.001&LT \\
57641.93&-&-&16.831&0.002&15.966&0.001&15.92&0.001&15.209&0.001&LT \\
57642.94&-&-&16.796&0.015&15.999&0.01&15.945&0.012&-&-&LCO (SAAO) \\
57644.92&-&-&16.864&0.002&16.015&0.002&15.982&0.001&15.259&0.001&LT \\
57647.90&-&-&16.903&0.008&16.048&0.002&16.027&0.002&15.304&0.002&LT \\
57650.32&-&-&16.871&0.013&16.05&0.005&16.043&0.004&-&-&LCO (Haleakala)* \\
57650.97&-&-&16.964&0.003&16.106&0.001&16.078&0.001&15.381&0.002&LT \\
57656.90&-&-&17.005&0.008&16.187&0.007&16.163&0.01&-&-&LCO (SAAO) \\
57657.26&-&-&16.988&0.016&16.172&0.007&16.174&0.007&-&-&LCO (Haleakala)* \\
57660.90&-&-&-&-&16.38&0.101&16.294&0.087&-&-&LCO (SAAO) \\
57661.47&-&-&17.08&0.021&16.212&0.008&16.224&0.008&-&-&LCO (Haleakala)* \\
57667.87&-&-&17.191&0.008&16.343&0.001&16.348&0.001&15.683&0.002&LT \\
57672.91&-&-&17.26&0.001&16.402&0.001&16.426&0.001&15.781&0.001&LT \\
57676.45&-&-&17.266&0.087&16.409&0.064&16.464&0.069&-&-&LCO (SSO)* \\
57680.12&-&-&17.318&0.017&16.508&0.013&-&-&-&-&LCO (CTIO) \\
57683.46&-&-&17.395&0.008&16.527&0.006&16.556&0.008&-&-&LCO (SSO) \\
57685.43&-&-&17.423&0.002&16.509&0.001&16.576&0.001&-&-&LCO (SSO)* \\
57692.44&-&-&17.517&0.005&16.626&0.001&16.712&0.002&-&-&LCO (SSO)* \\
57696.29&-&-&17.565&0.001&16.671&0.001&16.814&0.003&-&-&LCO (Haleakala)* \\
57699.76&-&-&17.636&0.015&16.71&0.011&16.786&0.012&-&-&LCO (SAAO) \\
57701.81&-&-&17.716&0.003&16.751&0.003&16.865&0.001&16.418&0.003&LT \\
57702.69&-&-&-&-&16.847&0.004&16.938&0.005&-&-&LJT \\
57703.56&-&-&-&-&16.899&0.225&16.997&0.005&-&-&LJT \\
57710.25&-&-&18.018&0.234&16.819&0.016&16.852&0.047&-&-&LCO (Haleakala)* \\
57711.83&-&-&17.889&0.001&16.886&0.001&16.999&0.001&16.612&0.001&LT \\
57716.14&-&-&17.883&0.021&16.931&0.015&16.98&0.02&-&-&LCO (McDonald) \\
57717.40&-&-&18.076&0.006&16.952&0.005&17.043&0.002&-&-&LCO (SSO)* \\
57727.80&-&-&18.183&0.001&17.121&0.001&17.227&0.001&16.997&0.001&LT \\
57729.80&-&-&18.236&0.012&17.058&0.008&17.106&0.010&17.228&0.011&LT \\
57732.14&-&-&18.256&0.027&17.195&0.019&17.257&0.026&-&-&LCO (McDonald) \\
57732.80&-&-&18.28&0.002&17.202&0.002&17.296&0.001&17.127&0.005&LT \\
57732.81&-&-&18.283&0.002&17.204&0.001&17.313&0.001&17.133&0.004&LT \\
57736.82&-&-&18.375&0.003&17.27&0.001&17.373&0.022&17.239&0.001&LT \\
57758.08&-&-&18.71&0.035&17.575&0.021&17.720&0.028&-&-&LCO (McDonald) \\
57759.06&-&-&-&-&17.589&0.023&17.686&0.02&-&-&LCO (McDonald) \\
57898.79&-&-&20.893&0.05&19.403&0.028&19.942&0.049&-&-&LCO (SSO) \\
57914.37&-&-&20.888&0.118&19.598&0.051&20.134&0.062&-&-&LCO (CTIO) \\
57995.92&-&-&21.597&0.034&20.661&0.021&21.324&0.030&-&-&LT \\

\hline
\multicolumn{12}{l}{*Denotes LCO data obtained independently of SNEx}\\
    \end{tabular}
    \label{tab:ugrizphot}
\end{table*}

\begin{table*}
	\caption{$UBVRI$ observations of SN 2016coi}
	\begin{tabular}{cccccccccccc}
 MJD & $U$ & $\delta U$& $B$ & $\delta B$& $V$ & $\delta V$& $R$ & $\delta R$& $I$ & $\delta I$ & Telescope \\
\hline
57536.98&-&-&15.958&0.073&15.314&0.044&14.875&0.036&14.912&0.024&Konkoly \\
57537.06&-&-&15.98&0.013&15.216&0.008&-&-&-&-&LCO (SAAO) \\
57539.12&-&-&15.501&0.010&14.661&0.017&-&-&-&-&LCO (SAAO) \\
57539.83&-&-&15.405&0.014&14.485&0.002&14.23&0.001&-&-&LJT \\
57540.02&-&-&15.31&0.095&14.506&0.051&14.159&0.042&14.256&0.021&Konkoly \\
57540.77&-&-&15.32&0.009&14.339&0.138&14.087&0.105&14.231&0.004&TNT \\
57541.75&14.9&0.021&15.15&0.005&14.16&0.004&13.92&0.005&14.077&0.006&TNT \\
57541.86&-&-&15.13&0.016&14.147&0.006&13.965&0.007&-&-&LJT \\
57542.12&-&-&15.076&0.004&14.13&0.003&-&-&-&-&LCO (SAAO) \\
57545.04&-&-&14.904&0.082&13.792&0.071&13.596&0.056&13.708&0.049&Konkoly \\
57545.10&-&-&14.907&0.005&13.876&0.006&-&-&-&-&LCO (SAAO) \\
57548.01&-&-&14.837&0.053&13.806&0.070&13.581&0.055&13.668&0.019&Konkoly \\
57549.72&-&-&14.859&0.004&13.698&0.003&13.447&0.003&13.572&0.004&TNT \\
57553.97&-&-&15.012&0.066&13.866&0.040&13.495&0.064&13.495&0.019&Konkoly \\
57557.15&-&-&15.287&0.006&13.893&0.003&-&-&-&-&LCO (SAAO) \\
57557.78&15.586&0.017&15.288&0.003&13.935&0.152&13.5&0.002&13.535&0.005&TNT \\
57558.04&-&-&15.454&0.117&13.954&0.084&13.505&0.063&13.569&0.080&Konkoly \\
57560.01&-&-&15.541&0.090&14.173&0.034&13.672&0.048&13.615&0.036&Konkoly \\
57561.04&-&-&15.656&0.182&14.199&0.041&13.656&0.066&13.553&0.039&Konkoly \\
57561.13&-&-&15.641&0.009&14.106&0.005&-&-&-&-&LCO (SAAO) \\
57562.05&-&-&15.775&0.104&14.266&0.089&13.816&0.102&13.573&0.047&Konkoly \\
57563.02&-&-&15.943&0.152&14.403&0.082&13.759&0.050&13.673&0.032&Konkoly \\
57564.15&-&-&15.903&0.015&14.299&0.006&-&-&-&-&LCO (SAAO) \\
57564.99&-&-&16.079&0.072&14.434&0.055&13.865&0.034&13.718&0.044&Konkoly \\
57568.14&-&-&16.174&0.011&14.552&0.005&-&-&-&-&LCO (SAAO) \\
57569.04&-&-&16.247&0.074&14.646&0.084&13.86&0.046&13.76&0.041&Konkoly \\
57570.98&-&-&16.237&0.138&14.727&0.065&13.835&0.078&13.773&0.065&Konkoly \\
57572.12&-&-&16.468&0.025&14.794&0.006&-&-&-&-&LCO (SAAO) \\
57573.98&-&-&16.457&0.054&14.919&0.071&14.171&0.031&13.999&0.027&Konkoly \\
57574.13&-&-&16.491&0.013&14.907&0.006&-&-&-&-&LCO (SAAO) \\
57574.70&17.065&0.046&16.424&0.008&14.935&0.005&14.172&0.003&13.993&0.004&TNT \\
57576.01&-&-&16.543&0.085&15.009&0.064&14.266&0.034&14.051&0.020&Konkoly \\
57576.40&-&-&16.588&0.009&15.004&0.004&-&-&-&-&LCO (CTIO) \\
57578.97&-&-&16.602&0.066&15.15&0.049&14.386&0.041&14.153&0.036&Konkoly \\
57579.98&-&-&16.629&0.097&15.178&0.049&14.472&0.066&14.186&0.034&Konkoly \\
57581.09&-&-&16.724&0.011&15.172&0.005&-&-&-&-&LCO (SAAO) \\
57587.11&-&-&16.901&0.141&15.396&0.015&-&-&-&-&LCO (SAAO) \\
57594.73&-&-&16.981&0.019&15.575&0.007&-&-&-&-&LCO (SSO) \\
57601.32&-&-&17.085&0.011&15.724&0.005&-&-&-&-&LCO (CTIO) \\
57602.67&-&-&-&-&15.77&1.776&15.055&0.003&-&-&LJT \\
57603.86&-&-&16.984&0.061&15.806&0.047&15.072&0.036&14.618&0.036&Konkoly \\
57603.90&-&-&17.069&0.091&15.795&0.053&15.071&0.036&14.647&0.028&Konkoly \\
57609.00&-&-&17.233&0.021&15.888&0.01&-&-&-&-&LCO (SAAO) \\
57613.00&-&-&17.139&0.065&15.953&0.064&15.27&0.037&14.783&0.021&Konkoly \\
57613.58&-&-&17.203&0.010&15.972&0.003&15.28&0.003&-&-&LJT \\
57617.75&-&-&17.09&0.040&16.058&0.014&15.275&0.058&-&-&LJT \\
57620.56&-&-&-&-&16.109&0.001&15.424&0.001&-&-&LJT \\
57621.01&-&-&17.205&0.061&16.022&0.032&-&-&-&-&LCO (SAAO) \\
57622.64&-&-&17.29&0.001&16.11&0.001&15.426&0.003&14.857&0.001&LJT \\
57623.68&-&-&17.302&0.012&16.118&0.003&15.444&0.001&-&-&LJT \\
57625.65&-&-&17.313&0.013&16.143&0.002&15.479&0.001&-&-&LJT \\
57625.90&-&-&17.238&0.085&16.165&0.041&15.501&0.046&15.003&0.025&Konkoly \\
57627.97&-&-&17.369&0.032&16.151&0.014&-&-&-&-&LCO (SAAO) \\
57631.88&-&-&17.329&0.073&16.234&0.049&15.625&0.054&15.088&0.040&Konkoly \\
57638.89&-&-&17.367&0.062&16.391&0.050&15.702&0.038&15.192&0.014&Konkoly \\
57642.94&-&-&17.565&0.041&16.387&0.016&-&-&-&-&LCO (SAAO) \\
57646.78&-&-&17.544&0.173&16.483&0.072&15.807&0.028&15.346&0.040&Konkoly \\
57649.92&-&-&17.435&0.083&16.503&0.031&-&-&-&-&LCO (SAAO) \\
57654.75&-&-&17.564&0.128&16.544&0.102&15.916&0.057&15.423&0.033&Konkoly \\
57656.89&-&-&17.724&0.027&16.595&0.011&-&-&-&-&LCO (SAAO) \\
\hline  
    \end{tabular}
    \label{tab:UBVRItable}
\end{table*}

\begin{table*}
	\contcaption{$UBVRI$ observations of SN 2016coi}
	\begin{tabular}{cccccccccccc}
 MJD & $U$ & $\delta U$& $B$ & $\delta B$& $V$ & $\delta V$& $R$ & $\delta R$& $I$ & $\delta I$ & Telescope \\
\hline
57660.81&-&-&17.639&0.048&16.709&0.066&16.031&0.034&15.577&0.028&Konkoly \\
57668.68&-&-&17.891&0.017&16.848&0.016&16.142&0.01&15.687&0.01&TNT \\
57669.53&-&-&17.805&0.012&16.807&0.01&16.11&0.007&15.661&0.007&TNT \\
57670.50&-&-&17.831&0.011&16.84&0.011&16.146&0.007&15.667&0.007&TNT \\
57671.53&-&-&-&-&16.916&0.029&16.143&0.016&15.683&0.012&TNT \\
57680.11&-&-&18.076&0.041&17.020&0.022&-&-&-&-&LCO (CTIO) \\
57683.46&-&-&18.023&0.015&17.04&0.01&-&-&-&-&LCO (SSO) \\
57683.72&-&-&17.935&0.088&17.109&0.051&16.316&0.030&15.948&0.027&Konkoly \\
57684.51&-&-&-&-&17.083&0.046&16.35&0.015&15.896&0.093&TNT \\
57685.64&-&-&18.074&0.028&17.133&0.018&16.369&0.009&15.978&0.128&TNT \\
57686.54&-&-&18.192&0.022&17.164&0.011&-&-&16&0.007&TNT \\
57687.57&-&-&-&-&17.184&0.096&-&-&-&-&TNT \\
57692.48&-&-&18.181&0.016&17.209&0.014&16.443&0.010&16.072&0.009&TNT \\
57696.52&-&-&18.196&0.021&17.273&0.01&-&-&16.107&0.008&TNT \\
57697.46&-&-&18.29&0.024&17.305&0.017&16.527&0.009&16.154&0.01&TNT \\
57699.46&-&-&-&-&17.305&0.035&16.489&0.019&16.093&0.026&TNT \\
57699.75&-&-&18.374&0.036&17.280&0.02&-&-&-&-&LCO (SAAO) \\
57700.49&-&-&-&-&17.362&0.013&16.536&0.007&16.2&0.009&TNT \\
57702.53&-&-&-&-&17.377&0.024&16.565&0.014&-&-&TNT \\
57703.56&-&-&18.407&0.004&17.435&0.002&16.587&0.003&-&-&LJT \\
57704.48&-&-&-&-&17.501&0.048&16.48&0.023&-&-&TNT \\
57706.42&-&-&18.41&0.041&17.449&0.025&16.6&0.013&20.350&0.014&TNT \\
57706.73&-&-&18.308&0.082&17.566&0.067&16.598&0.060&16.297&0.048&Konkoly \\
57707.44&-&-&-&-&17.456&0.036&16.624&0.014&-&-&TNT \\
57708.42&-&-&18.48&0.02&17.547&0.014&16.679&0.007&16.373&0.008&TNT \\
57711.52&-&-&18.530&0.040&17.647&0.309&16.687&0.018&16.395&0.019&TNT \\
57713.76&-&-&18.415&0.045&17.615&0.030&16.685&0.037&16.443&0.028&Konkoly \\
57714.45&-&-&18.271&0.362&17.452&0.066&16.679&0.058&16.403&0.034&TNT \\
57715.44&-&-&18.536&0.034&17.635&0.027&16.694&0.014&16.442&0.017&TNT \\
57716.02&-&-&-&-&17.596&0.036&-&-&-&-&LCO (CTIO) \\
57716.14&-&-&18.574&0.025&17.612&0.023&-&-&-&-&LCO (McDonald) \\
57716.56&-&-&-&-&17.677&0.016&16.721&0.009&16.463&0.010&TNT \\
57718.43&-&-&18.522&0.016&17.693&0.014&16.77&0.008&16.494&0.01&TNT \\
57719.49&-&-&18.543&0.023&17.697&0.018&16.806&0.008&16.525&0.019&TNT \\
57720.46&-&-&18.59&0.012&17.739&0.011&-&-&16.537&0.007&TNT \\
57722.46&-&-&18.641&0.020&17.765&0.017&16.834&0.007&-&-&TNT \\
57723.42&-&-&18.633&0.025&17.796&0.023&16.854&0.014&16.587&0.016&TNT \\
57725.67&-&-&18.668&0.108&17.88&0.052&16.834&0.065&16.625&0.035&Konkoly \\
57732.13&-&-&18.836&0.069&17.969&0.047&-&-&-&-&LCO (McDonald) \\
57738.75&-&-&18.887&0.100&18.193&0.061&17.047&0.047&16.906&0.028&Konkoly \\
57744.70&-&-&18.942&0.056&18.333&0.049&17.175&0.037&16.95&0.031&Konkoly \\
57759.06&-&-&19.092&0.051&18.462&0.049&-&-&-&-&LCO (McDonald) \\
57774.70&-&-&19.294&0.158&18.97&0.106&17.582&0.070&17.435&0.062&Konkoly \\

\hline
  
    \end{tabular}
\end{table*}

\begin{table*}
	\caption{DLT40 open filter (scaled to $r$) observations of SN 2016coi}
	\begin{tabular}{cccc}
 MJD & Open($r$) & $\delta$ Open($r$)& Telescope \\
\hline

57555.39&13.822&0.061&DLT40 \\
57557.38&13.879&0.090&DLT40 \\
57558.38&13.917&0.077&DLT40 \\
57559.37&13.937&0.069&DLT40 \\
57561.42&14.052&0.040&DLT40 \\
57562.38&14.089&0.093&DLT40 \\
57566.39&14.269&0.048&DLT40 \\
57568.38&14.364&0.055&DLT40 \\
57569.41&14.389&0.042&DLT40 \\
57575.34&14.612&0.051&DLT40 \\
57576.38&14.654&0.058&DLT40 \\
57577.35&14.730&0.055&DLT40 \\
57584.30&14.911&0.067&DLT40 \\
57585.32&14.967&0.079&DLT40 \\
57589.40&15.062&0.067&DLT40 \\
57590.29&15.081&0.076&DLT40 \\
57591.29&15.095&0.122&DLT40 \\
57592.38&15.116&0.064&DLT40 \\
57595.32&15.182&0.049&DLT40 \\
57597.27&15.255&0.061&DLT40 \\
57598.31&15.265&0.084&DLT40 \\
57599.26&15.295&0.066&DLT40 \\
57600.29&15.315&0.046&DLT40 \\
57601.25&15.345&0.089&DLT40 \\
57602.25&15.360&0.064&DLT40 \\
57603.25&15.357&0.059&DLT40 \\
57605.24&15.414&0.074&DLT40 \\
57606.24&15.424&0.041&DLT40 \\
57607.23&15.433&0.063&DLT40 \\
57608.23&15.466&0.064&DLT40 \\
57609.23&15.468&0.068&DLT40 \\
57610.23&15.494&0.088&DLT40 \\
57612.22&15.529&0.061&DLT40 \\
57611.22&15.491&0.057&DLT40 \\
57613.22&15.539&0.080&DLT40 \\
57614.22&15.555&0.053&DLT40 \\
57615.21&15.550&0.066&DLT40 \\
57616.21&15.571&0.065&DLT40 \\
57617.12&15.596&0.063&DLT40 \\
57618.21&15.620&0.074&DLT40 \\
57619.20&15.597&0.090&DLT40 \\
57620.20&15.534&0.067&DLT40 \\
57621.20&15.653&0.071&DLT40 \\
57622.2&15.624&0.054&DLT40 \\
57625.27&15.716&0.087&DLT40 \\
57626.25&15.738&0.055&DLT40 \\
57629.21&15.733&0.042&DLT40 \\
57630.21&15.792&0.059&DLT40 \\
57631.20&15.791&0.068&DLT40 \\
57632.21&15.857&0.095&DLT40 \\
57634.12&15.829&0.070&DLT40 \\
57635.12&15.853&0.072&DLT40 \\
57637.23&15.890&0.059&DLT40 \\
57638.06&15.914&0.052&DLT40 \\
57639.06&15.925&0.083&DLT40 \\
57640.06&15.924&0.056&DLT40 \\
57642.06&15.969&0.067&DLT40 \\

\hline
  
    \end{tabular}
    \label{tab:DLT40table}
\end{table*}

\begin{table*}
	\contcaption{DLT40 open filter (scaled to $r$) observations of SN 2016coi}
	\begin{tabular}{cccc}
 MJD & Open($r$) & $\delta$ Open($r$)& Telescope \\
\hline
57643.05&15.976&0.051&DLT40 \\
57644.03&16.008&0.115&DLT40 \\
57645.08&16.078&0.087&DLT40 \\
57647.03&16.071&0.054&DLT40 \\
57648.02&16.081&0.046&DLT40 \\
57649.01&16.055&0.039&DLT40 \\
57650.11&16.053&0.032&DLT40 \\
57651.02&16.037&0.048&DLT40 \\
57652.02&16.073&0.041&DLT40 \\
57653.04&16.110&0.050&DLT40 \\
57654.03&16.145&0.049&DLT40 \\
57655.03&16.178&0.052&DLT40 \\
57656.03&16.170&0.051&DLT40 \\
57657.02&16.188&0.033&DLT40 \\
57658.03&16.255&0.046&DLT40 \\
57659.11&16.269&0.045&DLT40 \\
57660.07&16.169&0.040&DLT40 \\
57661.02&16.252&0.046&DLT40 \\
57662.02&16.302&0.047&DLT40 \\
57663.02&16.287&0.053&DLT40 \\
57664.02&16.292&0.041&DLT40 \\
57666.02&16.337&0.036&DLT40 \\
57667.02&16.304&0.044&DLT40 \\
57668.02&16.329&0.047&DLT40 \\
57669.02&16.406&0.038&DLT40 \\
57670.08&16.330&0.054&DLT40 \\
57671.01&16.453&0.039&DLT40 \\
57672.03&16.445&0.039&DLT40 \\
57673.01&16.449&0.055&DLT40 \\
57675.01&16.452&0.047&DLT40 \\
57676.01&16.435&0.051&DLT40 \\

\hline
  
    \end{tabular}
\end{table*}
%


\bsp	
\label{lastpage}
\end{document}